\documentclass[preprint,12pt]{elsarticle}

\usepackage{subcaption}   
\usepackage{siunitx}
\usepackage{amssymb}
\usepackage{amsmath}
\usepackage{bm}
\usepackage{tabularx} 
\usepackage{booktabs}

\usepackage[section]{placeins}
\usepackage{caption}
\usepackage{graphicx}
\usepackage{makecell}

\begin{document}

\begin{frontmatter}



\title{A differentiable PCA–GPR surrogate framework for full-transient Bayesian inverse uncertainty quantification of black-box nuclear thermal–hydraulic systems }


\author[uw]{Zikang Yao\corref{cor1}}
\author[uw]{Eugene Yee}
\author[uw]{Fue-Sang Lien}
\cortext[cor1]{Corresponding author. Email: z24yao@uwaterloo.ca}

\affiliation[uw]{
    organization={Department of Mechanical and Mechatronics Engineering, University of Waterloo},
    addressline={200 University Avenue West},
    city={Waterloo},
    postcode={N2L 3G1},
    state={Ontario},
    country={Canada}
}

\begin{abstract}

Complex transient energy systems require uncertainty-quantified predictions, but Bayesian inverse uncertainty quantification (IUQ) is often impractical because posterior sampling with high-fidelity system codes requires many expensive full-transient simulations. This study develops a surrogate-assisted Bayesian IUQ framework that makes full-time-series calibration practical for the NACIE heavy-liquid-metal natural-circulation loop. The key innovation is to transform a black-box transient simulator into a differentiable probabilistic emulator by combining principal component analysis with Gaussian process regression (PCA--GPR). The surrogate reconstructs complete multichannel temperature and mass-flow histories from a compact latent representation, preserving dominant temporal and cross-channel correlations while enabling automatic differentiation of the log posterior. This allows both gradient-free ensemble slice sampling and gradient-informed sampling with the No-U-Turn Sampler and Microcanonical Langevin Monte Carlo. Five principal components capture more than 99\% of the output variance, and the surrogate achieves normalized root-mean-square errors below 2\%. Compared with direct Flownex SE evaluations, the surrogate reduces likelihood-evaluation cost by approximately 3000-fold while preserving posterior statistics and correlations. Posterior predictive analyses show reduced bias, narrower intervals, and improved agreement with measurements and independent validation.

\end{abstract}




\end{frontmatter}



\section{Introduction}

The global transition toward low-carbon energy systems has renewed the strategic importance of nuclear power as a reliable source of dispatchable electricity. For nuclear technologies to contribute safely and economically to this transition, reactor analysis tools must provide not only best-estimate predictions, but also defensible quantification of the uncertainties associated with those predictions. This requirement is particularly stringent for transient thermal--hydraulic (TH) phenomena, where coupled interactions among power input, flow resistance, heat transfer, structural thermal inertia, and measurement uncertainty can strongly influence safety-relevant system responses.

Modern nuclear safety analysis has therefore moved beyond purely conservative evaluation models toward best-estimate simulations supplemented by quantitative uncertainty assessment. This evolution is embodied in the Best Estimate Plus Uncertainty (BEPU) framework, which has become a central paradigm for risk-informed reactor safety assessment~\cite{dauria2012be,radaideh2023integrated}. In contrast to traditional conservative methodologies, which intentionally introduced pessimistic assumptions to preserve safety margins, BEPU seeks to retain physical realism while explicitly characterizing the uncertainty in model predictions~\cite{pourgol2010methodology}. Regulatory methodologies such as the Code Scaling, Applicability, and Uncertainty (CSAU) approach and the Evaluation Model Development and Assessment Procedure (EMDAP) formalized this philosophy by linking simulation credibility to experimental validation, model assessment, and uncertainty propagation~\cite{boyack1990quantifying}. Such statistically grounded approaches are especially important for advanced reactor concepts, for which limited operational experience and innovative design features require transparent, evidence-based safety arguments.

Forward uncertainty quantification (FUQ) is the most common route for propagating prescribed uncertainties in model inputs through a computational reactor model to obtain uncertainty bands for quantities of interest~\cite{oberkampf2010verification}. Although FUQ is indispensable for sensitivity analysis and margin evaluation, its reliability depends directly on the assumed prior distributions of uncertain parameters. In many TH applications, these distributions are inferred from sparse experiments, expert judgment, or engineering correlations whose applicability to integral transients may be uncertain~\cite{skorek2019premium}. Inverse uncertainty quantification (IUQ) addresses this limitation by using experimental observations to infer the posterior distribution of uncertain model parameters~\cite{wu2021survey}\cite{marrel2024probabilistic}\cite{kakhaia2023inverse}. Through Bayesian updating, IUQ provides a principled mechanism for reducing epistemic uncertainty, correcting systematic model bias, and improving the predictive credibility of reactor simulations.

Despite substantial methodological progress, most Bayesian IUQ and related
uncertainty-assessment studies in nuclear TH have remained confined to
closure relations, isolated physical processes, or steady-state subchannel
and bundle-scale benchmarks, such as BFBT/PSBT void-fraction data~\cite{radaideh2019integrated}. Separate-effect tests (SETs) are attractive
for calibration because they isolate dominant mechanisms and simplify
parameter identifiability~\cite{damblin2020bayesian,Perret03042022}.However, reactor transients of practical relevance are governed by strongly coupled phenomena distributed across multiple components, and the dominant uncertainty mechanisms cannot always be separated cleanly. This has motivated increasing use of combined-effect tests (CETs) and integral-effect tests (IETs), which better preserve system-level feedback mechanisms~\cite{baccou2019sapium}. International initiatives such as PREMIUM and SAPIUM have clarified both the promise and the difficulty of IUQ for TH codes, including the sensitivity of inferred results to user choices, calibration strategies, and numerical implementations~\cite{baccou2019sapium}. These findings underscore the need for IUQ frameworks that are robust, scalable, and suitable for integral system transients.

The Natural Circulation Experiment (NACIE) facility provides a particularly valuable benchmark for this purpose because it combines heavy liquid metal (HLM) natural circulation, rod-bundle heating, heat-exchanger dynamics, and transient loop feedback within a single experimental platform~\cite{coccoluto2011nacie}. Previous uncertainty studies on NACIE have demonstrated the relevance of system-scale analysis, but the inverse calibration of full transient NACIE responses remains considerably more demanding than conventional FUQ or steady-state calibration,especially when the observations are time-series outputs~\cite{perrin2020adaptive}. The challenge is amplified by the widespread use of commercial or proprietary system TH codes, such as Flownex SE~\cite{flownex2023}, RELAP5~\cite{relap5mod3}, TRACE~\cite{tracev5}, and ANSYS~\cite{ansysfluent2023}, which are effectively black-box simulators for most users: their internal residuals, source code, and differentiable solution operators are generally inaccessible. Direct Bayesian calibration of such models requires thousands to tens of thousands of model evaluations, making full-transient IUQ computationally prohibitive without additional acceleration~\cite{ohagan2006bayesian}. Surrogate-assisted inversion has been adopted in other engineering domains 
to reduce the computational cost of static and dynamic parameter 
identification for expensive numerical models~\cite{Li2023AEI}.

The present study is motivated by, and directly extends, the recent work of Yao et al.~\cite{YAO2026140531}, which introduced a Bayesian optimization (BO)-assisted IUQ framework for expensive black-box nuclear TH simulations. That earlier study demonstrated, for the first time for NACIE, that system-level Bayesian IUQ could be performed by combining BO-based informed initialization with efficient gradient-free ensemble MCMC sampling, particularly the Ensemble Slice Sampling Scheme (ENSS). It established the feasibility of calibrating uncertain physical parameters in a high-fidelity Flownex SE model and showed that posterior calibration can reduce predictive uncertainty and improve agreement with experimental responses. However, the central computational bottleneck remained unresolved: even with improved initialization and efficient gradient-free sampling, each posterior evaluation still required a high-fidelity transient Flownex SE simulation. Consequently, the earlier framework was constrained by the cost of repeated black-box model evaluations and by its reliance on gradient-free sampling strategies.

This paper generalizes that prior BO--ENSS framework by replacing repeated high-fidelity simulator calls with a rigorously validated surrogate representation of the complete transient system response. Specifically, we develop a principal component analysis--Gaussian process regression (PCA--GPR) surrogate for the NACIE facility. Rather than constructing independent emulators for individual scalar outputs or isolated time instants, the proposed approach first compresses the full multichannel time-series response into a low-dimensional set of orthogonal principal components and then learns the parameter-to-component mapping using Gaussian process regression. The trained surrogate reconstructs the complete spatiotemporal response of the system, preserving the dominant temporal and inter-channel correlations that are essential for full-transient IUQ.

This surrogate construction addresses a fundamental limitation of 
conventional emulation strategies for transient TH analysis. Direct 
Gaussian process modeling of every time step and every measurement channel 
is computationally inefficient and can become numerically ill-conditioned 
because transient responses are high-dimensional and strongly correlated. 
Nevertheless, simulation-driven GPR has demonstrated its effectiveness for 
learning nonlinear engineering-system behavior and supporting data-driven 
condition diagnosis~\cite{Jankovic2024AEI}. Non-intrusive polynomial chaos 
expansions and stochastic collocation methods have also been applied to 
propagate parametric uncertainty through complex CFD models
~\cite{xiu2002wiener,xiu2005high,Zhang2025AEI}. Although effective for 
forward uncertainty propagation, extending these approaches to complete 
multichannel transient responses can require constructing and evaluating 
a rapidly growing set of output-specific, time-dependent coefficients.

Neural-network-based surrogates~\cite{hornik1989multilayer} provide an 
alternative representation of nonlinear transient responses, but they 
often require comparatively large training datasets, specialized 
architectures, and careful regularization to achieve robust generalization. 
Bayesian physics-informed neural networks coupled with iterative ensemble 
Kalman inversion have recently enabled joint prediction and uncertainty 
quantification, including validation using nuclear circulating water pump 
data~\cite{Xie2026AEI}. Such physics-informed approaches are attractive 
when governing equations and physical residuals can be explicitly embedded 
in the learning process. In the present application, however, Flownex SE is 
treated as a commercial black-box simulator whose internal residuals and 
differentiable solution operators are inaccessible.

These considerations are particularly important in nuclear safety 
applications, where high-fidelity training data are expensive and model 
interpretability, uncertainty quantification, and traceability are 
essential. PCA provides a transparent dimensionality-reduction basis 
ordered by explained variance~\cite{shlens2014pca}, while GPR provides a 
probabilistic and differentiable mapping between uncertain model parameters 
and the retained latent response coordinates. Their combination has been 
shown to be effective for reducing the dimensionality of complex simulation 
outputs and accelerating reliability-oriented uncertainty analyses
~\cite{wu2018bison,wang2018bepu,marrel2024probabilistic,lu2024amfgp}. 
Related engineering-informatics studies have further demonstrated the value 
of distinguishing aleatory and epistemic contributions in Gaussian-process 
predictions~\cite{Liang2024AEI} and incorporating surrogate uncertainty 
into computationally intensive constrained engineering design
~\cite{Liu2025AEI}. In the present work, PCA--GPR is used not merely as a 
response-surface approximation, but as the computational core of a full 
Bayesian IUQ framework for system-level transient calibration.

The resulting framework advances the state of the art in three respects. First, it extends black-box system-level IUQ from direct high-fidelity sampling to surrogate-accelerated inference while retaining reconstruction of the complete transient response rather than only selected scalar metrics. Second, the PCA--GPR surrogate acts as an inexpensive-to-evaluate white-box emulator of the transient simulator: once trained, its algebraic operations are explicit and differentiable, so derivatives of the log-posterior with respect to uncertain parameters can be obtained efficiently using automatic differentiation. This capability is unavailable when the likelihood is evaluated directly through black-box simulators such as Flownex SE, for which internal residuals, solver states, and exact sensitivities are not exposed to the user. The white-box surrogate therefore enables gradient-informed posterior sampling strategies, such as the No-U-Turn Sampler (NUTS) and Microcanonical Langevin Monte Carlo (MCLMC), to exploit local geometric information in the posterior distribution rather than relying only on random-walk or ensemble proposals. This improves exploration of correlated posterior regions, reduces autocorrelation, increases effective sample generation, and makes full-transient Bayesian calibration substantially more computationally efficient. Third, the framework provides a generalizable methodology for integrating dimensionality reduction, probabilistic surrogate modeling, Bayesian calibration, automatic differentiation, and posterior predictive validation in complex reactor TH systems. In this sense, the present contribution is not a repetition of the earlier NACIE IUQ study~\cite{YAO2026140531}; it removes the dominant computational barrier identified in that work and broadens the framework from efficient black-box calibration to scalable, surrogate-enabled full-transient probabilistic inference.

The main original contributions of this work are summarized as follows:
\begin{itemize}
    \item A PCA--GPR surrogate is developed to reconstruct the full multichannel NACIE transient response from a compact set of principal components, preserving dominant spatiotemporal correlations while reducing the cost of model evaluation by orders of magnitude.
    \item The surrogate is embedded within a Bayesian IUQ framework to calibrate key uncertain physical parameters against experimental time-series data from the NACIE facility, thereby enabling full-transient posterior inference at system scale.
    \item The framework generalizes the BO--ENSS black-box IUQ strategy of Yao et al.~\cite{YAO2026140531} by replacing repeated high-fidelity simulator calls with an inexpensive white-box surrogate that supports automatic differentiation and thereby enables both gradient-free and gradient-informed posterior samplers.
    \item Posterior predictive validation is performed using calibrated parameter distributions propagated through both the PCA--GPR surrogate and the reference Flownex SE model, allowing direct assessment of surrogate fidelity, uncertainty reduction, and generalization to responses not used directly in calibration.
\end{itemize}

The remainder of this paper is organized to follow the complete surrogate-assisted IUQ workflow. Section~\ref{sec:nacie} introduces the NACIE experimental facility, the corresponding Flownex SE system model, the uncertain calibration parameters, and the transient and steady-state data used for calibration and validation. Section~\ref{sec:methodology} develops the PCA--GPR surrogate, including the dimensionality-reduction and Gaussian-process regression steps used to emulate the full transient response. Section~\ref{sec:uq-framework} formulates the Bayesian uncertainty-quantification problem, including prior propagation, likelihood construction, inverse updating, and posterior predictive analysis. Section~\ref{sec:computational-methods} then describes the computational inference strategy, with emphasis on Bayes\-ian-optimization-based initialization, ensemble slice sampling, NUTS, MC\-LMC, and automatic differentiation of the surrogate posterior. Section~\ref{sec:results} reports the numerical results, progressing from surrogate validation to posterior consistency, sampler-efficiency comparisons, data-size effects, and prior/pos\-ter\-ior predictive validation. Finally, Section~\ref{sec:conclusion} summarizes the main conclusions and discusses implications for surrogate-accelerated Bayesian calibration in reactor safety analysis.

\section{NACIE Benchmark}
\label{sec:nacie}

\subsection{NACIE Facility and Natural-Circulation Benchmark}
\label{sec:nacie-overview}

\begin{figure}[t!]
\centering
\includegraphics[width=\linewidth]{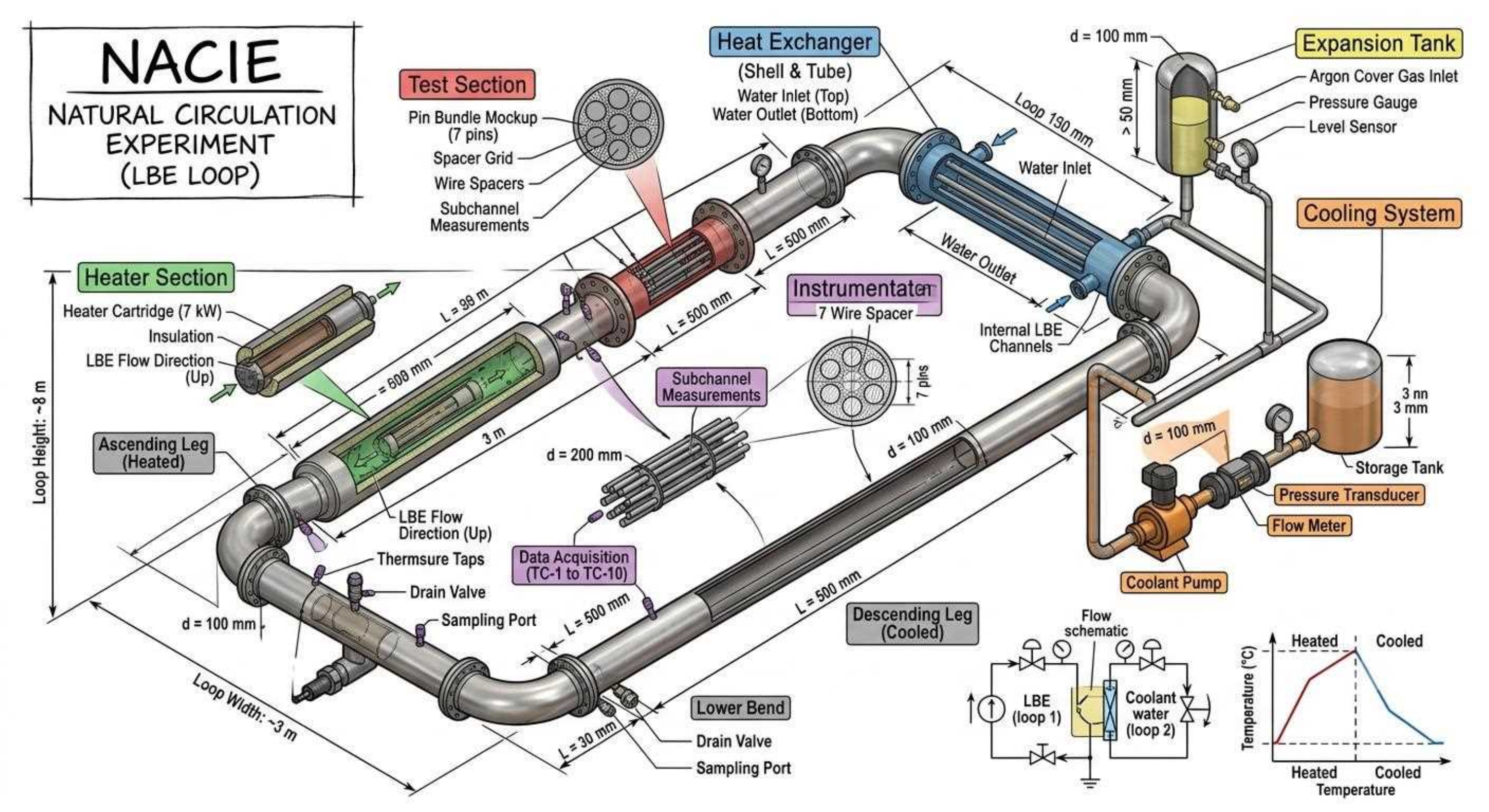}
\caption{Schematic layout of the NACIE integral-effect natural-circulation loop used as the experimental benchmark in this study. The facility couples electrical heat addition, buoyancy-driven LBE circulation, heat removal through the heat exchanger, and distributed hydraulic resistance, thereby providing system-level transient responses for Bayesian calibration and uncertainty quantification. Graphic produced from information in~\cite{coccoluto2011nacie}.}
\label{fig:nacie_layout}
\end{figure}

The Natural Circulation Experiment (NACIE) facility, developed at the ENEA Brasimone Research Centre, is an integral-effect experimental platform for studying thermal--hydraulic behavior in heavy-liquid-metal-cooled nuclear systems~\cite{coccoluto2011nacie}. As shown schematically in Fig.~\ref{fig:nacie_layout}, NACIE is designed to reproduce the coupled buoyancy, heat-addition, heat-removal, and loop-inertia mechanisms that govern natural circulation in lead-cooled fast reactors (LFRs) and accelerator-driven systems (ADS). The facility is therefore well suited for system-level uncertainty quantification because its measured responses arise from interactions among multiple components rather than from an isolated closure relation or a single separate-effect phenomenon.

The loop has a rectangular configuration comprising vertical riser and downcomer branches connected by upper and lower horizontal sections. This arrangement provides a controlled representation of the main circulation path in pool-type heavy-liquid-metal systems while preserving the essential elevation difference needed to generate buoyancy-driven flow. The vertical legs have an internal diameter of 62.7~mm, and the total vertical extent of the loop is approximately 9.2~m. These geometric characteristics establish a sufficiently large hydrostatic driving head and thermal inventory for observing slow natural-circulation transients with reactor-relevant time scales.

\begin{figure}[t!]
\centering
\includegraphics[width=0.5\linewidth]{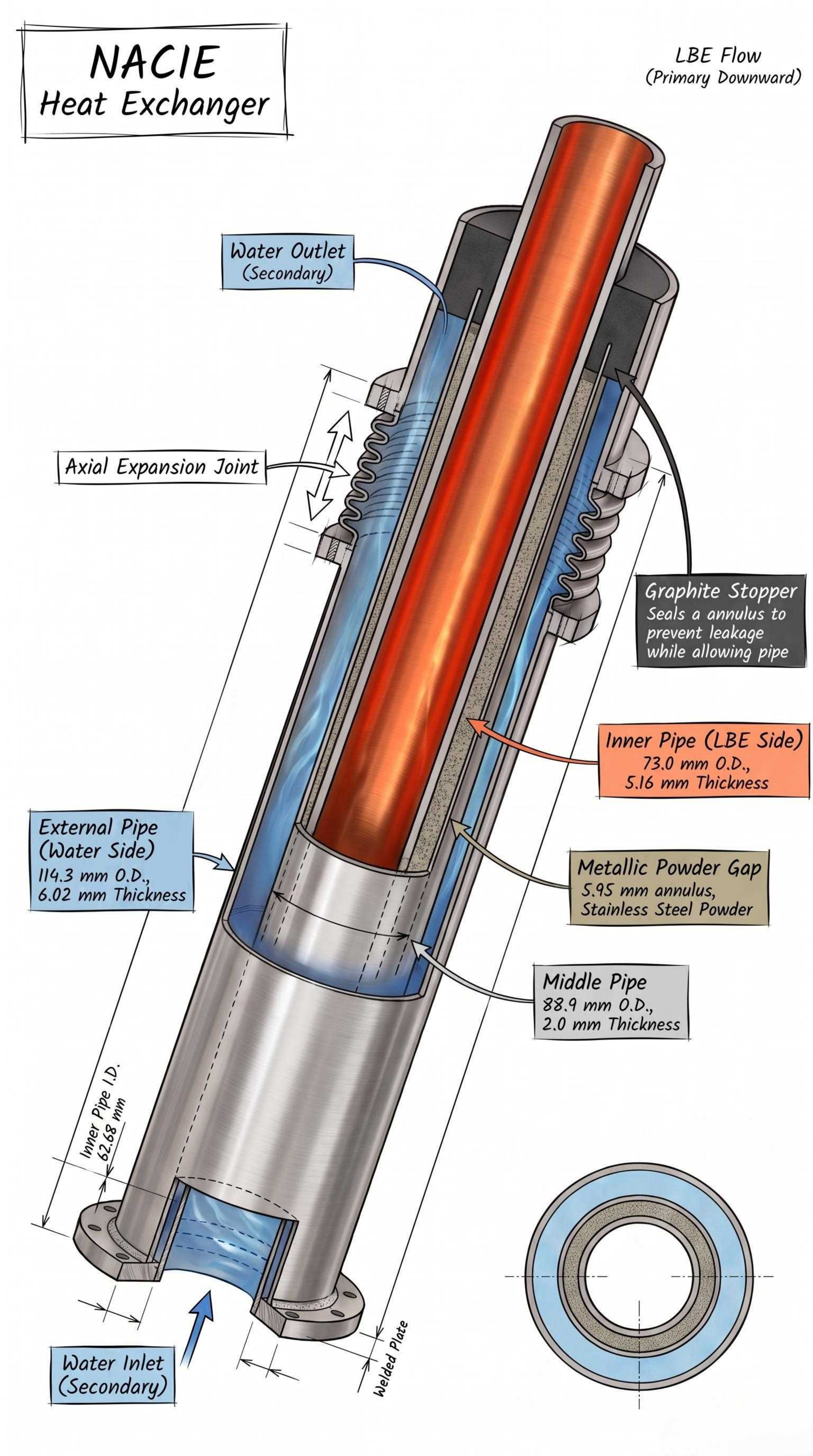}
\caption{Heat-exchanger configuration in the NACIE facility. The heat exchanger provides the primary heat-removal boundary for the natural-circulation loop and couples the LBE-side thermal response to the secondary cooling-water flow, making its inlet and outlet temperatures key observables for transient calibration and posterior predictive validation. Graphic produced from information in~\cite{coccoluto2011nacie}.}
\label{fig:nacie_hx}
\end{figure}


Thermal power is supplied by electrically heated fuel-pin simulators located near the lower part of the riser. Heat removal is provided by a counterflow tube-in-tube heat exchanger located in the upper downcomer, as illustrated in Fig.~\ref{fig:nacie_hx}. Lead--bismuth eutectic (LBE) is used as the primary working fluid, with an inventory of approximately 1000~kg. The large LBE inventory gives the loop substantial thermal inertia, which is essential for assessing transient heat-up, cooldown, and post-shutdown natural-circulation behavior.

The transient considered in this study follows a controlled power-history representative of natural-circulation operation and shutdown. The heater power is initiated at approximately 100~s and increased to a nominal thermal power of about 25~kW. The powered phase is followed by an intentional shutdown at approximately 4400~s, representing a scram-like reduction in heat input. During the test, the inlet temperature of the secondary cooling water is maintained close to \SI{80}{\celsius}, so that the measured primary-side response is governed primarily by the coupled evolution of LBE circulation, heater heat addition, heat-exchanger removal, and distributed heat losses.

The principal measured quantities used for model assessment are the LBE mass flow rate and the temperature histories at selected locations along the circulation path. Temperature measurements are obtained using type-K thermocouples, for which the relevant tolerance is approximately 0.75\% under the present operating range~\cite{IEC60584-2}. A direct LBE flow-rate measurement was not available in the experiment; consequently, the mass flow rate is inferred from an energy balance over the heater section. This indirect evaluation makes the mass-flow response particularly useful for validation, but also introduces additional uncertainty because it depends on heater power, temperature differences, and heat-loss assumptions.

The experimental uncertainty budget is therefore dominated by both instrumentation limitations and operating-condition variability. The main reported uncertainty sources are thermocouple tolerance, heater-power fluctuations of approximately $\pm5$\%, secondary-side water-flow regulation uncertainty of approximately $\pm1$\%, and an estimated LBE mass-flow uncertainty of approximately $\pm5$\% associated with the indirect energy-balance reconstruction~\cite{coccoluto2011nacie}. Additional epistemic uncertainty arises from structural heat storage, distributed ambient heat losses, and imperfect knowledge of heat-transfer properties in the heat-exchanger region. These features make NACIE a stringent benchmark for inverse uncertainty quantification: the calibration data are transient, mutually correlated, and affected by both measurement uncertainty and model-form uncertainty.

\subsection{Uncertain Model Physical Parameters}
\label{sec:nacie-parameters}

The calibration problem is formulated in terms of four dominant uncertain parameters identified in the previous NACIE IUQ study~\cite{YAO2026140531}: the heater power bias $P_b$, the hydraulic resistance coefficient $C_k$, the ambient heat transfer coefficient $h_a$, and the powder-gap thermal conductivity $\lambda_p$. These parameters are selected because they directly control the principal mechanisms governing the transient response. The heater power bias modifies the effective heat input to the primary loop; the hydraulic resistance coefficient controls the pressure-loss balance and therefore the natural-circulation mass flow rate; the ambient heat transfer coefficient represents distributed thermal losses to the surroundings; and the powder-gap thermal conductivity affects heat transfer through the heat-exchanger structure. Together, these parameters define a compact but physically interpretable calibration space for assessing the ability of the PCA--GPR-assisted Bayesian framework to infer system-level thermal--hydraulic uncertainties from transient NACIE data.

\section{PCA--GPR Surrogate Construction for Transient Responses}
\label{sec:methodology}

\begin{figure}[t!]
\centering
\includegraphics[width=0.85\linewidth]{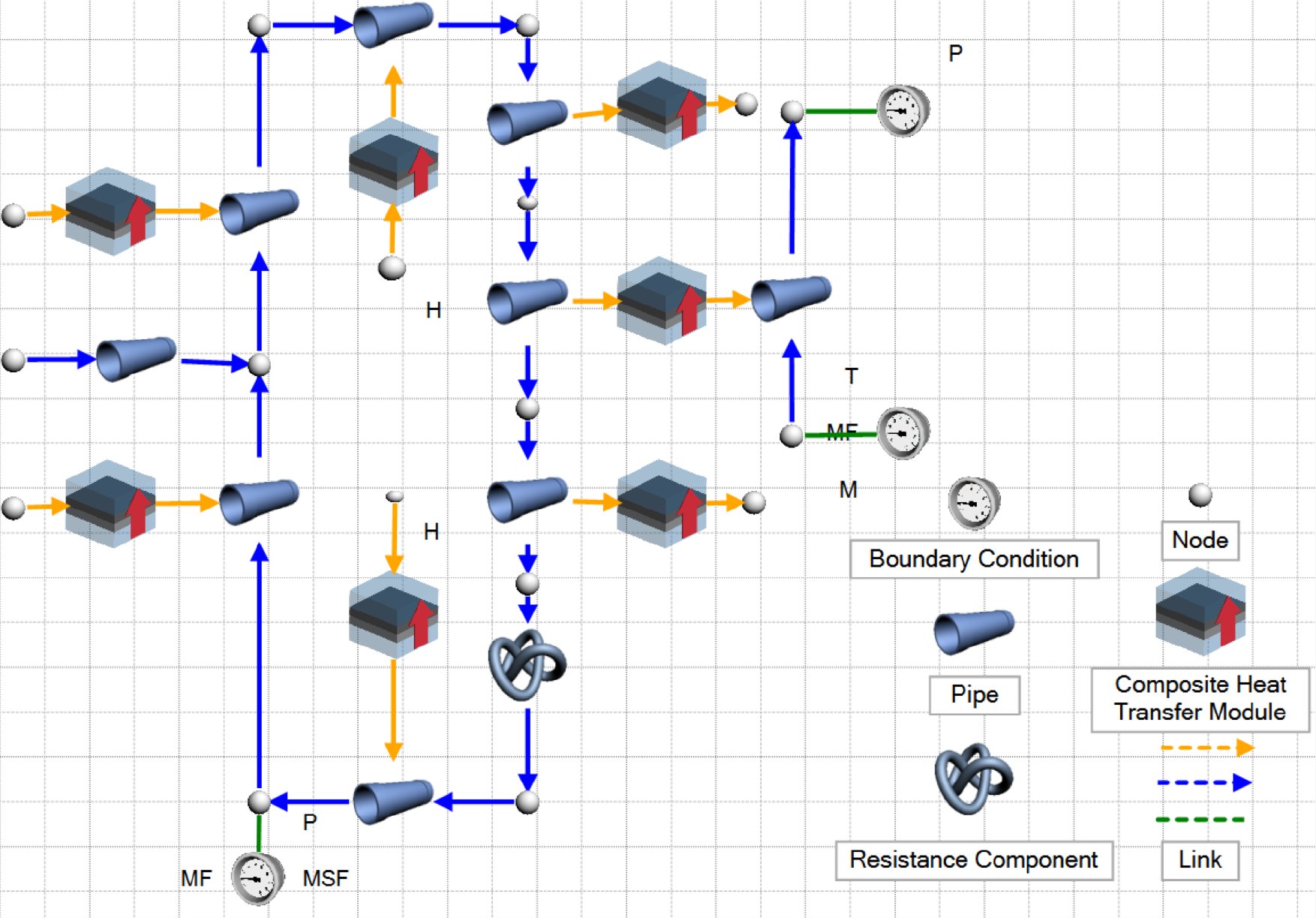}
\caption{Flownex SE representation of the NACIE loop used as the high-fidelity thermal--hydraulic reference model. The network model resolves the coupled circulation path, heat-source section, heat-exchanger section, hydraulic-resistance elements, and measurement locations used to generate transient training data, evaluate prior predictions, and define the reference likelihood for Bayesian inverse uncertainty quantification.}
\label{fig:FLOWNEX-MODEL}
\end{figure}

As shown in Fig.~\ref{fig:FLOWNEX-MODEL}, the NACIE loop is represented in Flownex SE~\cite{flownex2023} as a one-dimensional system-level thermal--hydraulic network composed of pipe-flow elements, heat-transfer components, heat-source and heat-sink modules, and hydraulic resistance elements. This network formulation solves the coupled conservation equations for mass, momentum, and energy across the circulation loop, thereby providing a physics-based reference model for the transient evolution of LBE flow rate and temperature. In the present study, Flownex SE is used as the high-fidelity simulator from which training and validation data are generated for the PCA--GPR surrogate. Although the code provides a detailed component-based representation of the NACIE facility, it is treated here as an expensive-to-evaluate black-box model because the internal solver states, residual equations, and exact parameter sensitivities are not directly accessible during Bayesian calibration. Consequently, repeated direct evaluations of the Flownex SE model are computationally burdensome for posterior sampling, motivating the surrogate-assisted framework developed in this work.

Although the Flownex SE model provides physically detailed predictions based on established thermal--hydraulic conservation principles and component-level constitutive models, its computational cost becomes a major limitation in probabilistic calibration. For the NACIE configuration considered here, a single full-transient simulation requires approximately 6--15 minutes, depending on the sampled parameter values and the numerical stiffness of the resulting transient. This cost is manageable for deterministic analysis or a limited number of validation calculations, but it becomes prohibitive in Bayesian IUQ, where posterior exploration typically requires thousands to tens of thousands of likelihood evaluations. Moreover, because Flownex SE is used as a black-box executable, gradient information required by efficient Hamiltonian or Langevin-type samplers cannot be obtained directly from the governing residuals. The computational burden therefore has two coupled consequences: it limits the number of high-fidelity model evaluations that can be afforded, and it restricts direct inference to gradient-free sampling strategies with comparatively slower posterior exploration. These considerations motivate the construction of a surrogate model that preserves the dominant transient response characteristics of the Flownex SE simulator while reducing the cost of each likelihood evaluation by several orders of magnitude.

\subsection{Principal Component Analysis for Dimensionality Reduction}
\label{sec:pca-dimensionality-reduction}

The high-fidelity NACIE simulations generate multivariate transient responses rather than isolated scalar quantities of interest. Each simulation produces temperature histories and mass-flow responses at multiple measurement locations, and these responses are strongly correlated in time and across output channels because they are governed by the same loop-scale conservation laws and thermal inertia. Directly constructing a surrogate for every output value at every time instant would therefore be inefficient and statistically fragile: the output dimension is large, adjacent time points are highly redundant, and different channels have different physical units and characteristic magnitudes. Principal component analysis (PCA) is used here to exploit this correlation structure by identifying a low-dimensional orthogonal basis that captures the dominant modes of variation in the full transient response.

Let \(N\) denote the number of high-fidelity simulation runs used for surrogate training, \(d\) the number of uncertain input parameters, \(M\) the number of output channels, and \(T\) the number of time points retained for each channel. The dimension of the concatenated transient response is then \(p=MT\). For the \(i\)-th simulation run, the uncertain input vector is written as
\[
\boldsymbol{\theta}_i
= \left(\theta_{i,1},\theta_{i,2},\ldots,\theta_{i,d} \right)^{\top} \in \mathbb{R}^{d},
\qquad i=1,\ldots,N .
\]
In this study, the training inputs are generated using a Sobol low-discrepancy design over the prescribed prior bounds of the uncertain parameters. Compared with independent pseudo-random sampling, the Sobol design provides more uniform coverage of the admissible parameter domain, which is important for learning a stable global surrogate from a limited number of expensive Flownex SE evaluations.

For a sampled input \(\boldsymbol{\theta}_i\), the simulator returns a multichannel time series. After ordering the response channel by channel and then in time, the complete output vector is written as
\begin{align*}
    \mathbf{y}_i 
    &=
    \big(
    y_{i,1}(t_1),\ldots,y_{i,1}(t_T),
    y_{i,2}(t_1),\ldots,y_{i,2}(t_T),
    \ldots,\\
    &\qquad
    y_{i,M}(t_1),\ldots,y_{i,M}(t_T)
    \big)^{\top}
    \in \mathbb{R}^{p}\ ,\qquad i=1,2,\ldots,N\ ,
\end{align*}
where \(y_{i,m}(t_j)\) denotes the value of the \(m\)-th output channel at time \(t_j\) for the \(i\)-th simulation. Stacking the responses from all training simulations gives the output data matrix
\[
\mathbf{Y}
=
\begin{bmatrix}
\mathbf{y}_1^{\top} \\
\mathbf{y}_2^{\top} \\
\vdots \\
\mathbf{y}_N^{\top}
\end{bmatrix}
\in \mathbb{R}^{N \times p}\ .
\]
Thus, each row of \(\mathbf{Y}\) represents one complete NACIE transient, while each column represents one specific channel--time output component across the training ensemble.

Prior to PCA, the output matrix is standardized to prevent variables with large numerical magnitudes or different engineering units from dominating the decomposition. The sample mean vector is
\[
\bar{\mathbf{y}} = \frac{1}{N}\sum_{i=1}^{N}\mathbf{y}_i \in \mathbb{R}^{p},
\]
and the component-wise sample standard deviation vector is
\[
\mathbf{s}
=
\left[
\frac{1}{N-1}
\sum_{i=1}^{N}
(\mathbf{y}_i-\bar{\mathbf{y}})\bigodot(\mathbf{y}_i-\bar{\mathbf{y}})
\right]^{1/2}
\in \mathbb{R}^{p},
\]
where \(\bigodot\) denotes the Hadamard product. Defining the diagonal scaling matrix \(\mathbf{S}=\mathrm{diag}(s_1,s_2,\ldots,s_p)\in\mathbb{R}^{p\times p}\), the standardized response matrix is
\[
\tilde{\mathbf{Y}}
=
\left(
\mathbf{Y}
-
\mathbf{1}_N\bar{\mathbf{y}}^{\top}
\right)
\mathbf{S}^{-1}
\in \mathbb{R}^{N\times p}\ ,
\]
where \(\mathbf{1}_N\in\mathbb{R}^{N}\) is a column vector of ones. This standardization makes the PCA basis reflect coherent variations in the transient response rather than differences in dimensional units or measurement scale.

PCA is then applied to \(\tilde{\mathbf{Y}}\) through the empirical covariance matrix
\[
\mathbf{C}
=
\frac{1}{N-1}
\tilde{\mathbf{Y}}^{\top}
\tilde{\mathbf{Y}}
\in \mathbb{R}^{p\times p}\ .
\]
The matrix \(\mathbf{C}\) encodes the covariance among all standardized channel--time components of the NACIE transient. Solving the eigenvalue problem
\[
\mathbf{C}\mathbf{W}=\mathbf{W}\boldsymbol{\Lambda}
\]
yields the orthonormal eigenvector matrix \(\mathbf{W}=(\mathbf{w}_1,\mathbf{w}_2,\ldots,\mathbf{w}_p)\in\mathbb{R}^{p\times p}\) and the diagonal eigenvalue matrix \(\boldsymbol{\Lambda}=\mathrm{diag}(\lambda_1,\lambda_2,\ldots,\lambda_p)\in\mathbb{R}^{p\times p}\). The eigenvector \(\mathbf{w}_k\) defines the \(k\)-th principal direction in the standardized transient-output space, and the corresponding eigenvalue \(\lambda_k\) quantifies the variance represented by that direction. The eigenvalues are ordered as \(\lambda_1\geq\lambda_2\geq\cdots\geq\lambda_p\). Because the covariance matrix is estimated from \(N\) mean-centered simulations, no more than \(N-1\) principal directions can have nonzero variance in practice.

Dimensionality reduction is achieved by retaining only the first \(r\) principal directions,
\[
\mathbf{W}_r = \left[\mathbf{w}_1,\mathbf{w}_2,\ldots,\mathbf{w}_r\right]
\in \mathbb{R}^{p\times r},
\qquad r\ll p .
\]
The retained dimension \(r\) is selected using the cumulative explained-variance criterion
\[
\frac{\sum_{k=1}^{r}\lambda_k}{\sum_{k=1}^{p}\lambda_k} \geq \eta,
\]
where \(\eta\) is a prescribed variance-retention threshold. In the present application, choosing a high value of \(\eta\) ensures that the reduced representation preserves the dominant transient dynamics while discarding low-energy components that are more likely to represent numerical noise, weakly informative variability, or highly localized fluctuations.

Projection of the standardized simulation outputs onto the retained basis gives the PCA score matrix
\[
\mathbf{Z}=\tilde{\mathbf{Y}}\mathbf{W}_r
\in \mathbb{R}^{N\times r}.
\]
The \(i\)-th row, denoted \(\mathbf{z}_i^{\top}\), contains the reduced coordinates of the \(i\)-th full transient response, while the \(j\)-th column, denoted \(\mathbf{z}_j\in\mathbb{R}^{N}\), contains the training scores associated with the \(j\)-th retained principal component. These PCA scores, rather than the original high-dimensional output vectors, are used as the regression targets in the subsequent Gaussian process models. This transforms the surrogate-learning task from approximating \(p\) correlated output components to approximating \(r\) dominant latent coordinates.

The standardized transient response can be reconstructed from the retained scores according to \(\tilde{\mathbf{Y}}\approx\mathbf{Z}\mathbf{W}_r^{\top}\). Transforming this approximation back to the original physical output space gives
\begin{equation}
\mathbf{Y} \approx
\mathbf{1}_N\bar{\mathbf{y}}^{\top} + \mathbf{Z}\mathbf{W}_r^{\top}\mathbf{S}\ .
\label{eq:invPCA}
\end{equation}
Equation~\eqref{eq:invPCA} shows that the full multichannel NACIE transient can be reconstructed from a compact set of PCA scores together with the learned mean, scaling, and basis matrices. The resulting representation is physically useful for the present IUQ problem because it preserves the dominant temporal and cross-channel correlations of the system while providing a low-dimensional output space in which probabilistic surrogate modeling is computationally tractable.

\subsection{Gaussian Process Regression in the Reduced PCA Space}
\label{sec:gpr-reduced-pca-space}

After PCA has reduced the simulator output from the original channel--time dimension \(p=MT\) to \(r\) dominant latent coordinates, the surrogate-learning problem becomes the approximation of the parameter-to-score map. This reformulation is central to the present framework. Rather than fitting a separate emulator to every temperature or mass-flow value at every time instant, Gaussian process regression (GPR) is applied to the PCA scores, which encode the principal transient response modes. The resulting model preserves the dominant dynamic structure of the NACIE simulations while substantially reducing the number of regression targets.

The input design matrix is defined as
\[
\mathbf{X}
=
\big(\boldsymbol{\theta}_1,\boldsymbol{\theta}_2,\ldots,\boldsymbol{\theta}_N\big)^{\top}
\in \mathbb{R}^{N\times d},
\]
where the \(i\)-th row corresponds to the uncertain parameter vector used in the \(i\)-th high-fidelity simulation. For each retained PCA mode \(j=1,\ldots,r\), the corresponding score is represented as an unknown scalar function of the input parameters,
\(
z_j=f_j(\boldsymbol{\theta})
\).
Here, \(z_j\) is the \(j\)-th PCA score and \(f_j:\mathbb{R}^d\rightarrow\mathbb{R}\) is the latent response surface in parameter space. Each latent function is assigned a Gaussian process prior,
\[
f_j(\boldsymbol{\theta})
\sim
\mathcal{GP}\left(m_j(\boldsymbol{\theta}),
 k_j(\boldsymbol{\theta},\boldsymbol{\theta}')\right)\ ,
\qquad j=1,\ldots,r\ ,
\]
where \(m_j(\boldsymbol{\theta})\) is the prior mean function and \(k_j(\boldsymbol{\theta},\boldsymbol{\theta}')\) is the covariance kernel. The kernel determines the smoothness, correlation length scales, and amplitude of the score response over the uncertain-parameter domain. In this work, the retained PCA scores are modeled using independent GPR models. This assumption is computationally efficient and is consistent with the orthogonality of the PCA coordinates in the training output space, while still allowing each dominant response mode to have its own input--output dependence.

For the \(j\)-th PCA mode, the training targets are the entries of \(\mathbf{z}_j\in\mathbb{R}^{N}\), i.e., the \(j\)-th column of the score matrix \(\mathbf{Z}\). Given the training pair \((\mathbf{X},\mathbf{z}_j)\), the posterior predictive distribution of the score at a new parameter vector \(\boldsymbol{\theta}_*\in\mathbb{R}^{d}\) is Gaussian:
\[
z_j(\boldsymbol{\theta}_*)\mid(\mathbf{X},\mathbf{z}_j)
\sim
\mathcal{N}\left(
\mu_j(\boldsymbol{\theta}_*),
\sigma_j^2(\boldsymbol{\theta}_*)
\right)\ .
\]
The predictive mean is
\begin{equation}
\mu_j(\boldsymbol{\theta}_*)
=
\mathbf{k}_{*,j}^{\top}
\left(
\mathbf{K}_j+\sigma_{n,j}^2\mathbf{I}_N
\right)^{-1}
\mathbf{z}_j\ ,
\qquad j=1,2,\ldots,r\ ,
\label{eq:mean_scores}
\end{equation}
and the predictive variance is
\begin{equation}
\sigma_j^2(\boldsymbol{\theta}_*)
=
 k_j(\boldsymbol{\theta}_*,\boldsymbol{\theta}_*)
-
\mathbf{k}_{*,j}^{\top}
\left(
\mathbf{K}_j+\sigma_{n,j}^2\mathbf{I}_N
\right)^{-1}
\mathbf{k}_{*,j}\ ,
\qquad j=1,2,\ldots,r\ .
\label{eq:variance_scores}
\end{equation}
In Eqs.~\eqref{eq:mean_scores} and \eqref{eq:variance_scores}, \(\mathbf{K}_j\in\mathbb{R}^{N\times N}\) is the covariance matrix for the \(j\)-th PCA score, with entries
\[
[\mathbf{K}_j]_{ab}=k_j(\boldsymbol{\theta}_a,\boldsymbol{\theta}_b)\ ,
\qquad a,b=1,\ldots,N\ ,
\]
and \(\mathbf{k}_{*,j}\in\mathbb{R}^{N}\) is the covariance vector between the prediction point and the training inputs,
\[
\mathbf{k}_{*,j}
=
\left[
 k_j(\boldsymbol{\theta}_*,\boldsymbol{\theta}_1),
 k_j(\boldsymbol{\theta}_*,\boldsymbol{\theta}_2),
 \ldots,
 k_j(\boldsymbol{\theta}_*,\boldsymbol{\theta}_N)
\right]^{\top}\ .
\]
The parameter \(\sigma_{n,j}^2\) represents either an assumed observation/noise variance for the \(j\)-th score or, in deterministic computer experiments, a numerical nugget used to improve conditioning of the kernel matrix. The identity matrix \(\mathbf{I}_N\in\mathbb{R}^{N\times N}\) has the same dimension as \(\mathbf{K}_j\).

Collecting the posterior means of all retained scores gives the reduced-space predictive mean vector
\[
\boldsymbol{\mu}_z(\boldsymbol{\theta}_*)
=
\left[
\mu_1(\boldsymbol{\theta}_*),
\mu_2(\boldsymbol{\theta}_*),
\ldots,
\mu_r(\boldsymbol{\theta}_*)
\right]^{\top}
\in \mathbb{R}^{r}.
\]
The corresponding standardized transient response is reconstructed using the retained PCA basis,
\[
\hat{\tilde{\mathbf{y}}}(\boldsymbol{\theta}_*)
=
\mathbf{W}_r\boldsymbol{\mu}_z(\boldsymbol{\theta}_*)
\in \mathbb{R}^{p}\ .
\]
Applying the inverse standardization defined in Eq.~\eqref{eq:invPCA} yields the predictive mean in the original physical output space:
\[
\hat{\mathbf{y}}(\boldsymbol{\theta}_*)
=
\bar{\mathbf{y}}
+
\mathbf{S}\hat{\tilde{\mathbf{y}}}(\boldsymbol{\theta}_*)
=
\bar{\mathbf{y}}
+
\mathbf{S}\mathbf{W}_r\boldsymbol{\mu}_z(\boldsymbol{\theta}_*)
\in \mathbb{R}^{p}\ .
\]
Thus, the predicted PCA scores are mapped back to physically interpretable temperature and mass-flow histories through the same scaling and PCA basis used during training.

The GPR predictive variances provide a local measure of surrogate uncertainty in the reduced latent space. Under the conditional independence approximation for the retained PCA-score emulators, the covariance matrix of the predicted scores is
\[
\boldsymbol{\Sigma}_z(\boldsymbol{\theta}_*)
=
\mathrm{diag}
\left(
\sigma_1^2(\boldsymbol{\theta}_*),
\sigma_2^2(\boldsymbol{\theta}_*),
\ldots,
\sigma_r^2(\boldsymbol{\theta}_*)
\right)
\in \mathbb{R}^{r\times r}\ .
\]
Propagation through the PCA reconstruction gives the approximate predictive covariance in the standardized output space,
\[
\boldsymbol{\Sigma}_{\tilde{y}}(\boldsymbol{\theta}_*)
=
\mathbf{W}_r\boldsymbol{\Sigma}_z(\boldsymbol{\theta}_*)\mathbf{W}_r^{\top}
\in \mathbb{R}^{p\times p}\ ,
\]
and the corresponding covariance in the original physical output space is
\[
\boldsymbol{\Sigma}_{y}(\boldsymbol{\theta}_*)
=
\mathbf{S}\boldsymbol{\Sigma}_{\tilde{y}}(\boldsymbol{\theta}_*)\mathbf{S}
\in \mathbb{R}^{p\times p}\ .
\]
This covariance accounts for uncertainty in the learned PCA-score maps and propagates it consistently through the linear reconstruction. It does not, by itself, represent all sources of physical model discrepancy or experimental measurement error; those additional uncertainty sources are treated separately in the Bayesian calibration framework.

The PCA--GPR construction therefore replaces a high-dimensional, strong\-ly correlated transient-output regression problem with \(r\) low-dimensional probabilistic regression problems in an orthogonal latent space. This design is advantageous for the present IUQ application because it reduces the training burden, stabilizes the surrogate learning problem, preserves the dominant temporal and cross-channel response patterns, and produces predictive uncertainty estimates that can be incorporated into posterior inference and posterior predictive validation.

\subsection{Integrated PCA--GPR Surrogate Modeling Framework}
\label{sec:integrated-pca-gpr}

The integrated PCA--GPR surrogate combines the dimensionality-re\-duc\-tion step described in Section~\ref{sec:pca-dimensionality-reduction} with the probabilistic regression model introduced in Section~\ref{sec:gpr-reduced-pca-space}. Its purpose is to replace repeated evaluations of the expensive Flownex SE transient simulator with an inexpensive emulator that preserves the dominant system dynamics and provides a quantifiable approximation uncertainty. This structure is particularly important for Bayesian IUQ because posterior sampling requires repeated likelihood evaluations over the uncertain-parameter space.

The workflow begins by generating a space-filling training design for the uncertain parameter vector \(\boldsymbol{\theta}\) using Sobol low-discrepancy sampling over the prescribed prior bounds. Each sampled input vector is propagated through the Flownex SE model to obtain the corresponding multichannel transient response, yielding the training dataset \((\mathbf{X},\mathbf{Y})\). The output matrix \(\mathbf{Y}\) is then standardized and decomposed using PCA. Retaining the first \(r\) principal components gives the reduced score matrix \(\mathbf{Z}\) and basis matrix \(\mathbf{W}_r\), which together provide a compact representation of the dominant temporal and cross-channel variations in the NACIE responses.

In the regression stage, independent GPR models are trained for the mappings \(\boldsymbol{\theta}\mapsto z_j\), \(j=1,\ldots,r\), where each \(z_j\) is a retained PCA score. For a new input parameter vector \(\boldsymbol{\theta}_*\), the trained GPR models return the predictive score mean vector
\[
\hat{\mathbf{z}}_*
=
\boldsymbol{\mu}_z(\boldsymbol{\theta}_*)
=
\left[
\mu_1(\boldsymbol{\theta}_*),
\ldots,
\mu_r(\boldsymbol{\theta}_*)
\right]^{\top}\ .
\]
The corresponding mean transient response is reconstructed by applying the inverse PCA and inverse standardization transformations,
\[
\hat{\mathbf{y}}_*
=
\bar{\mathbf{y}}
+
\mathbf{S}\mathbf{W}_r\hat{\mathbf{z}}_*\ .
\]
This expression is consistent with Eq.~\eqref{eq:invPCA} and maps the reduced latent prediction back to the original physical output space containing the temperature and mass-flow histories.

The same reconstruction operator can be applied to the GPR predictive covariance in the PCA-score space, thereby propagating surrogate uncertainty to the physical output variables. Consequently, the PCA--GPR model provides both a fast deterministic prediction of the transient response and a local measure of emulator uncertainty. The surrogate therefore serves as a computationally tractable replacement for the high-fidelity simulator in the subsequent uncertainty quantification workflow: it reduces the cost of repeated model evaluations by orders of magnitude, preserves the dominant correlated response structure identified from the training simulations, and supplies uncertainty information that can be incorporated into Bayesian calibration and posterior predictive assessment.

\FloatBarrier   
\section{Uncertainty Quantification Framework Based on PCA--GPR}
\label{sec:uq-framework}

\subsection{Forward Uncertainty Quantification}
\label{sec:forward-uq}

Forward uncertainty quantification (FUQ) is used to characterize the prior predictive behavior of the NACIE thermal--hydraulic model before experimental information is assimilated. In this setting, the uncertain input vector \(\boldsymbol{\theta}\) is treated as a random variable with prescribed prior density \(p(\boldsymbol{\theta}\mid I)\), where \(I\) denotes the available prior information, including engineering judgment, experimental tolerances, and parameter bounds. The deterministic simulator, denoted here by \(\mathcal{M}(\boldsymbol{\theta})\), maps each parameter realization to a multichannel transient response containing the relevant temperatures and LBE mass-flow history. The forward problem therefore consists of propagating \(p(\boldsymbol{\theta}\mid I)\) through \(\mathcal{M}\) to obtain the prior predictive distribution of the quantities of interest.

In practice, this propagation is performed by drawing an ensemble of parameter samples \(\{\boldsymbol{\theta}^{(q)}\}_{q=1}^{N_f}\) from the prescribed prior distributions and evaluating the NACIE model for each realization. Depending on the analysis stage, \(\mathcal{M}\) may correspond either to the reference Flownex SE model shown in Fig.~\ref{fig:FLOWNEX-MODEL} or to the validated PCA--GPR surrogate described in Section~\ref{sec:integrated-pca-gpr}. The resulting ensemble of transient outputs,
\[
\mathbf{y}^{(q)} = \mathcal{M}\left(\boldsymbol{\theta}^{(q)}\right)\ ,
\qquad q=1,\ldots,N_f\ ,
\]
provides a Monte Carlo approximation of the prior predictive response distribution. Statistical summaries of this ensemble, including the pointwise mean, variance, and credible or percentile intervals, quantify the expected system response and the spread induced solely by uncertainty in the input parameters.

Within the present study, FUQ is applied to principal NACIE observables, including the LBE mass flow rate and the temperature histories at selected measurement locations. These responses are system-level quantities because they depend simultaneously on heat addition, hydraulic resistance, heat-exchanger performance, distributed heat losses, and the thermal inertia of the LBE loop. The FUQ ensemble therefore establishes how the assumed prior uncertainty in the physical parameters translates into uncertainty in observable transient behavior.

This prior predictive analysis serves two roles in the overall Bayesian workflow. First, it provides a baseline uncertainty envelope against which posterior predictive distributions can be compared after inverse calibration. A substantial reduction in bias or predictive spread after conditioning on data indicates that the experimental observations contain information capable of constraining the uncertain parameters. Second, FUQ exposes the limitations of relying on prior propagation alone: if the prior distributions are poorly centered or overly broad, the predictive ensemble may either miss the measured transient response or produce uncertainty intervals that are too diffuse for decision support. The forward analysis therefore motivates the inverse uncertainty quantification step, in which experimental data are used to update \(p(\boldsymbol{\theta}\mid I)\) into a posterior distribution and thereby obtain data-informed predictive uncertainty estimates.

\subsection{Inverse Uncertainty Quantification}
\label{sec:inverse-uq}

Inverse uncertainty quantification (IUQ) uses experimental observations to update uncertain model parameters and thereby convert prior uncertainty into data-informed posterior uncertainty. Whereas FUQ propagates assumed parameter distributions through the simulator, IUQ addresses the inverse problem: it infers which parameter values are most consistent with the measured transient response of the NACIE loop. This distinction is important for system-level thermal--hydraulic analysis because several influential parameters, including hydraulic resistance coefficients, heat-loss coefficients, and effective heat-transfer properties, cannot be measured directly during the experiment and therefore constitute epistemic uncertainty.

\begin{figure}[t!]
        \centering
        \includegraphics[width=1\linewidth]{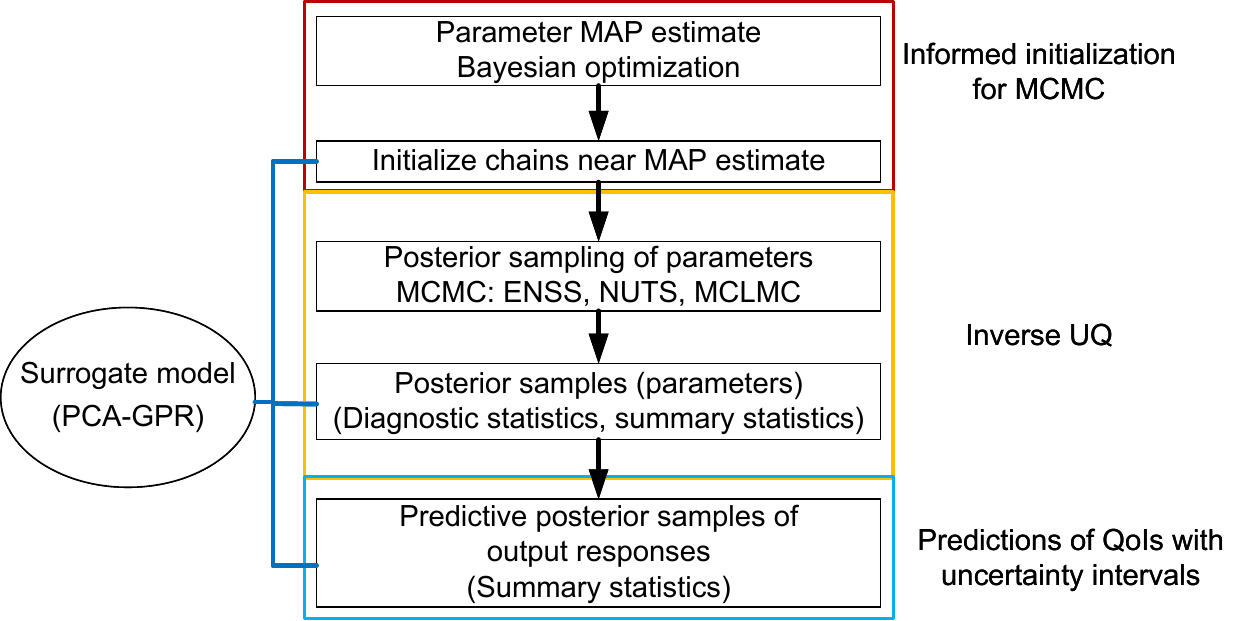}
        \caption{Schematic of the surrogate-assisted uncertainty quantification framework used for Bayesian calibration and posterior predictive assessment.}
        \label{fig:enter-label}
    \end{figure}

The present IUQ framework extends the BO-assisted Bayesian calibration strategy introduced in the previous NACIE study~\cite{YAO2026140531}. In that earlier work, posterior inference relied primarily on gradient-free MCMC sampling with repeated evaluations of an expensive black-box simulator. Here, the calibrated PCA--GPR surrogate provides an inexpensive white-box emulator of the transient response, enabling both gradient-free and gradient-informed posterior sampling. As illustrated in Fig.~\ref{fig:enter-label}, the workflow consists of three stages: (i) Bayesian-optimization-based identification of a high-posterior-probability region for initialization, (ii) Bayesian posterior sampling of the uncertain physical and nuisance parameters, and (iii) posterior predictive propagation to quantify the calibrated uncertainty in the NACIE quantities of interest.

\subsubsection{Stage 1: Informed initialization using Bayesian optimization}
\label{sec:bo-initialization}

Efficient posterior sampling requires Markov chains to reach the high-probability region of the posterior distribution within a limited burn-in period. Poor initialization is particularly problematic for expensive simulators and correlated parameter spaces because many early likelihood evaluations may be spent exploring regions with negligible posterior support. To reduce this inefficiency, Bayesian optimization (BO) is used to identify an informed initial parameter vector near a dominant posterior mode.

In the present context, the BO objective is the log-likelihood, or equivalently the log-posterior when the physical-parameter priors are weakly informative over the admissible region. The resulting optimizer provides an approximation to the maximum a posteriori (MAP) point,
\[
\bm{\theta}_{\mathrm{MAP}}
=
\arg\max_{\bm{\theta}}\, p(\bm{\theta}\mid D,I)
\approx
\arg\max_{\bm{\theta}}\, p(D\mid\bm{\theta},I)\ ,
\]
where the approximation is exact only when the prior is uniform over the explored region. BO treats the objective as an expensive black-box function and constructs a Gaussian-process surrogate for it,
\[
\ell(\bm{\theta})
\equiv
\log p(D\mid\bm{\theta},I)
\sim
\mathcal{GP}\big(m(\bm{\theta}),k(\bm{\theta},\bm{\theta}')\big)\ ,
\]
with a radial basis function covariance kernel. An expected-improvement acquisition function is then used to balance exploration of uncertain regions against exploitation of parameter values that are likely to improve the objective. The resulting MAP-informed initialization is not used as a substitute for posterior sampling; rather, it places the subsequent MCMC chains in a region where efficient exploration of the target distribution can begin.

\subsubsection{Stage 2: Bayesian inference}
\label{sec:bayesian-inference}

Let the experimental dataset be denoted by
\[
D \equiv \bigl\{d_j(t_k)\,\big|\,j=1,2,\ldots,M;\ k=1,2,\ldots,T\bigr\}\ ,
\]
where \(d_j(t_k)\) is the observed response of the \(j\)-th measurement channel at time \(t_k\). The calibration data therefore contain \(N_d=MT\) scalar observations after the time-series downselection described below. For a given parameter vector \(\bm{\theta}\), the simulator or surrogate prediction for the same channel and time is denoted by \(\eta_j(t_k;\bm{\theta})\). The statistical observation model is
\begin{equation} 
d_j(t_k)
=
\eta_j(t_k;\bm{\theta})
+
\epsilon_{jk}\ ,
\qquad j=1,2,\ldots,M;\quad k=1,2,\ldots,T\ .
\label{eq:SNRmodel}
\end{equation}
Here, \(\epsilon_{jk}\) is a residual term representing the combined effect of measurement error, unresolved model discrepancy, and surrogate/emulator approximation error not otherwise represented explicitly.

Bayesian inference updates the prior distribution of the uncertain parameters using Bayes' rule,
\begin{equation}
p(\bm{\theta}\mid D,I)
\propto
p(\bm{\theta}\mid I)\,
p(D\mid\bm{\theta},I)\ ,
\label{eq:BayesRule}
\end{equation}
where \(p(\bm{\theta}\mid D,I)\) is the posterior distribution, \(p(\bm{\theta}\mid I)\) is the prior distribution, \(p(D\mid\bm{\theta},I)\) is the likelihood, and \(I\) denotes the background information and modeling assumptions. The likelihood defines how agreement between the model prediction and the experimental transient data is quantified.

The residuals in Eq.~\eqref{eq:SNRmodel} are modeled as zero-mean Gaussian errors with variance
\[
\sigma_{jk}^2(\bm{\theta})
=
\sigma_{o;jk}^2+
\sigma_{m;jk}^2(\bm{\theta})\ ,
\]
where \(\sigma_{o;jk}^2\) is the experimental measurement-error variance and \(\sigma_{m;jk}^2\) represents additional model/emulator uncertainty. A proportional model-error formulation is adopted,
\[
\sigma_{m;jk}(\bm{\theta})
=
f\,\left|\eta_j(t_k;\bm{\theta})\right|\ ,
\]
where \(f>0\) is a nuisance parameter controlling the relative magnitude of the discrepancy. This heteroscedastic formulation is more appropriate than a constant additive variance for transient thermal--hydraulic responses whose magnitudes vary substantially over time and across channels. Under the working assumption that the downsampled residuals are conditionally independent, the likelihood becomes
\begin{equation}
p(D\mid\bm{\theta},I)
=
\prod_{j=1}^{M}
\prod_{k=1}^{T}
\frac{1}{\sqrt{2\pi\sigma_{jk}^2(\bm{\theta})}}
\exp\left[-
\frac{\bigl(d_j(t_k)-\eta_j(t_k;\bm{\theta})\bigr)^2}
{2\sigma_{jk}^2(\bm{\theta})}
\right]\ .
\label{eq:iuq_likelihood}
\end{equation}
The negative sign in the exponential enforces the usual Gaussian penalty, so parameter values that produce larger weighted residuals receive lower likelihood.

The independence assumption in Eq.~\eqref{eq:iuq_likelihood} is an approximation. Transient thermal--hydraulic measurements are correlated in time and across locations because the loop dynamics are governed by coupled conservation laws and shared thermal inertia. If every available time step were included in the likelihood, these correlations would artificially inflate the effective information content of the dataset and could produce overly narrow posterior credible intervals. To reduce this effect while retaining the dominant transient information, the original time series, containing approximately 20,000 time steps, is downsampled before calibration. Specifically, one point every 30 time steps is retained in the likelihood. This decimation reduces short-lag temporal correlation and yields a more conservative and computationally efficient calibration dataset.

The prior distribution \(p(\bm{\theta}\mid I)\) encodes physically admissible parameter ranges before the NACIE observations are assimilated. Assuming prior independence among the calibrated parameters, the joint prior is factorized as
\[
p(\bm{\theta}\mid I)=\prod_j p(\theta_j\mid I)\ .
\]
Uniform priors are assigned to the physical parameters over engineering-feasible ranges derived from design information, empirical correlations, previous NACIE analyses, and expert judgment. The relative model-error scale \(f\) is assigned a Jeffreys-type prior, \(p(f)\propto 1/f\), which is uniform in \(\log(f)\) and expresses scale invariance. Because this prior is improper on an unbounded domain, finite bounds are imposed on \(\log(f)\) to obtain a proper prior distribution.

\begin{table} [t!]
    \centering
    \caption{Prior bounds assigned to the uncertain calibration parameters used in the Bayesian IUQ analysis. The physical parameters represent heater power bias $P_b$, hydraulic resistance coefficient $C_k$, ambient heat-transfer coefficient $h_a$, and powder-gap thermal conductivity $\lambda_p$; the nuisance parameter $\log f$ controls the residual model-discrepancy scale in the likelihood.}
    \begin{tabular}{l c l l l}
    \hline
    Parameter & Units & Min & Max & Description \\ [0.5ex]
    \hline
    $P_b$   & kW       & $-2.5$ & 2.5 & power bias \\
    $C_k$   & ---   & $2\cdot 10^5$ & $5\cdot 10^5$ & resistance coefficient\\
    $h_a$   & W m$^{-2}$ K$^{-1}$ & 1 & 20 & heat transfer coefficient \\
    $\lambda_p$ & W m$^{-1}$ K$^{-1}$ & 1.5 & 6.5 & powder gap conductivity \\ 
    $\log(f)$ & --- & $-10$ & 5 & model error scale factor \\ [1ex]
    \hline
    \end{tabular}
    \label{tab:PriorAssignment}
\end{table}

The prior bounds used in this work are summarized in Table~\ref{tab:PriorAssignment}. The range of the heater power bias \(P_b\) reflects uncertainty in heater control and power measurement. The hydraulic resistance coefficient \(C_k\) accounts for lumped pressure-loss uncertainty associated with pipes, fittings, and valves. The ambient heat-transfer coefficient \(h_a\) represents uncertain distributed heat loss to the surroundings, and the powder-gap thermal conductivity \(\lambda_p\) captures uncertainty in the effective heat-transfer path through the heat-exchanger powder region. The bounds are intentionally broad enough to avoid artificially constraining the posterior; consequently, posterior concentration away from the bounds indicates that the calibrated distributions are primarily informed by the NACIE measurements rather than by prior truncation.

\subsubsection{Stage 3: Posterior predictive distribution}
\label{sec:posterior-predictive}

After posterior samples have been obtained, calibrated predictive uncertainty is assessed through the posterior predictive distribution. For a future or validation response \(\mathbf{y}_*\), this distribution is
\[
p(\mathbf{y}_*\mid D,I)
=
\int p(\mathbf{y}_*\mid\bm{\theta},I)\,
p(\bm{\theta}\mid D,I)\,d\bm{\theta}\ .
\]
In practice, the integral is approximated by propagating posterior samples \(\{\bm{\theta}^{(s)}\}_{s=1}^{S}\) through the PCA--GPR surrogate, or through the high-fidelity Flownex SE model for selected validation cases. The resulting ensemble
\[
\mathbf{y}_*^{(s)}=\mathcal{M}\left(\bm{\theta}^{(s)}\right)\ ,
\qquad s=1,\ldots,S\ ,
\]
provides an empirical approximation of \(p(\mathbf{y}_*\mid D,I)\). Pointwise means, credible intervals, and predictive distributions for quantities such as transient temperatures and LBE mass flow rate are then computed from this ensemble.

This posterior predictive step is essential for evaluating calibration performance. It determines whether the inferred parameters not only fit the calibration data but also reduce predictive uncertainty and improve agreement for responses or operating conditions not directly used in the likelihood. In summary, the proposed IUQ framework combines PCA--GPR surrogate modeling, Bayesian optimization, MCMC posterior sampling, and posterior predictive propagation to provide a computationally efficient and statistically principled calibration methodology for full-transient nuclear thermal--hydraulic systems.

\section{Posterior Sampling and Differentiable Inference Methods}
\label{sec:computational-methods}

The Bayesian IUQ formulation in Section~\ref{sec:inverse-uq} defines the posterior distribution \(p(\bm{\theta}\mid D,I)\) over the calibrated physical parameters and nuisance parameters. Quantities of interest such as posterior means, credible intervals, effective sample sizes, and posterior predictive summaries require integration with respect to this posterior distribution. Because these integrals are high-dimensional and analytically intractable, Markov chain Monte Carlo (MCMC) methods are used to generate dependent but asymptotically representative samples from \(p(\bm{\theta}\mid D,I)\).

The computational efficiency of MCMC depends critically on the cost of each likelihood evaluation and on the ability of the sampler to explore correlated posterior geometry. Direct use of the high-fidelity Flownex SE simulator is impractical for extensive posterior sampling because each likelihood evaluation requires a full transient calculation. The PCA--GPR surrogate introduced in Section~\ref{sec:integrated-pca-gpr} substantially reduces this cost while preserving the dominant transient response modes of the NACIE system. In addition, the surrogate defines a smooth differentiable mapping from \(\bm{\theta}\) to the predicted response, which allows gradients of the log-posterior to be obtained through automatic differentiation. This feature broadens the feasible sampling strategies from purely gradient-free methods to gradient-informed samplers.

Three MCMC algorithms are considered in this work: the Ensemble Slice Sampler (ENSS), the No-U-Turn Sampler (NUTS), and Microcanonical Langevin Monte Carlo (MCLMC). ENSS provides a robust gradient-free reference that is appropriate when the simulator is treated as a black box. NUTS and MCLMC exploit gradients of the log-posterior and are therefore enabled by the differentiable PCA--GPR emulator. Comparing these samplers provides a direct assessment of how surrogate differentiability affects convergence, autocorrelation, and effective sample generation in full-transient nuclear thermal--hydraulic IUQ.
 
\subsection{Ensemble Slice Sampler}
\label{sec:enss}

The Ensemble Slice Sampler (ENSS)~\cite{KaramanisBeutler2021,Karamanis2021} combines the robustness of slice sampling with the geometric adaptation available from an ensemble of walkers. Unlike gradient-based methods, ENSS requires only evaluations of the unnormalized posterior density and is therefore applicable to black-box likelihoods. This property makes it a natural baseline for comparison with surrogate-enabled gradient-informed samplers.

Let \(S=\{\bm{\theta}_1,\bm{\theta}_2,\ldots,\bm{\theta}_K\}\) denote an ensemble of \(K\) walkers. For the \(k\)-th walker, proposal directions are constructed from the complementary ensemble \(S_{[k]}=S\setminus\{\bm{\theta}_k\}\). A commonly used differential move has the form
\[
\bm{\zeta}_k=\mu\left(\bm{\theta}_l-\bm{\theta}_m\right)\ ,
\]
where \(\bm{\theta}_l\) and \(\bm{\theta}_m\) are two distinct walkers selected from \(S_{[k]}\), and \(\mu\) is a scale factor controlling the move length. Sampling then proceeds by performing one-dimensional slice sampling along the direction \(\bm{\zeta}_k\), producing a candidate state
\[
\bm{\theta}_k^{*}=\bm{\theta}_k+X'\bm{\zeta}_k\ ,
\]
where \(X'\) is sampled from the admissible slice interval. Because the proposal directions are informed by the current ensemble, ENSS can adapt to posterior elongation and parameter correlation without explicit gradient information.

The ensemble construction also supports parallel updates of complementary walker subsets while preserving detailed balance. For the present application, ENSS is valuable because it remains applicable to both the original Flownex SE black-box simulator and the PCA--GPR surrogate. Its main limitation is that, like other gradient-free methods, it does not exploit local differential information in the posterior; consequently, it may require more iterations than gradient-informed samplers when the posterior is smooth and strongly correlated.

\subsection{No-U-Turn Sampler}
\label{sec:nuts}

The No-U-Turn Sampler (NUTS)~\cite{hoffman2014nuts} is an adaptive variant of Hamiltonian Monte Carlo (HMC). It augments the parameter vector \(\bm{\theta}\in\mathbb{R}^{D_p}\)\footnote{The dimension $D_p$ of the parameter vector here includes both the physical parameters, as well as the nuisance parameters that arise from the statistical observation model for the data (see Eq.~(\ref{eq:SNRmodel})).} with an auxiliary momentum vector \(\mathbf{p}\in\mathbb{R}^{D_p}\) and explores the target posterior through approximate Hamiltonian dynamics. Let \(\mathcal{L}(\bm{\theta})\) denote the log-posterior density up to an additive constant. With a standard Gaussian momentum distribution, the joint density is
\[
p(\bm{\theta},\mathbf{p})
\propto
\exp\left\{
\mathcal{L}(\bm{\theta})
-
\frac{1}{2}\mathbf{p}^{\mathsf T}\mathbf{p}
\right\}\ ,
\]
and the corresponding Hamiltonian is
\[
H(\bm{\theta},\mathbf{p})
=
-\mathcal{L}(\bm{\theta})
+
\frac{1}{2}\mathbf{p}^{\mathsf T}\mathbf{p}\ .
\]
The Hamiltonian flow is approximated using the leapfrog integrator,
\[
\begin{aligned}
\mathbf{p}_{t+\frac{\varepsilon}{2}} &= \mathbf{p}_t + \frac{\varepsilon}{2}\nabla_{\bm{\theta}}\mathcal{L}(\bm{\theta}_t)\ , \\
\bm{\theta}_{t+\varepsilon} &= \bm{\theta}_t + \varepsilon\,\mathbf{p}_{t+\frac{\varepsilon}{2}}\ , \\
\mathbf{p}_{t+\varepsilon} &= \mathbf{p}_{t+\frac{\varepsilon}{2}} + \frac{\varepsilon}{2}\nabla_{\bm{\theta}}\mathcal{L}(\bm{\theta}_{t+\varepsilon})\ ,
\end{aligned}
\]
where \(\varepsilon\) is the integration step size. These updates are symplectic and approximately energy preserving, which enables long-distance proposals with high acceptance probability when the gradient is accurate and the step size is well adapted.

Classical HMC requires the user to prescribe the number of leapfrog steps in each trajectory. NUTS removes this tuning requirement by adaptively expanding a binary tree of leapfrog states until the trajectory begins to double back on itself. If \((\bm{\theta}_{-},\mathbf{p}_{-})\) and \((\bm{\theta}_{+},\mathbf{p}_{+})\) denote the two endpoints of the current trajectory, the no-U-turn condition is detected when
\[
(\bm{\theta}_{+}-\bm{\theta}_{-})^{\mathsf T}\mathbf{p}_{-}<0
\quad \text{or} \quad
(\bm{\theta}_{+}-\bm{\theta}_{-})^{\mathsf T}\mathbf{p}_{+}<0\ .
\]
This criterion terminates trajectories that would otherwise retrace previously explored regions of parameter space.

To preserve detailed balance, NUTS introduces a slice variable associated with the joint Hamiltonian density. For the current state \((\bm{\theta},\mathbf{p})\),
\[
u\sim \mathrm{Uniform}\left(
0,
\exp\left\{
\mathcal{L}(\bm{\theta})
-
\frac{1}{2}\mathbf{p}^{\mathsf T}\mathbf{p}
\right\}
\right)\ ,
\]
and the next state is selected from candidate states \((\bm{\theta}',\mathbf{p}')\) satisfying
\[
u\leq
\exp\left\{
\mathcal{L}(\bm{\theta}')
-
\frac{1}{2}{\mathbf{p}'}^{\mathsf T}\mathbf{p}'
\right\}\ .
\]
During warm-up, the leapfrog step size is adapted, commonly through dual averaging, to achieve a target acceptance probability. After warm-up, the adapted step size is fixed for stationary posterior sampling.

NUTS is well suited to the PCA--GPR posterior because the surrogate makes \(\nabla_{\bm{\theta}}\mathcal{L}(\bm{\theta})\) available. By using gradient information, NUTS can suppress random-walk behavior and efficiently traverse correlated posterior regions. Its main cost is that each iteration may require multiple gradient evaluations due to recursive tree construction. Moreover, NUTS requires a continuous differentiable target density; it is not appropriate for discontinuous likelihoods, discrete parameters, or simulator-only black-box settings where gradients are unavailable.

\subsection{Microcanonical Langevin Monte Carlo Sampler}
\label{sec:mclmc}

The Microcanonical Langevin Monte Carlo (MCLMC) algorithm~\cite{robnik2024fluctuation} is a gradient-informed sampler that evolves the parameters using microcanonical dynamics in an extended phase space. Let the target posterior density be written as
\[
p(\bm{\theta}\mid D,I)
\equiv p^{*}(\bm{\theta})
\propto
\exp[-S(\bm{\theta})]\ ,
\]
where \(S(\bm{\theta})=-\log p^{*}(\bm{\theta})\) is the potential function, i.e., the negative log-posterior up to an additive constant. The parameter vector \(\bm{\theta}\in{\Theta}_{p}\subseteq\mathbb{R}^{D_p}\) is augmented with a velocity variable \(\mathbf{u}\in\mathbb{R}^{D_p}\) constrained to the unit sphere, \(\mathbf{u}^{\mathsf T}\mathbf{u}=1\). The extended state therefore lies on \({\Theta}_{p}\times\mathcal{S}^{D_p-1}\).

The deterministic microcanonical flow is governed by
\[
\dot{\bm{\theta}}=\mathbf{u}\ ,
\qquad
\dot{\mathbf{u}}=P(\mathbf{u})\,\mathbf{f}(\bm{\theta})\ ,
\qquad
\mathbf{f}(\bm{\theta})=-\frac{\nabla_{\bm{\theta}}S(\bm{\theta})}{D_p-1}\ ,
\]
where
\[
P(\mathbf{u})=\mathbf{I}-\mathbf{u}\mathbf{u}^{\mathsf T}
\]
is the projection matrix onto the tangent space of the unit sphere at \(\mathbf{u}\). This projection maintains the unit-norm velocity constraint throughout the dynamics.

The invariant density of the extended system is
\[
\rho_{\infty}(\bm{\theta},\mathbf{u})
\propto
\exp[-S(\bm{\theta})]\sqrt{g(\boldsymbol{\vartheta})}\ ,
\]
where \(\boldsymbol{\vartheta}\) denotes coordinates on \(\mathcal{S}^{D_p-1}\) and \(g(\boldsymbol{\vartheta})\) is the determinant of the metric tensor on the sphere. Marginalizing over \(\mathbf{u}\) recovers the desired posterior density \(p^{*}(\bm{\theta})\).

Stochasticity is introduced in the velocity dynamics while preserving the spherical constraint:
\[
d\bm{\theta}=\mathbf{u}\,dt\ ,
\qquad
 d\mathbf{u}=P(\mathbf{u})\,\mathbf{f}(\bm{\theta})\,dt+\,\eta\,P(\mathbf{u})\,d\mathbf{W}\ ,
\]
where \(t\) is artificial sampling time, \(\eta>0\) is the noise amplitude, and \(\mathbf{W}\) is a \(D_p\)-dimensional standard Wiener process. The associated Fokker--Planck equation for the joint density \(\rho(\bm{\theta},\mathbf{u},t)\) is
\[
\frac{\partial \rho}{\partial t}
= -\nabla_{\bm{\theta}}\cdot(\rho\,\mathbf{u})
  -\nabla_{\mathbf{u}}\cdot\left(\rho\,P(\mathbf{u})\,\mathbf{f}(\bm{\theta})\right)
  + \frac{\eta^{2}}{2}\,\Delta_{\mathcal{S}^{D_p-1}}\rho\ ,
\]
where \(\Delta_{\mathcal{S}^{D_p-1}}\) is the Laplace--Beltrami operator on the unit sphere. The deterministic drift and stochastic diffusion preserve the same invariant density, so the dominant practical source of bias is the numerical discretization of the dynamics.

MCLMC is attractive for smooth Bayesian inverse problems because it can use posterior gradients to move persistently through parameter space without a random-walk proposal mechanism. For the present surrogate-assisted IUQ problem, this is advantageous because the differentiable PCA--GPR emulator provides the gradients needed to evaluate \(\nabla_{\bm{\theta}}S(\bm{\theta})\). The method is less appropriate for discontinuous or non-differentiable targets, and its trajectory integration is inherently sequential. Nevertheless, for smooth correlated posteriors, MCLMC can yield rapid decorrelation and high effective sample generation per unit computational cost.

\subsection{Automatic Differentiation and Gradient-Based MCMC Sampling}
\label{sec:automatic-differentiation}

Automatic differentiation (AD) provides the computational mechanism that makes gradient-informed posterior sampling practical in the present surrogate-assisted IUQ framework~\cite{baydin2018automatic,margossian2019review}. Unlike finite-difference differentiation, which approximates derivatives through repeated perturbations of each input parameter, AD applies the chain rule exactly to the elementary operations that define a computer program. As a result, it evaluates derivatives to machine precision, up to floating-point roundoff, with a computational cost that is typically only a small multiple of the cost of evaluating the original function. This property is especially important for Bayesian inverse problems, where NUTS and MCLMC require repeated evaluations of \(\nabla_{\bm{\theta}}\mathcal{L}(\bm{\theta})\) or, equivalently, \(\nabla_{\bm{\theta}}S(\bm{\theta})\).

In this work, the differentiable target density is obtained by combining the PCA--GPR surrogate with the Gaussian likelihood introduced in Eq.~\eqref{eq:iuq_likelihood}. For a parameter vector \(\bm{\theta}\), the surrogate predicts the retained PCA scores through the GPR mappings \(\bm{\theta}\mapsto\boldsymbol{\mu}_z(\bm{\theta})\), reconstructs the full transient response through the linear PCA inverse map, and evaluates the residuals between the reconstructed response and the experimental data. These operations define an explicit computational graph,
\[
\bm{\theta}
\longmapsto
\boldsymbol{\mu}_z(\bm{\theta})
\longmapsto
\hat{\mathbf{y}}(\bm{\theta})
\longmapsto
\log p(D\mid\bm{\theta},I)
\longmapsto
\log p(\bm{\theta}\mid D,I)\ ,
\]
where each transformation is differentiable under the smooth GPR kernel used in the surrogate. AD can therefore propagate sensitivities backward through the likelihood, the PCA reconstruction, and the GPR score predictors to obtain the gradient of the log-posterior with respect to all calibrated parameters.

The availability of these gradients is a key distinction between the PCA--GPR surrogate and the original Flownex SE simulator. Flownex SE is used as a high-fidelity black-box model: although it solves physically meaningful conservation equations internally, its residual equations, solver states, and exact derivatives with respect to calibration parameters are not exposed to the Bayesian sampler. Consequently, direct gradient-based MCMC with Flownex SE would require finite-difference approximations, which scale poorly with parameter dimension and would multiply the already high cost of each transient simulation. By contrast, the PCA--GPR surrogate is an inexpensive white-box emulator whose algebraic structure is accessible to AD. The gradient information needed by NUTS and MCLMC is therefore obtained without additional high-fidelity simulator calls.

For NUTS, AD supplies \(\nabla_{\bm{\theta}}\mathcal{L}(\bm{\theta})\), which determines the Hamiltonian trajectories used to propose long-distance moves through posterior space. For MCLMC, AD supplies \(\nabla_{\bm{\theta}}S(\bm{\theta})\), which enters the projected microcanonical dynamics through the force term \(\mathbf{f}(\bm{\theta})\). In both cases, accurate gradients allow the sampler to exploit local posterior geometry, move efficiently along correlated directions, and reduce random-walk behavior. This typically decreases integrated autocorrelation times and increases the number of effective posterior samples obtained per unit wall-clock time.

The use of AD also improves numerical reliability and reproducibility. Finite-difference gradients depend on user-selected perturbation sizes and can become inaccurate when parameters have different scales or when the likelihood surface is locally curved. AD avoids this step-size sensitivity and provides gradients that are consistent with the implemented surrogate and likelihood. For the present NACIE application, this consistency is important because the posterior combines parameters with different units and magnitudes, including power bias, hydraulic resistance, heat-transfer coefficients, thermal conductivity, and the nuisance error scale. Overall, AD transforms the PCA--GPR surrogate from a fast response surface into a differentiable probabilistic simulator, enabling a rigorous comparison between gradient-free and gradient-informed MCMC strategies within the same Bayesian calibration framework.


\section{Results}
\label{sec:results}

\subsection{PCA--GPR Training Configuration and Surrogate Validation}
\label{sec:numerical-config-pcagpr}

This subsection describes the construction and validation of the PCA--GPR surrogate used in the subsequent Bayesian IUQ analysis. The objective is to replace the high-dimensional transient mapping generated by the Flownex SE model with a compact probabilistic emulator that accurately reconstructs the NACIE system response over the prescribed uncertain-para\-met\-er domain. The workflow follows the methodology developed in Sections~\ref{sec:pca-dimensionality-reduction}--\ref{sec:integrated-pca-gpr}: high-fidelity transient responses are first compressed using PCA, and the retained PCA scores are then modeled as functions of the uncertain input parameters using GPR.

A total of 800 Flownex SE simulations were generated by sampling the uncertain physical parameter vector
\[
\bm{\theta}=(P_b,C_k,h_a,\lambda_p)
\]
from the prior ranges summarized in Table~\ref{tab:PriorAssignment}. Of these simulations, 720 were used to train the PCA--GPR surrogate and the remaining 80 were reserved for out-of-sample validation. This split allows the surrogate to be assessed on parameter realizations not used during training, thereby testing interpolation performance across the prior domain rather than simply measuring in-sample reconstruction accuracy.

For each simulation, the multichannel transient outputs were concatenated into a single response vector \(\mathbf{y}_i\). The output channels comprise five quantities: the inlet and outlet temperatures of the heater section (HS), the inlet and outlet temperatures of the heat exchanger (HX), and the LBE mass flow rate. Each channel was sampled every 10~s over a total transient duration of 20,000~s, giving \(T=2000\) time points per channel. With \(M=5\) channels, the dimensionality of each flattened response vector is therefore \(p=MT=10{,}000\). Thus, direct surrogate modeling would require learning a mapping from a four-dimensional parameter space to a 10,000-dimensional transient output space, which motivates the PCA-based reduction used here. Prior to PCA, each output component was standardized to remove differences in physical units, offsets, and magnitudes across temperature and mass-flow channels.

\begin{figure}[t!]
    \centering
    \includegraphics[width=0.9\linewidth]{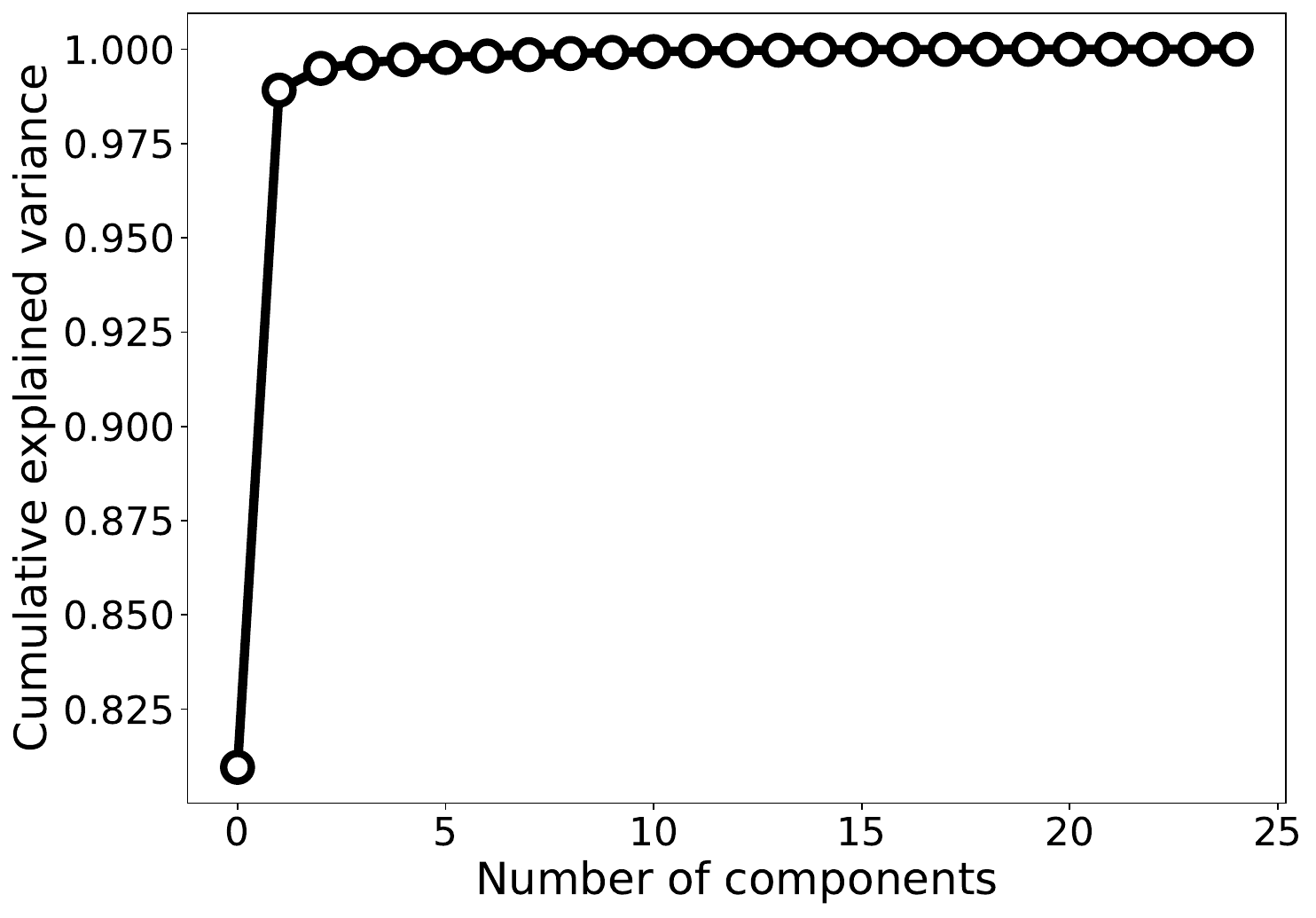}
    \caption{Cumulative explained variance of the PCA representation used for dimensionality reduction of the NACIE transient response. The first four modes capture approximately 98.2\% of the standardized output variance, while the five retained modes capture more than 99\%, reducing the regression target from the full spatiotemporal output dimension to a five-dimensional score vector for subsequent GPR surrogate construction.}
    \label{fig:pca_variance}
\end{figure}

Figure~\ref{fig:pca_variance} shows the cumulative explained variance of the PCA decomposition. The first four principal components account for approximately \(98.2\%\) of the total standardized output variance, indicating that the dominant transient behavior of the NACIE loop is concentrated in a very low-dimensional manifold. Five principal components were retained in the final surrogate, capturing more than \(99\%\) of the total variance. Retaining the fifth component provides additional robustness by preserving small but systematic response variations that may be relevant during posterior calibration, while still reducing the output dimension from \(p=10{,}000\) to \(r=5\). This represents a reduction of three orders of magnitude in the dimension of the regression target.

Let \(\mathbf{W}_r\in\mathbb{R}^{p\times r}\) denote the retained PCA basis and \(\mathbf{Z}\in\mathbb{R}^{N\times r}\) the corresponding score matrix. The columns of \(\mathbf{Z}\), denoted \(\mathbf{z}_j\) for \(j=1,\ldots,r\), define the training targets for the GPR models. Each retained score \(z_j=f_j(\bm{\theta})\) was modeled independently using a Gaussian process. The surrogate prediction in physical output space was then recovered using the inverse PCA transformation in Eq.~\eqref{eq:invPCA}.

To represent the nonlinear dependence of the principal-component scores on the uncertain thermal--hydraulic parameters, a composite covariance kernel was used. The kernel combines a squared exponential component, a Mat\'ern-\(5/2\) component, and a rational quadratic component, scaled by an overall signal variance. For the \(j\)-th PCA score, the covariance function is written as
\begin{equation}
\begin{aligned}
k_j(\bm{\theta}, \bm{\theta}')
&= \sigma_c^2 \Bigg[
   \exp\!\left(-\frac{\rho^2}{2\ell_{\mathrm{SE}}^2}\right) \\
&\quad + \left(1+\frac{\sqrt{5}\rho}{\ell_{\mathrm{M}}}+\frac{5\rho^2}{3\ell_{\mathrm{M}}^2}\right)
   \exp\!\left(-\frac{\sqrt{5}\rho}{\ell_{\mathrm{M}}}\right) \\
&\quad + \left(1+\frac{\rho^2}{2\alpha\ell_{\mathrm{RQ}}^2}\right)^{-\alpha}
\Bigg],
\end{aligned}
\label{eq:composite_kernel}
\end{equation}
where \(\rho=\|\bm{\theta}-\bm{\theta}'\|\) is the Euclidean distance between two input parameter vectors. The squared exponential term captures smooth large-scale variations in the PCA scores, the Mat\'ern-\(5/2\) term allows moderately rougher behavior associated with nonlinear transient transitions, and the rational quadratic term represents a mixture of characteristic length scales. This composite structure is useful for the NACIE response because the transient output combines slow thermal inertia, flow-resistance effects, and localized changes associated with heat-up and shutdown phases.

The hyperparameter vector
\[
\boldsymbol{\zeta}=(\sigma_c,\ell_{\mathrm{SE}},\ell_{\mathrm{M}},\ell_{\mathrm{RQ}},\alpha,\sigma_n)
\]
was estimated for each retained PCA score by maximizing the Gaussian-process marginal likelihood, or equivalently by minimizing the negative log marginal likelihood. The optimization was performed using a limited-memory Broyden--Fletcher--Goldfarb--Shanno algorithm with bound con\-straints (L-BFGS-B). This quasi-Newton method is appropriate here because it incorporates curvature information efficiently without forming the full Hessian, which is advantageous when fitting multiple GPR models.

Surrogate accuracy was evaluated on the independent validation set using the root-mean-square error (RMSE) and normalized RMSE (NRMSE) for each output channel. For channel \(j\), these metrics are defined as
\begin{equation}
\mathrm{RMSE}_j =
\left(\frac{1}{N_v}\sum_{i=1}^{N_v}
\left\|
\mathbf{y}_{i,j}^{\mathrm{sim}}-
\mathbf{y}_{i,j}^{\mathrm{pred}}
\right\|_2^2\right)^{1/2}\ ,
\qquad j=1,2,\ldots,M\ ,
\label{eq:rmse_validation}
\end{equation}
and
\[
\mathrm{NRMSE}_j
=
\frac{\mathrm{RMSE}_j}{y_j^{\max}-y_j^{\min}}\ .
\]
Here, \(N_v=80\) is the number of validation samples, \(\mathbf{y}_{i,j}^{\mathrm{sim}}\) is the Flownex SE response for the \(j\)-th channel and \(i\)-th validation case, and \(\mathbf{y}_{i,j}^{\mathrm{pred}}\) is the corresponding PCA--GPR prediction. The normalization range \(y_j^{\max}-y_j^{\min}\) is computed over the validation responses for channel \(j\), so the NRMSE expresses the surrogate error relative to the dynamic range of each measured quantity.

\begin{figure}[t!]
    \centering
    \includegraphics[width=1\linewidth]{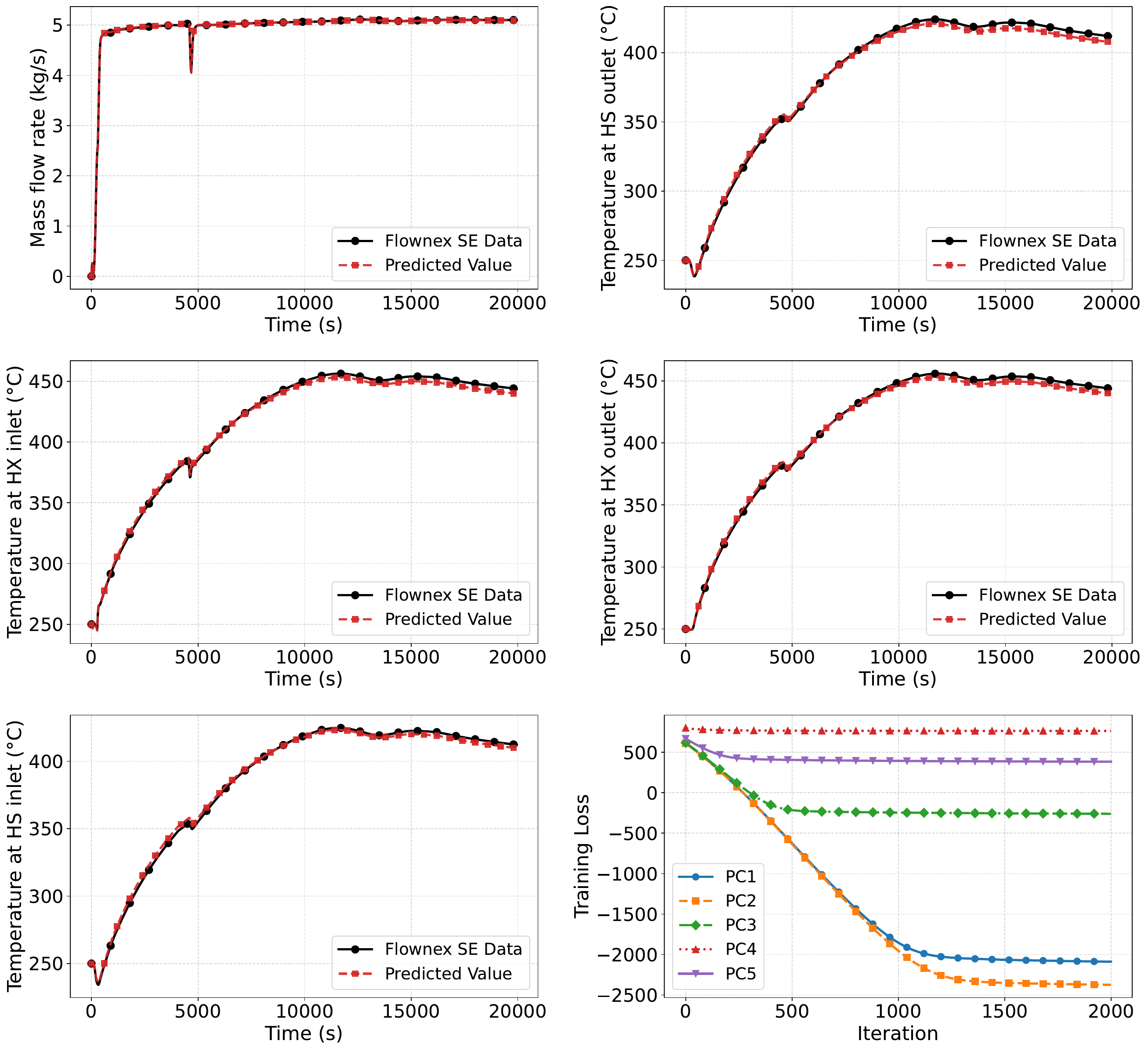}
    \caption{Out-of-sample validation of the PCA--GPR surrogate against the Flownex SE reference model. The response panels show the worst-case validation trajectory among the reserved test simulations for the retained NACIE output channels, comparing surrogate reconstructions with high-fidelity transient temperatures and LBE mass flow rate. The accompanying loss histories show convergence of the independently trained GPR models for the retained PCA scores, confirming stable surrogate training before Bayesian calibration.}
    \label{fig:pca_gpr_validation}
\end{figure}

The channel-averaged validation error,
\[
\overline{\mathrm{NRMSE}}
=
\frac{1}{M}\sum_{j=1}^{M}\mathrm{NRMSE}_j,
\]
remained below approximately \(5\%\) for the most challenging validation trajectory and below approximately \(2\%\) when averaged over the full validation set. These results indicate that the surrogate accurately preserves the dominant transient response characteristics of the Flownex SE model across the validation cases. Figure~\ref{fig:pca_gpr_validation} presents the worst-case validation comparison for the output channels. Even in these most challenging cases, the surrogate reproduces the principal timing, amplitude, and relaxation behavior of the temperature and LBE mass-flow transients. This agreement is important because Bayesian calibration depends not only on pointwise accuracy but also on preserving the correlated temporal structure of the response used in the likelihood.

The lower-right panel of Fig.~\ref{fig:pca_gpr_validation} reports the training loss for the GPR models associated with the retained principal components. The loss corresponds to the negative marginal log-likelihood,
\[
\mathcal{L}_k
=
-\log p(\mathbf{z}_k\mid\mathbf{X},\boldsymbol{\zeta})\ ,
\]
where \(\mathbf{z}_k\) is the score vector of the \(k\)-th retained principal component, \(\mathbf{X}\in\mathbb{R}^{N\times d}\) is the training input matrix, and \(\boldsymbol{\zeta}\) denotes the GPR hyperparameters. The decreasing and eventually plateauing loss curves indicate stable hyperparameter optimization. The higher-order principal components exhibit smaller loss reductions because they contain lower-variance residual structures that are less strongly controlled by the uncertain input parameters. This behavior is expected: once the dominant coherent modes have been captured by the leading components, later components tend to represent weaker transient features, numerical residuals, or small-scale variations that are intrinsically more difficult to predict.

Overall, the validation results confirm that the PCA--GPR surrogate provides an accurate, differentiable, and computationally efficient representation of the NACIE transient simulator. The reduction from a 10,000-dimensional output vector to five latent PCA coordinates makes surrogate training tractable, while the inverse PCA reconstruction preserves the physical time histories required for likelihood evaluation. Compared with direct Flownex SE simulation, the surrogate reduces computational cost by more than three orders of magnitude, thereby enabling the extensive posterior sampling and gradient-informed MCMC analyses reported in the following sections.

\FloatBarrier  

\subsection{Inverse UQ Analysis}

The inverse uncertainty quantification (IUQ) analysis was configured to remain directly comparable with the sampler implementation reported in our previous work~\cite{YAO2026140531}. For the gradient-free ensemble slice sampler (ENSS), posterior exploration used $K=4D_p$ walkers with the differential move, following the ensemble-sampler recommendation of Karamanis {\em et al.}~\cite{Karamanis2021}, where $D_p$ denotes the dimension of the calibration vector $\bm\theta\in \mathbb{R}^{D_p}$---the parameter vector includes both the physical and nuisance parameters. The likelihood function assimilated four transient responses from the NACIE loop: the heat-section (HS) outlet temperature, heat-exchanger (HX) inlet temperature, HX outlet temperature, and lead--bismuth eutectic (LBE) mass flow rate. Although the experimental and simulated responses were available at one-second resolution, each time series was downsampled by a factor of 30 before likelihood evaluation. This reduction preserves the dominant transient evolution while limiting redundant temporal correlation in the data vector and reducing the cost of repeated posterior evaluations.

Bayesian optimization (BO) was used as a deterministic preconditioning step to locate an approximate maximum-a-posteriori (MAP) estimate of $\bm{\theta}$. The BO surrogate was a Gaussian process regression model with a radial-basis-function kernel and automatic relevance determination, trained to emulate the log-posterior over the hyper-rectangle defined by the prior bounds. Candidate evaluations were selected using the expected-improvement acquisition function, thereby balancing exploration of uncertain regions with exploitation near high-posterior-density candidates.

The resulting MAP estimate was used to initialize the MCMC chains. For ensemble sampling, walkers were placed in a small neighborhood of the MAP point, so that the initial population occupied a physically plausible, high-posterior-density region rather than diffuse prior space. This strategy reduces the initial transient phase of the chains and makes the subsequent comparison of sampler efficiency less sensitive to arbitrary starting values.

\subsubsection{MCMC Convergence Diagnostics}

Convergence and sampling efficiency were evaluated using diagnostics that separately address two requirements for reliable Bayesian inference: the Markov chains must sample the stationary posterior distribution, and the retained samples must contain sufficient independent information to estimate posterior summaries accurately. Accordingly, the analysis focused on the effective sample size, which measures the loss of information caused by serial correlation, and on the Monte Carlo standard error, which quantifies the numerical uncertainty in posterior mean estimates.

The primary efficiency metric was the effective sample size (ESS), $N_{\mathrm{eff}}$, which expresses a correlated Markov-chain sample as the equivalent number of independent posterior draws. For an ensemble of $M$ chains with $N_s$ retained samples per chain after burn-in removal, the ESS is defined as
\begin{equation}
N_{\mathrm{eff}}=\frac{M N_s}{\tau_I}\ ,
\label{eq:ess}
\end{equation}
where $\tau_I$ is the integrated autocorrelation time, \(\tau_I = 1 + 2\sum_{k=1}^{\infty}\rho_k \), and $\rho_k$ is the lag-$k$ autocorrelation of the sampled parameter sequence. A value of $\tau_I=1$ corresponds to independent sampling, whereas larger values indicate stronger serial dependence and a smaller number of effectively independent draws for a fixed chain length. In this study, ESS was computed using the multi-chain formulation proposed by Gelman {\it et al.}~\cite{GelmanEtal2013} and refined by Vehtari {\it et al.}~\cite{vehtari2021rank}. This formulation incorporates both within-chain and between-chain variability and is therefore more appropriate for comparing independent chains or ensemble walkers than a single-chain estimate. To avoid unstable estimates caused by noisy large-lag autocorrelations, the summation in $\tau_I$ was truncated using an initial monotone sequence criterion.

For ensemble-based samplers such as ENSS, the $K$ walkers generate parallel Markov trajectories. The autocorrelation function (ACF) was therefore computed independently for each walker before any averaging was performed. The mean autocorrelation across walkers was calculated as
\begin{equation}
\bar{\rho}_k = \frac{1}{K} \sum_{i=1}^{K} \rho_k^{(i)}\ ,
\end{equation}
where $\rho_k^{(i)}$ denotes the lag-$k$ autocorrelation for walker $i$ ($i=1,2,\ldots,K$). The walker trajectories were not concatenated before computing the ACF. Concatenation would introduce artificial transitions between statistically distinct walkers and would not represent the temporal evolution of any individual Markov chain. It would therefore tend to underestimate autocorrelation and overstate sampler efficiency. Computing the ACF at the walker level and then averaging preserves the intrinsic temporal dependence of each trajectory while providing a representative ensemble-level diagnostic.

The averaged ACF provides a direct measure of the mixing rate of the ensemble sampler. Rapid decay toward zero indicates that successive states lose memory of previous states over a short lag, producing a larger ESS and more precise posterior summaries for a fixed computational budget. Slow ACF decay, in contrast, indicates persistent dependence between samples and implies that longer chains are required to achieve the same statistical accuracy.

In addition to ESS, the Monte Carlo standard error (MCSE) was monitored to quantify the numerical uncertainty introduced by finite MCMC sampling. For the $j$-th parameter, the MCSE of the posterior mean estimate is defined as
\begin{equation}
\mathrm{MCSE}(\hat{\theta}_j)
=
\frac{\hat{\sigma}_j}{\sqrt{N_{\mathrm{eff}}}}\ ,
\qquad j=1,\ldots,D_p\ ,
\label{eq:mcse}
\end{equation}
where $\hat{\sigma}_j$ is the estimated posterior standard deviation of parameter $\theta_j$ and $D_p$ is the number of calibrated parameters. The MCSE therefore links posterior dispersion with sampling efficiency: even a well-mixed chain requires a sufficiently large $N_{\mathrm{eff}}$ when the posterior uncertainty of a parameter is large. Smaller MCSE values indicate that posterior mean estimates are insensitive to additional MCMC sampling.

The retained chain length was selected using two complementary criteria. First, every calibrated parameter was required to satisfy $N_{\mathrm{eff}}\gtrsim500$, ensuring that posterior summaries were not based on a small number of effectively independent draws. Second, posterior-mean estimates were required to be numerically stable, with convergence for parameter $j$ declared when $\mathrm{MCSE}(\hat\theta_j)<0.01$. These criteria jointly ensure that the posterior summaries derived from $p(\bm{\theta}|D,I)$ are statistically reliable, reproducible to the reported precision, and not dominated by residual Markov-chain autocorrelation.

\subsubsection{Burn-in and Thinning of Markov Chains}

An initial burn-in phase was removed before computing posterior summaries. This step eliminates the non-stationary portion of each chain during which the sampler is still moving from its initialization region toward the typical set of the posterior distribution. The burn-in length was determined from trace-plot inspection by identifying the point after which the sampled parameters fluctuated around stable posterior levels without systematic drift. This criterion is particularly important in the present IUQ problem because BO-informed initialization places chains near high-posterior-density regions, but stationarity must still be verified rather than assumed.

After burn-in removal, thinning was applied only for diagnostic visualization and for posterior summaries where approximately independent retained draws were desired. The thinning interval was selected using the parameter-wise ACFs and the corresponding integrated autocorrelation time $\tau_I$; specifically, every $k$-th stationary state was retained, where $k$ denotes the chosen thinning interval. This procedure reduces visual and tabular redundancy caused by highly correlated consecutive samples while preserving the target posterior distribution. Importantly, thinning was not used as a substitute for convergence: the unthinned chains were first required to satisfy the ESS and MCSE criteria described above, and thinning was then used only to obtain a compact representation of the stationary sample.

\FloatBarrier   

\subsubsection{Performance of MCMC Samplers for IUQ}

\begin{figure}[t!]
    \centering

    \begin{subfigure}[b]{0.45\linewidth}
        \centering
        \includegraphics[width=\linewidth]{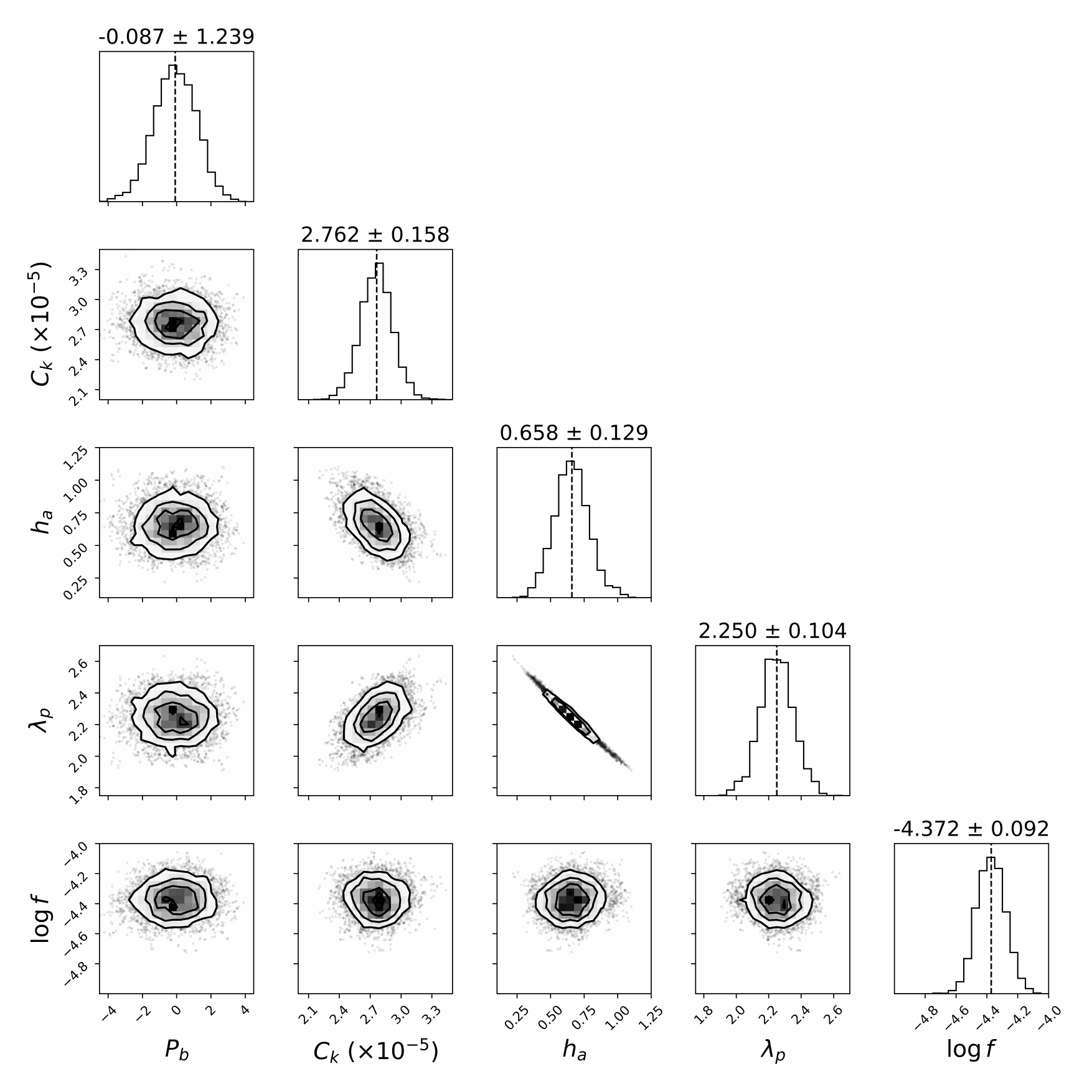}
        \caption{Flownex SE/ENSS}
        \label{fig:post_flownex_enss}
    \end{subfigure}
    \hfill
    \begin{subfigure}[b]{0.45\linewidth}
        \centering
        \includegraphics[width=\linewidth]{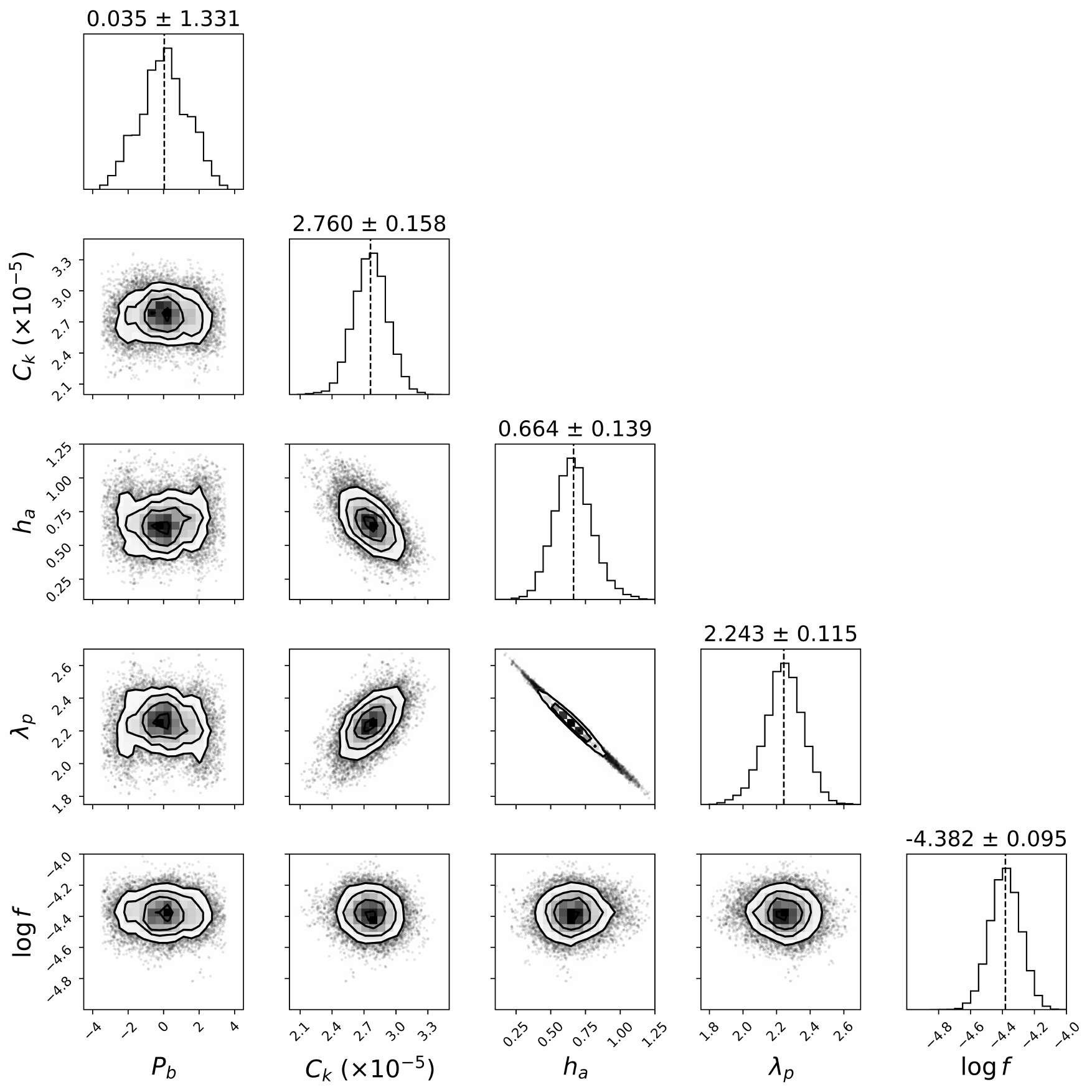}
        \caption{PCA--GPR/ENSS}
        \label{fig:post_pca_enss}
    \end{subfigure}

    \vspace{0.3cm}

    \begin{subfigure}[b]{0.45\linewidth}
        \centering
        \includegraphics[width=\linewidth]{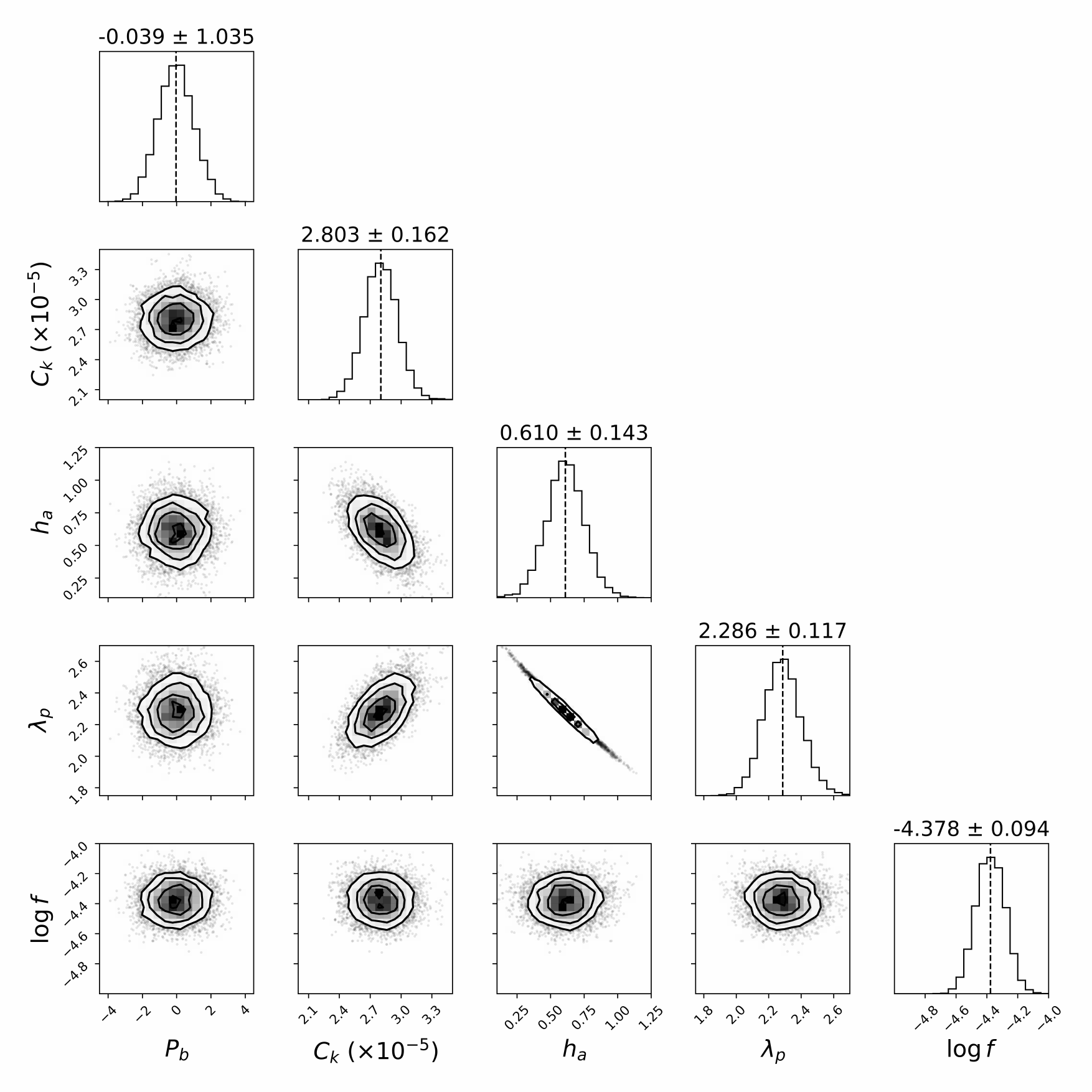}
        \caption{PCA--GPR/NUTS}
        \label{fig:post_pca_nuts}
    \end{subfigure}
    \hfill
    \begin{subfigure}[b]{0.45\linewidth}
        \centering
        \includegraphics[width=\linewidth]{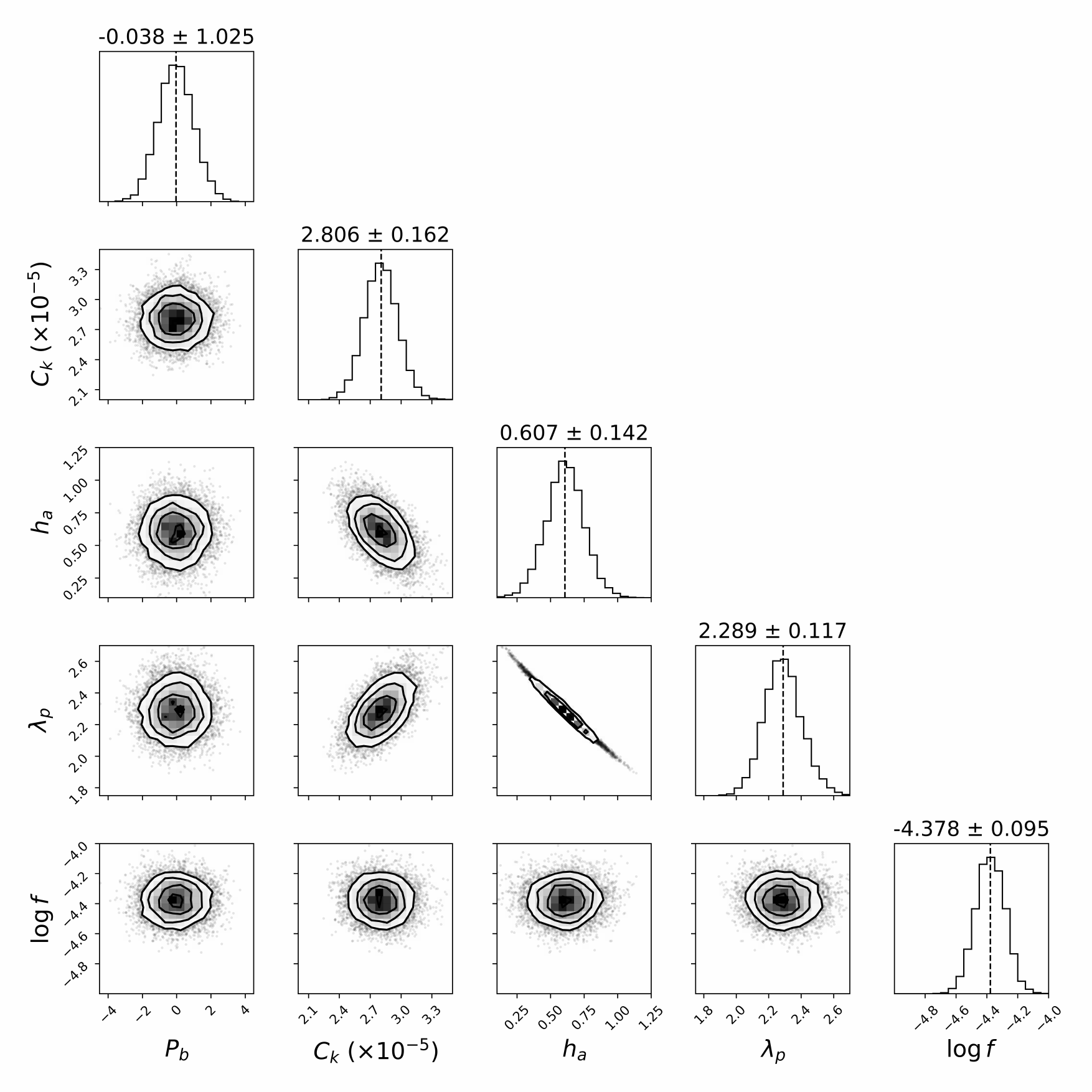}
        \caption{PCA--GPR/MCLMC}
        \label{fig:post_pca_mclmc}
    \end{subfigure}

    \caption{Marginal and pairwise posterior distributions of the calibration vector \(\bm{\theta}\) for the four simulator--sampler configurations. Diagonal panels show one-dimensional marginal densities, and off-diagonal panels show bivariate posterior contours that reveal parameter correlations and compensation effects. The Flownex SE/ENSS case serves as the high-fidelity reference, while the PCA--GPR/ENSS, PCA--GPR/NUTS, and PCA--GPR/MCLMC cases test whether surrogate likelihood evaluation and gradient-informed sampling preserve the same posterior location, spread, and dependence structure. Contour levels enclose 11.8\%, 39.3\%, 67.5\%, and 86.5\% of the bivariate posterior probability mass.}
    \label{fig:sampler_comparison_logp}
\end{figure}

Figure~\ref{fig:sampler_comparison_logp} compares the marginal posterior distributions obtained with four simulator--sampler configurations. The Flownex SE/ENSS case, shown in Fig.~\ref{fig:sampler_comparison_logp}(a), is used as the high-fidelity reference because the likelihood is evaluated directly with the original black-box thermal--hydraulic simulator. Figures~\ref{fig:sampler_comparison_logp}(b)--(d) show the corresponding posterior distributions obtained when the likelihood is evaluated with the PCA--GPR surrogate and sampled using ENSS, NUTS, and MCLMC, respectively. This comparison therefore tests two distinct requirements of the surrogate-assisted IUQ framework: the surrogate must preserve the posterior geometry induced by the high-fidelity model, and the MCMC sampler must explore that posterior without introducing algorithm-dependent distortions.

The surrogate-based posterior distributions are in close agreement with the Flownex SE/ENSS reference. For the hydraulic resistance coefficient \(C_k\), ambient heat-transfer coefficient \(h_a\), powder-gap thermal conductivity \(\lambda_p\), and discrepancy-scale parameter \(\log f\), the posterior locations, spreads, skewness, and bivariate contour shapes remain nearly unchanged across the four configurations. This agreement indicates that the PCA--GPR surrogate preserves the dominant likelihood structure of the high-fidelity simulator in the region of posterior support. Consequently, the use of the surrogate does not merely reproduce point predictions; it also retains the uncertainty and correlation information required for Bayesian calibration.

The only visible difference occurs for the power-bias parameter \(P_b\), whose surrogate-based posterior mean is slightly shifted relative to the Flownex SE/ENSS reference. This deviation is limited in magnitude and does not materially change the posterior spread or the qualitative dependence structure. It is consistent with the weak identifiability of \(P_b\) discussed later in Fig.~\ref{fig:datasize_three_rows}: when the assimilated responses are only weakly sensitive to a parameter, small surrogate errors, finite-sample variability, or differences in high-density-region exploration can produce modest shifts in its posterior mean. The shift should therefore be interpreted as a sensitivity of a weakly constrained parameter rather than as evidence of a systematic surrogate bias.

The bivariate posterior contours provide additional physical insight into parameter compensation. A pronounced negative correlation between \(h_a\) and \(\lambda_p\) appears in all simulator--sampler combinations. This trend is physically consistent: stronger ambient heat removal can be offset by reduced internal conductive transport through the powder gap, and vice versa, while producing similar transient temperature histories. A positive association between \(C_k\) and \(\lambda_p\) is also visible, indicating that hydraulic resistance and thermal transport jointly influence the coupled mass-flow and temperature response of the natural-circulation loop. In contrast, \(\log f\) remains only weakly correlated with the physical calibration parameters. This weak dependence suggests that the inferred discrepancy scale is not compensating for a particular physical parameter, but instead represents residual model--data mismatch after the dominant thermal--hydraulic parameters have been calibrated.

The close agreement among PCA--GPR/ENSS, PCA--GPR/NUTS, and PCA--GPR/MCLMC demonstrates that the posterior distribution is not an artifact of a particular sampling algorithm. The gradient-based samplers recover the same posterior region and correlation structure as the gradient-free ensemble sampler, while the PCA--GPR/ENSS result remains consistent with the Flownex SE/ENSS reference. These observations support the central methodological claim of this work: the PCA--GPR surrogate enables efficient posterior inference without sacrificing the posterior structure generated by the high-fidelity simulator.

\begin{figure}[t!]
    \centering

    \begin{subfigure}[b]{0.45\linewidth}
        \centering
        \includegraphics[width=\linewidth]{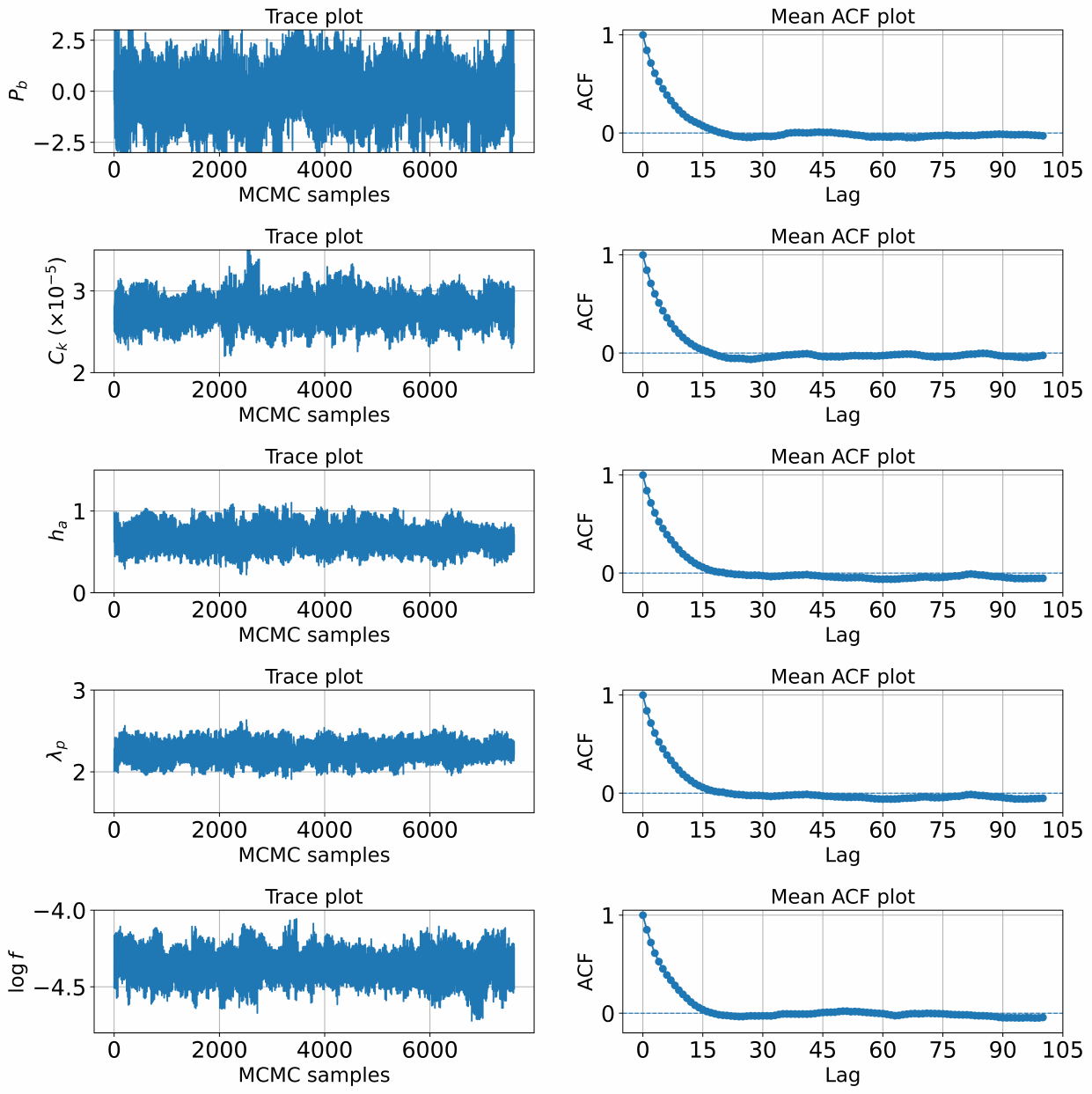}
        \caption{Flownex SE/ENSS}
        \label{fig:acf_enss_flownex}
    \end{subfigure}
    \hfill
    \begin{subfigure}[b]{0.45\linewidth}
        \centering
        \includegraphics[width=\linewidth]{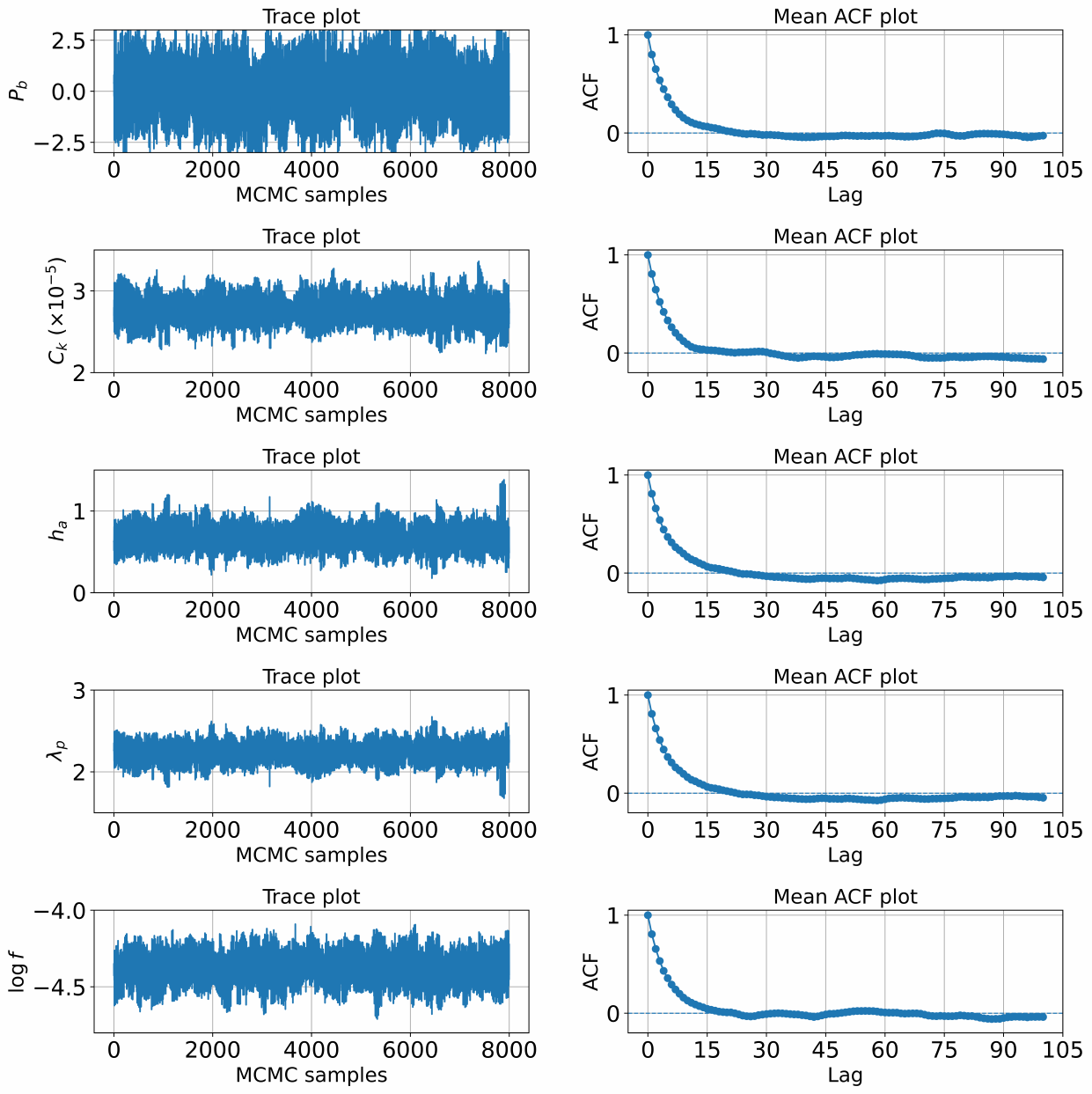}
        \caption{PCA--GPR/ENSS}
        \label{fig:acf_enss_pca}
    \end{subfigure}

    \vspace{0.3cm}

    \begin{subfigure}[b]{0.45\linewidth}
        \centering
        \includegraphics[width=\linewidth]{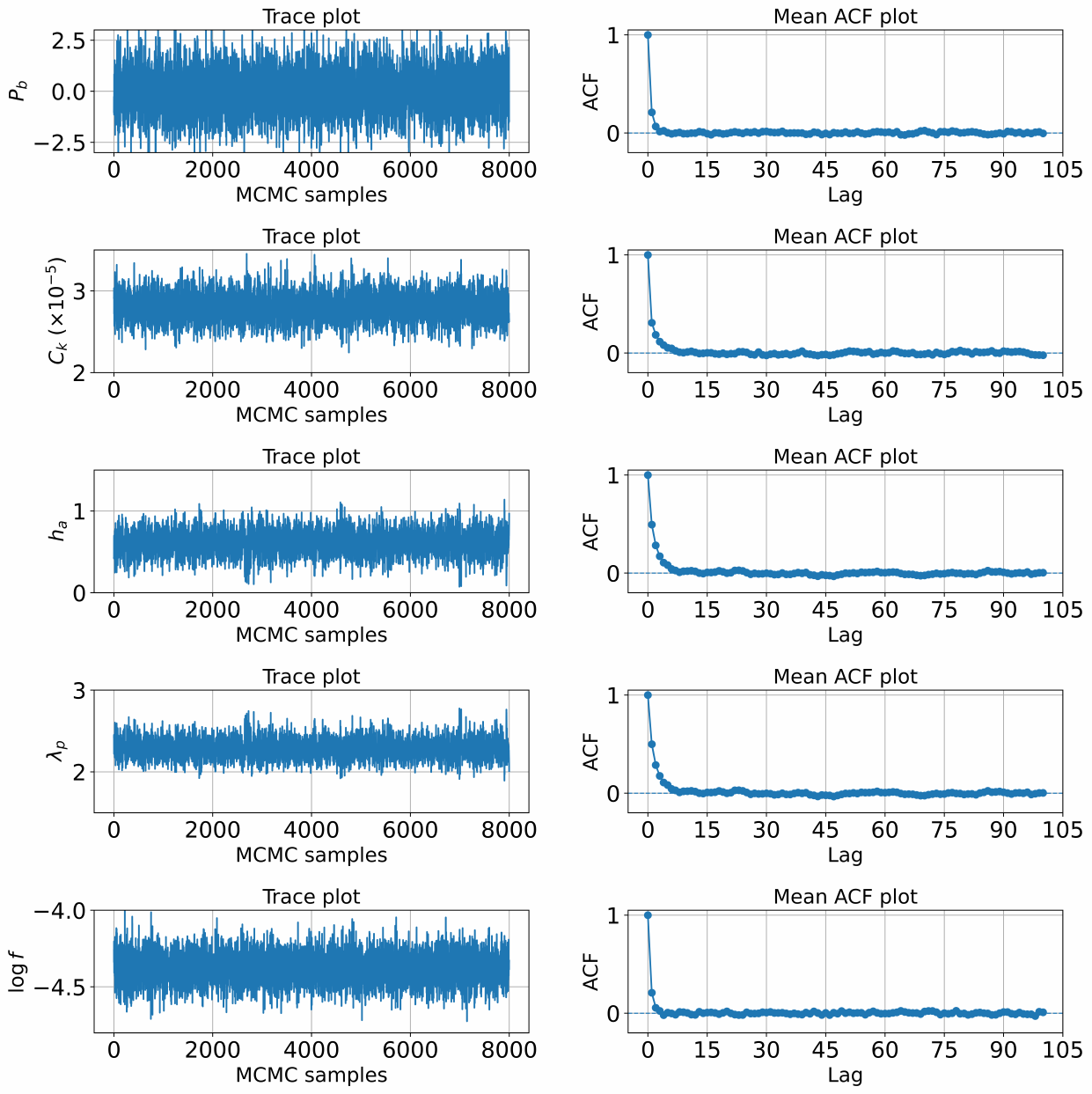}
        \caption{PCA--GPR/NUTS}
        \label{fig:acf_nuts_pca}
    \end{subfigure}
    \hfill
    \begin{subfigure}[b]{0.45\linewidth}
        \centering
        \includegraphics[width=\linewidth]{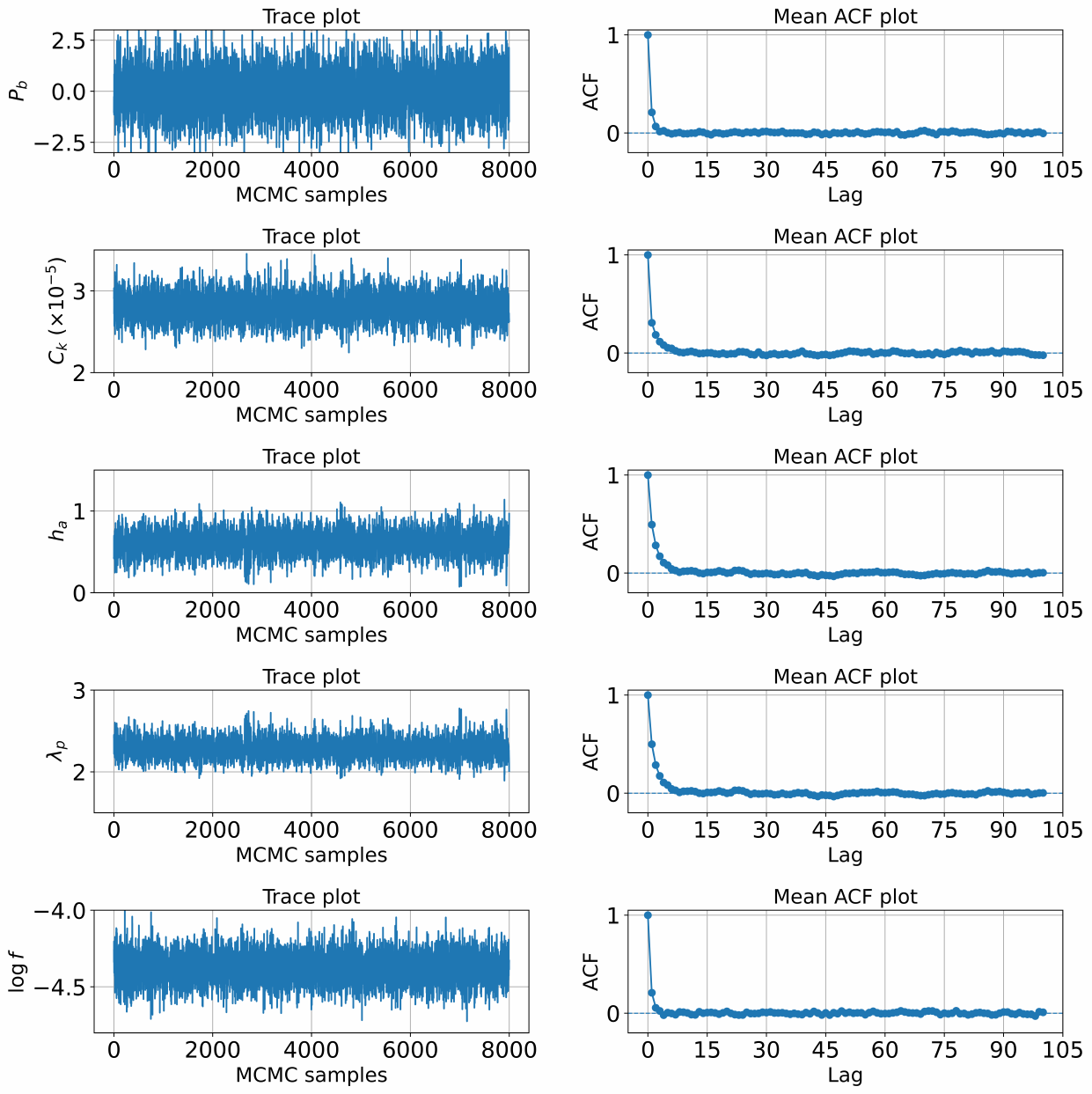}
        \caption{PCA--GPR/MCLMC}
        \label{fig:acf_mclmc_pca}
    \end{subfigure}

    \caption{Trace and autocorrelation diagnostics for posterior samples of the calibration vector \(\bm{\theta}\) under the high-fidelity Flownex SE/ENSS reference and the PCA--GPR surrogate sampling schemes. For each simulator--sampler combination, the left panels show post-burn-in chains used to assess stationarity and posterior exploration, whereas the right panels show autocorrelation functions used to evaluate serial dependence and effective-sample generation. The diagnostics show that the surrogate preserves the mixing behavior of ENSS while enabling substantially faster decorrelation with the gradient-informed NUTS and MCLMC samplers.}
    \label{fig:sampler_comparison_acf}
\end{figure}

Figure~\ref{fig:sampler_comparison_acf} evaluates the dynamic efficiency of the same four configurations using trace plots and autocorrelation functions. Only post-burn-in samples are shown. For ensemble samplers, the trace plots are displayed by concatenating walker trajectories for visualization, whereas the ACF is computed at the walker level and then averaged, as described in Section~6.2.1. This treatment prevents artificial decorrelation caused by transitions between independent walkers and provides a more conservative estimate of the true serial dependence.

The trace plots show stable fluctuation around stationary posterior levels for all configurations, with no visible long-term drift after burn-in. This behavior supports the use of the retained samples for posterior estimation. The ACFs, however, reveal substantial differences in sampling efficiency. The Flownex SE/ENSS reference exhibits gradual autocorrelation decay, indicating persistent dependence between successive states. The PCA--GPR/ENSS configuration shows a similar ACF profile, confirming that the surrogate changes the computational cost of the likelihood evaluation but does not fundamentally change the random-walk-like mixing behavior of the gradient-free sampler.

By contrast, PCA--GPR/NUTS and PCA--GPR/MCLMC exhibit much faster autocorrelation decay. For most parameters, the ACF approaches zero within a small number of lags, implying shorter integrated autocorrelation times and larger effective sample sizes for the same number of retained iterations. This improvement arises because both NUTS and MCLMC use gradients of the differentiable PCA--GPR log-posterior to propose coherent moves through parameter space. NUTS achieves this through adaptive Hamiltonian trajectories that reduce diffusive random-walk behavior, whereas MCLMC uses microcanonical dynamics with stochastic momentum refreshments to traverse high-probability regions efficiently while preserving the target posterior distribution.

These diagnostics show that the surrogate and the sampler contribute complementary sources of acceleration. The PCA--GPR surrogate reduces the cost of each likelihood evaluation relative to Flownex SE, while gradient-based MCMC reduces the number of correlated iterations required to obtain a prescribed number of effective posterior samples. The combined PCA--GPR/MCLMC and PCA--GPR/NUTS configurations therefore improve both computational efficiency and statistical efficiency. In practical IUQ terms, this means that posterior uncertainty can be resolved with fewer high-cost model evaluations, lower MCSE, and more reliable estimates of parameter correlations.

Overall, Figs.~\ref{fig:sampler_comparison_logp} and~\ref{fig:sampler_comparison_acf} demonstrate that the PCA--GPR surrogate preserves the high-fidelity posterior distribution while enabling more efficient exploration when coupled with gradient-based samplers. ENSS remains a robust gradient-free option and provides a useful reference for black-box inference, but the differentiability of the surrogate makes NUTS and MCLMC substantially more efficient for the present calibration problem. Among the tested schemes, MCLMC exhibits the fastest decorrelation and therefore provides the most favorable balance between posterior fidelity and effective-sample generation.

\FloatBarrier   

\subsubsection{Posterior Consistency and MCMC Sampling Efficiency}

Table~\ref{tab:posterior_summary_all} reports posterior summary statistics for each calibrated component of \(\bm\theta\) and for each simulator--sampler configuration. The table includes the posterior mean, posterior standard deviation, and selected percentiles \(P_\alpha\), with \(\alpha=5,25,50,75,95\). The median is given by \(P_{50}\), and the remaining percentiles provide a compact description of posterior asymmetry and tail behavior. For any component \(\theta_j\) and confidence level \(\beta\in(0,1)\), a Bayesian credible interval \(I_\beta\) satisfies
\begin{equation}
p(\theta_j\in I_\beta|D,I) = 1-\beta\ ,
\end{equation}
so intervals constructed from the reported percentiles quantify posterior probability mass rather than repeated-sampling coverage. These summaries therefore allow the simulator and sampler choices to be compared in terms of posterior location, dispersion, and distributional shape.

\begin{table}[t!]
\centering
\caption{Posterior summary statistics for the calibrated parameter vector $\bm\theta$ obtained using the high-fidelity Flownex SE/ENSS reference and the PCA--GPR surrogate with ENSS, NUTS, and MCLMC sampling. The reported mean, standard deviation, and selected posterior percentiles quantify agreement in posterior location, uncertainty width, tail behavior, and parameter identifiability across simulator--sampler combinations.}
\label{tab:posterior_summary_all}

\resizebox{\textwidth}{!}{%
\begin{tabular}{
    l
    S[table-format=-1.6]
    S[table-format=1.6]
    S[table-format=-1.6]
    S[table-format=-1.6]
    S[table-format=-1.6]
    S[table-format=1.6]
    S[table-format=1.6]
}
\toprule
\multicolumn{8}{c}{Power bias $P_b$} \\
\midrule
{Method} & {Mean} & {Std} & {P5} & {P25} & {P50} & {P75} & {P95} \\
\midrule
Flownex SE/ENSS & -0.0872 & 1.23863 & -2.1421 & -0.9112 & -0.0776 & 0.76821 & 1.90075 \\
PCA--GPR/ENSS & 0.04292 & 1.32041 & -2.2088 & -0.8056 & 0.05390 & 0.93162 & 2.21462 \\
PCA--GPR/NUTS & -0.0524 & 1.28086 & -2.1309 & -0.9185 & -0.0541 & 0.82608 & 2.05783 \\
PCA--GPR/MCLMC & -0.0591 & 1.27888 & -2.1650 & -0.9433 & -0.0665 & 0.82846 & 2.04302 \\
\midrule

\multicolumn{8}{c}{Resistance coefficient $C_k$ ($\times 10^{-5}$)} \\
\midrule
{Method} & {Mean} & {Std} & {P5} & {P25} & {P50} & {P75} & {P95} \\
\midrule
Flownex SE/ENSS & 2.76203 & 0.15774 & 2.50357 & 2.65722 & 2.76265 & 2.86304 & 3.02490 \\
PCA--GPR/ENSS & 2.76042 & 0.15723 & 2.50479 & 2.65590 & 2.76139 & 2.86417 & 3.01965 \\
PCA--GPR/NUTS & 2.78016 & 0.15283 & 2.53204 & 2.67738 & 2.77602 & 2.88110 & 3.03812 \\
PCA--GPR/MCLMC & 2.77902 & 0.15214 & 2.53287 & 2.67426 & 2.77816 & 2.88100 & 3.03264 \\
\midrule

\multicolumn{8}{c}{Ambient heat transfer coefficient $h_a$($\times 10$)} \\
\midrule
{Method} & {Mean} & {Std} & {P5} & {P25} & {P50} & {P75} & {P95} \\
\midrule
Flownex SE/ENSS & 0.65781 & 0.12920 & 0.44750 & 0.57220 & 0.65423 & 0.73879 & 0.87157 \\

PCA--GPR/ENSS & 0.66340 & 0.13924 & 0.44365 & 0.57130 & 0.65797 & 0.74838 & 0.90177 \\
PCA--GPR/NUTS & 0.64458 & 0.11757 & 0.45028 & 0.56538 & 0.64547 & 0.72412 & 0.83693 \\
PCA--GPR/MCLMC & 0.64838 & 0.12035 & 0.44935 & 0.56710 & 0.64875 & 0.73012 & 0.84223 \\
\midrule

\multicolumn{8}{c}{Powder gap conductivity $\lambda_p$} \\
\midrule
{Method} & {Mean} & {Std} & {P5} & {P25} & {P50} & {P75} & {P95} \\
\midrule
Flownex SE/ENSS & 2.24957 & 0.10424 & 2.08115 & 2.18171 & 2.24938 & 2.31713 & 2.42412 \\
PCA--GPR/ENSS & 2.24305 & 0.11491 & 2.04981 & 2.17264 & 2.24520 & 2.31807 & 2.42786 \\
PCA--GPR/NUTS & 2.25657 & 0.09536 & 2.10269 & 2.19251 & 2.25324 & 2.31851 & 2.41694 \\
PCA--GPR/MCLMC & 2.25375 & 0.09759 & 2.10130 & 2.18588 & 2.25151 & 2.31815 & 2.41972 \\
\midrule

\multicolumn{8}{c}{Logarithm of model discrepancy factor $\log(f)$} \\
\midrule
{Method} & {Mean} & {Std} & {P5} & {P25} & {P50} & {P75} & {P95} \\
\midrule
Flownex SE/ENSS & -4.3725 & 0.09237 & -4.5189 & -4.4370 & -4.3737 & -4.3090 & -4.2201 \\
PCA--GPR/ENSS & -4.3820 & 0.09533 & -4.5387 & -4.4462 & -4.3814 & -4.3167 & -4.2259 \\
PCA--GPR/NUTS & -4.3805 & 0.09635 & -4.5373 & -4.4460 & -4.3798 & -4.3153 & -4.2213 \\
PCA--GPR/MCLMC & -4.3795 & 0.09501 & -4.5342 & -4.4448 & -4.3798 & -4.3149 & -4.2238 \\
\bottomrule
\end{tabular}
}
\end{table}

The posterior summaries in Table~\ref{tab:posterior_summary_all} show close quantitative agreement across all simulator--sampler combinations. For \(C_k\), \(h_a\), \(\lambda_p\), and \(\log(f)\), the posterior means, medians, standard deviations, and percentile ranges are nearly identical for the Flownex SE/ENSS reference and for the PCA--GPR-based samplers. This agreement provides a stringent validation of the surrogate-assisted IUQ workflow because it shows that the surrogate preserves not only the posterior mode or mean, but also the uncertainty width and tail behavior induced by the high-fidelity likelihood.

The largest method-to-method variation occurs for the power-bias parameter \(P_b\), whose posterior is substantially broader than those of the other calibrated parameters. The posterior standard deviation of \(P_b\) remains of order unity for all schemes, and the 5th--95th percentile range spans several units. This broad posterior is consistent with the weak data-driven identifiability of \(P_b\): the assimilated transient responses constrain the heat-transfer and hydraulic-resistance parameters more strongly than they constrain a global power-bias correction. Consequently, modest differences among samplers or between the surrogate and high-fidelity model can shift the posterior mean of \(P_b\) without changing the scientific conclusion that this parameter remains weakly informed by the present calibration data.

The thermal parameters \(h_a\) and \(\lambda_p\) exhibit particularly consistent posterior summaries across the four configurations. This is important because their bivariate posterior contours in Fig.~\ref{fig:sampler_comparison_logp} show a pronounced compensatory relationship. Agreement in their marginal statistics therefore implies that both the high-fidelity simulator and the PCA--GPR surrogate recover the same balance between external convective heat removal and internal conductive transport. Similarly, the consistency of \(C_k\) indicates that the hydraulic-resistance contribution to the coupled mass-flow response is preserved by the surrogate model.

The nuisance parameter \(\log(f)\) also shows only minor method-to-method variation, with all schemes producing nearly identical posterior means and percentile ranges. This result indicates that the inferred residual discrepancy scale is stable with respect to both simulator replacement and sampler choice. In physical terms, the dominant model--data mismatch is addressed through identifiable changes in the calibrated thermal--hydraulic parameters, while \(\log(f)\) captures the remaining unresolved discrepancy, including measurement uncertainty, surrogate approximation error, and modeling simplifications not represented by the calibrated parameters.

Overall, Table~\ref{tab:posterior_summary_all} demonstrates posterior fidelity: the PCA--GPR surrogate reproduces the principal posterior characteristics of the Flownex SE reference, and ENSS, NUTS, and MCLMC all target the same calibrated distribution when applied to the surrogate. The subsequent comparison therefore focuses on efficiency rather than posterior correctness.

\begin{table}[t!]
  \centering
  \caption{Computational and statistical efficiency of the high-fidelity Flownex SE/ENSS reference and PCA--GPR surrogate sampling schemes. The comparison reports the aggregate effective sample size \(\overline{\mathrm{ESS}}\), mean wall-clock time per effective sample, total CPU time, and nominal sample count, highlighting the separate gains from surrogate acceleration and gradient-informed posterior sampling.}
  \label{tab:sampler_performance}
  \begin{tabularx}{\linewidth}{
    >{\centering\arraybackslash}p{3.9cm}
    >{\centering\arraybackslash}p{1.5cm}
    >{\centering\arraybackslash}p{2.6cm}
    >{\centering\arraybackslash}p{2.0cm}
    >{\centering\arraybackslash}p{1.5cm}
  }
    \hline
    \vspace{5pt}Scheme &
    \vspace{5pt}\(\overline{\mathrm{ESS}}\) &
    Mean time per effective sample (s) &
    \vspace{0.5pt}Total time (CPU h) &
    \vspace{0.5pt}Total samples \\
    \hline
    Flownex SE/ENSS & 791 & 1517.06 & 332 & 8000 \\
    PCA--GPR/ENSS   & 812 & 0.48    & 0.11 & 8000 \\
    PCA--GPR/NUTS   & 4350 & 0.101  & 0.111 & 8000 \\
    PCA--GPR/MCLMC  & 5542 & 0.064  & 0.1 & 8000 \\
    \hline
  \end{tabularx}
\end{table}

Table~\ref{tab:sampler_performance} compares the computational and statistical efficiency of the four schemes using the same nominal chain length of 8000 posterior samples. The reported aggregate effective sample size is
\begin{equation}
\overline{\mathrm{ESS}} = \frac{1}{D_p}\sum_{j=1}^{D_p}\mathrm{ESS}_j\ ,
\end{equation}
where \(\mathrm{ESS}_j\) is the effective sample size of the \(j\)-th component of \(\bm\theta\), and \(D_p\) is the dimension of the calibration vector. Thus, \(\overline{\mathrm{ESS}}\) measures the average number of effectively independent posterior samples generated across the full parameter space. The mean time per effective sample combines model-evaluation cost and Markov-chain correlation, making it a more relevant IUQ performance metric than raw wall-clock time per iteration.

The direct Flownex SE/ENSS configuration is the computational baseline. It produces \(\overline{\mathrm{ESS}}=791\) from 8000 samples, but requires 1517.06~s per effective sample and 332~CPU~h in total. Replacing Flownex SE with the PCA--GPR surrogate while retaining the same gradient-free ENSS sampler reduces the mean time per effective sample to 0.48~s and the total cost to 0.11~CPU~h. This reduction isolates the effect of surrogate acceleration: the statistical efficiency of ENSS changes only modestly, but the cost of each likelihood evaluation is reduced by several orders of magnitude.

The gradient-based surrogate samplers improve the second component of efficiency: sample independence. PCA--GPR/NUTS increases \(\overline{\mathrm{ESS}}\) to 4350 and reduces the mean time per effective sample to 0.101~s, while PCA--GPR/MCLMC further increases \(\overline{\mathrm{ESS}}\) to 5542 and reduces the mean time per effective sample to 0.064~s. Because the total CPU costs of PCA--GPR/NUTS and PCA--GPR/MCLMC remain comparable to PCA--GPR/ENSS, these gains arise primarily from reduced autocorrelation rather than from cheaper individual surrogate evaluations. The performance data are therefore consistent with the ACF analysis in Fig.~\ref{fig:sampler_comparison_acf}: gradient-informed proposals explore the posterior more coherently and generate more effectively independent samples for a fixed number of retained iterations.

Relative to the high-fidelity Flownex SE/ENSS baseline, the combination of PCA--GPR/MCLMC simultaneously reduces the total computational cost from 332 to 0.1~CPU~h and increases \(\overline{\mathrm{ESS}}\) from 791 to 5542. This represents a change from a computationally restrictive black-box IUQ calculation to a statistically efficient surrogate-assisted calibration workflow. The improvement is not solely a speed-up in simulator evaluation; it is a compound gain obtained by combining a low-cost differentiable surrogate with a sampler that exploits posterior-gradient information.

In summary, Tables~\ref{tab:posterior_summary_all} and~\ref{tab:sampler_performance} establish two complementary conclusions. First, the posterior summaries confirm that the PCA--GPR surrogate and all tested samplers recover essentially the same calibrated distribution as the high-fidelity reference. Second, the performance metrics show that the surrogate enables orders-of-magnitude computational acceleration, while NUTS and especially MCLMC further increase statistical efficiency by reducing Markov-chain autocorrelation. Among the tested schemes, the combination PCA--GPR/MCLMC provides the most favorable balance between posterior fidelity, computational cost, and effective-sample generation.

The present study is conducted within the BO/IUQ framework proposed by~\cite{YAO2026140531}, where Bayesian optimization is used to identify high-posterior-density regions of the parameter space and to provide informed initial states for subsequent MCMC sampling. This initialization strategy is most valuable for gradient-free samplers applied to expensive black-box simulators because it places the chains closer to posterior modes and reduces the length of the initial exploration phase. For gradient-based samplers such as NUTS and MCLMC, however, BO initialization is less critical. These algorithms exploit gradients of the log-posterior and can therefore move efficiently toward high-probability regions even from less favorable initial states. The BO stage consequently provides its largest practical benefit for gradient-free black-box sampling, whereas for gradient-based surrogate sampling its additional cost should be weighed against the comparatively smaller reduction in burn-in.

\subsubsection{Effect of Data Size on Posterior Uncertainty and Identifiability}

\begin{figure}[t!]
    \centering

    \begin{subfigure}[b]{0.475\linewidth}
        \centering
        \includegraphics[width=\linewidth]{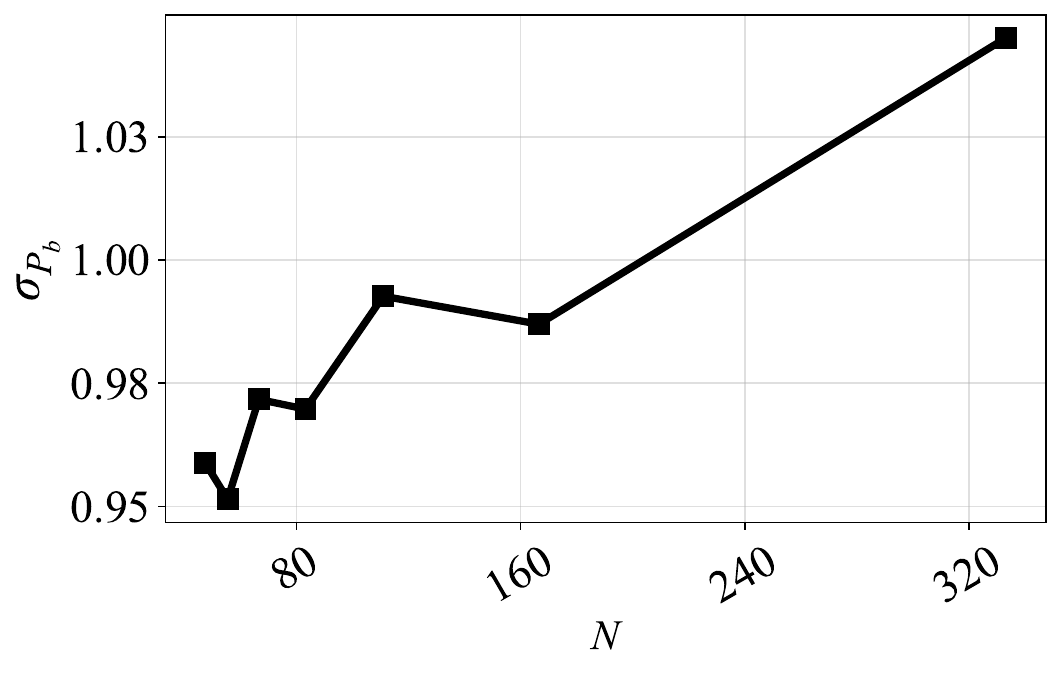}
        \caption{}
        \label{fig:datasize_Pb}
    \end{subfigure}
    \hfill
    \begin{subfigure}[b]{0.475\linewidth}
        \centering
        \includegraphics[width=\linewidth]{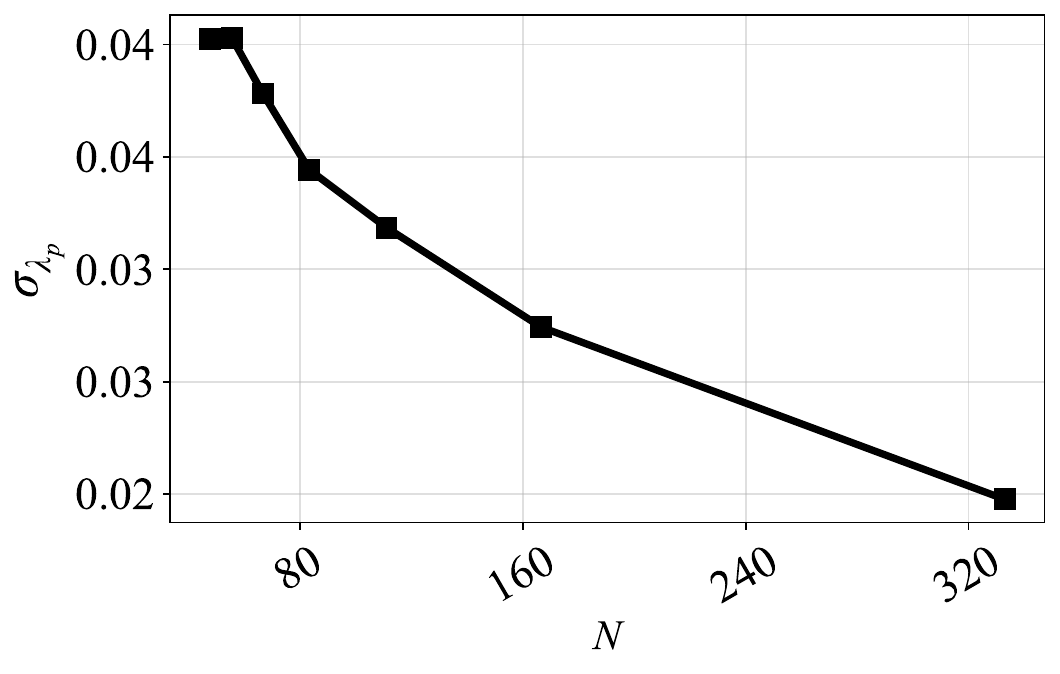}
        \caption{}
        \label{fig:datasize_lambdap}
    \end{subfigure}

    \vspace{0.25cm}
    \begin{subfigure}[b]{0.475\linewidth}
        \centering
        \includegraphics[width=\linewidth]{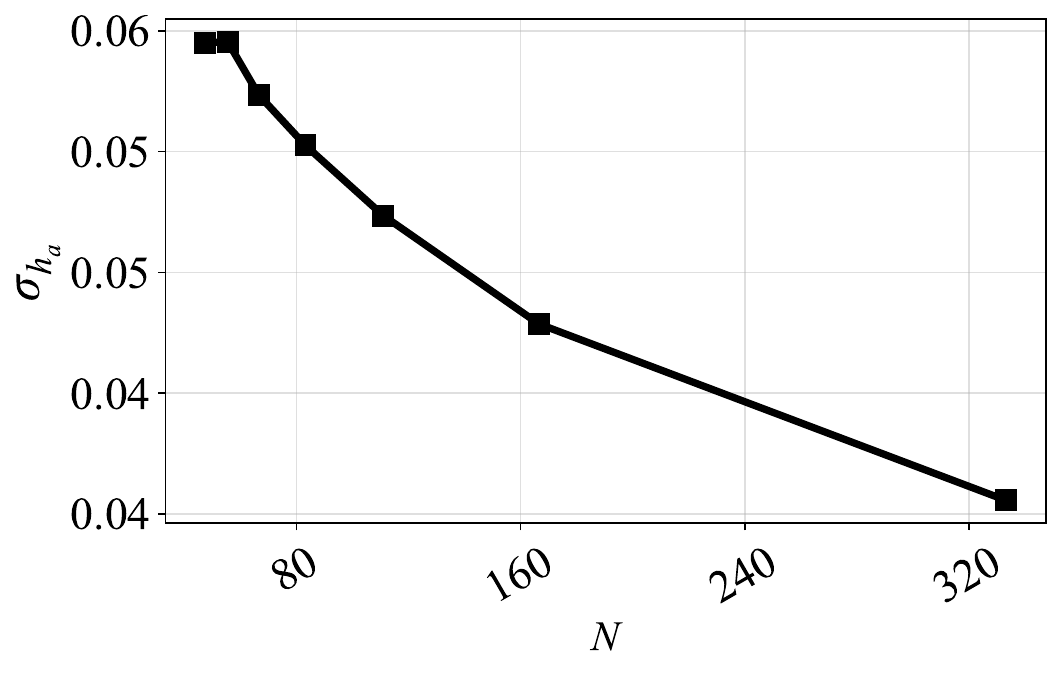}
        \caption{}
        \label{fig:datasize_ha}
    \end{subfigure}
    \hfill
    \begin{subfigure}[b]{0.475\linewidth}
        \centering
        \includegraphics[width=\linewidth]{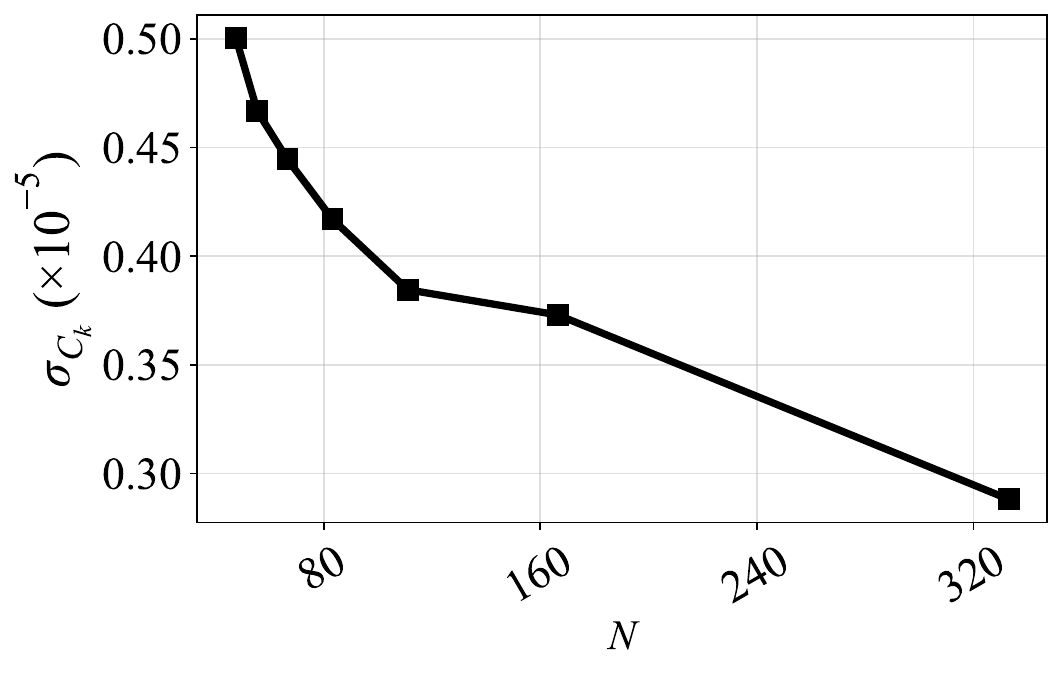}
        \caption{}
        \label{fig:datasize_Ck}
    \end{subfigure}

    \vspace{0.25cm}
    \begin{subfigure}[b]{0.5\linewidth}
        \centering
        \includegraphics[width=\linewidth]{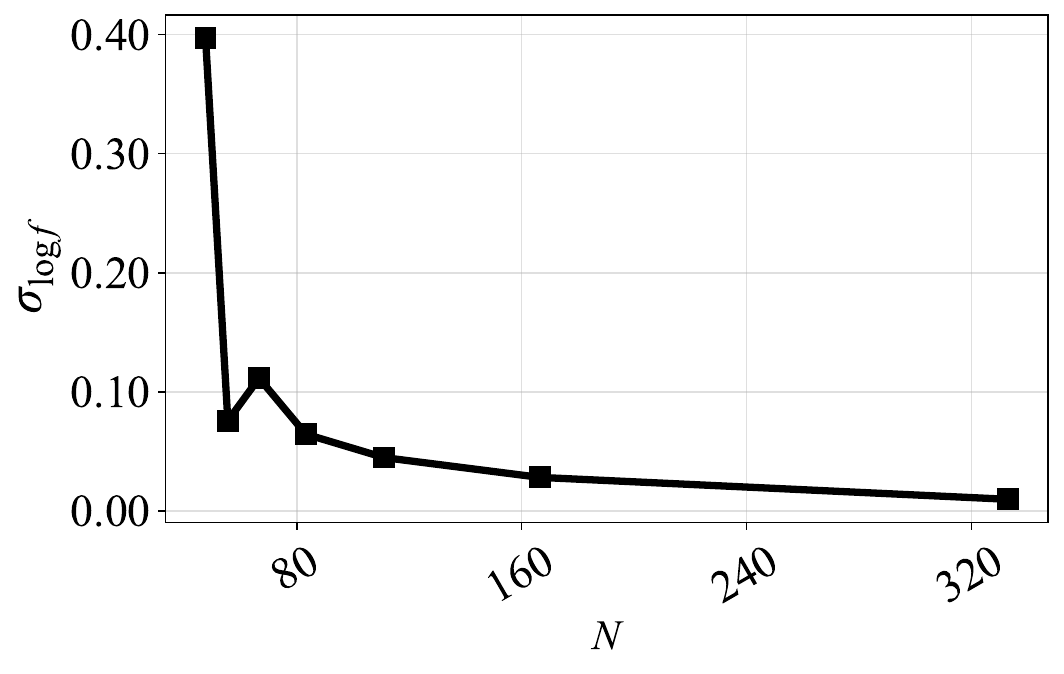}
        \caption{}
        \label{fig:datasize_overall}
    \end{subfigure}
    \caption{
        Effect of the number of assimilated transient response data points on posterior parameter uncertainty. Each panel reports the estimated posterior standard deviation as a function of the number of likelihood observations \(N\) for (a) power bias \(P_b\), (b) powder-gap thermal conductivity \(\lambda_p\), (c) ambient heat-transfer coefficient \(h_a\), (d) hydraulic resistance coefficient \(C_k\), and (e) logarithmic residual model-error scale \(\log f\). Decreasing standard deviation indicates posterior contraction and improved parameter identifiability as additional time-resolved experimental information is included.}
    \label{fig:datasize_three_rows}
\end{figure}

The amount of experimental information assimilated into the likelihood directly affects posterior concentration and parameter identifiability. To quantify this effect, Fig.~\ref{fig:datasize_three_rows} reports the posterior standard deviation of each calibrated quantity as a function of the number of experimental response data points included in the likelihood. The standard deviation is used here as a scalar measure of posterior uncertainty: decreasing values indicate that the data provide stronger information about the corresponding parameter, whereas nearly constant values indicate weak identifiability under the selected observation set.

When the likelihood is constructed from only a limited number of measurement points, the posterior distributions remain broad because the data contain insufficient temporal information to distinguish among correlated parameter combinations. As additional transient observations are assimilated, the likelihood becomes more concentrated and the posterior standard deviations generally decrease. This contraction demonstrates that the time-resolved response contains information that separates the effects of heat transfer, hydraulic resistance, and residual model discrepancy, thereby reducing epistemic uncertainty in the calibrated parameters.

Among the calibrated physical parameters, \(P_{\mathrm{b}}\) exhibits the weakest data-driven contraction. Its posterior standard deviation changes only modestly as the number of assimilated data points increases, indicating that the measured responses used in this calibration are not strongly sensitive to the global power-bias parameter over the considered transient window. In this setting, \(P_{\mathrm{b}}\) acts primarily as a compensating parameter: variations in power bias can be partially offset by changes in \(h_a\), \(C_k\), and \(\lambda_p\) while preserving similar temperature and mass-flow histories. This compensation is most pronounced for sparse datasets, where limited temporal information allows a wider range of correlated parameter combinations to reproduce the observed transient behavior.

The remaining physical parameters show clearer posterior contraction as the dataset expands. The decreasing standard deviations of \(h_a\), \(\lambda_p\), and \(C_k\) indicate that the additional temporal observations improve the separation of their distinct thermal--hydraulic roles. In particular, \(h_a\) controls heat exchange with the surroundings, \(\lambda_p\) affects conductive transport through the powder gap, and \(C_k\) influences the hydraulic resistance and hence the natural-circulation mass-flow response. Although these parameters remain correlated, the enriched time series provides enough information to constrain their individual contributions more tightly.

The behavior of \(\log f\) provides an additional diagnostic of model--data discrepancy. A reduction in its posterior standard deviation with increasing data size indicates that the residual error scale becomes more precisely inferred as the likelihood incorporates more observations. This does not imply that all model discrepancy is eliminated; rather, it indicates that the magnitude of the remaining unresolved discrepancy is more sharply quantified once the physical parameters have been calibrated against a sufficiently informative dataset.

Overall, Fig.~\ref{fig:datasize_three_rows} shows that posterior uncertainty is governed not only by the number of assimilated observations, but also by their sensitivity to each parameter. Increasing the data volume sharpens the likelihood and improves identifiability for parameters that leave distinct signatures in the measured QoIs. Parameters with weak direct sensitivity, such as \(P_{\mathrm{b}}\), remain only weakly informed unless the calibration dataset includes responses or transient regimes that more directly excite their physical influence. This result underscores an important practical implication for IUQ experiment design: adding more data is most valuable when the additional observations increase parameter sensitivity and reduce compensatory correlations, rather than merely increasing the number of statistically redundant time points.


\subsection{Prior-to-Posterior Predictive Assessment of Transient QoIs}

A forward uncertainty quantification (FUQ) analysis was first performed to establish the prior predictive baseline before assimilating experimental data. Samples of the uncertain physical parameters, \(P_b\), \(C_k\), \(h_a\), and \(\lambda_p\), were drawn from their engineering  prior distributions and propagated through the Flownex SE model to generate ensembles of transient predictions for the principal quantities of interest (QoIs): HS outlet temperature, HX inlet temperature, HX outlet temperature, and LBE mass flow rate. The resulting prior predictive distribution represents the uncertainty implied by the assumed parameter priors alone and therefore provides the reference against which the effect of Bayesian calibration can be assessed.

\begin{figure}[t!]
    \centering
    \includegraphics[width=1\linewidth]{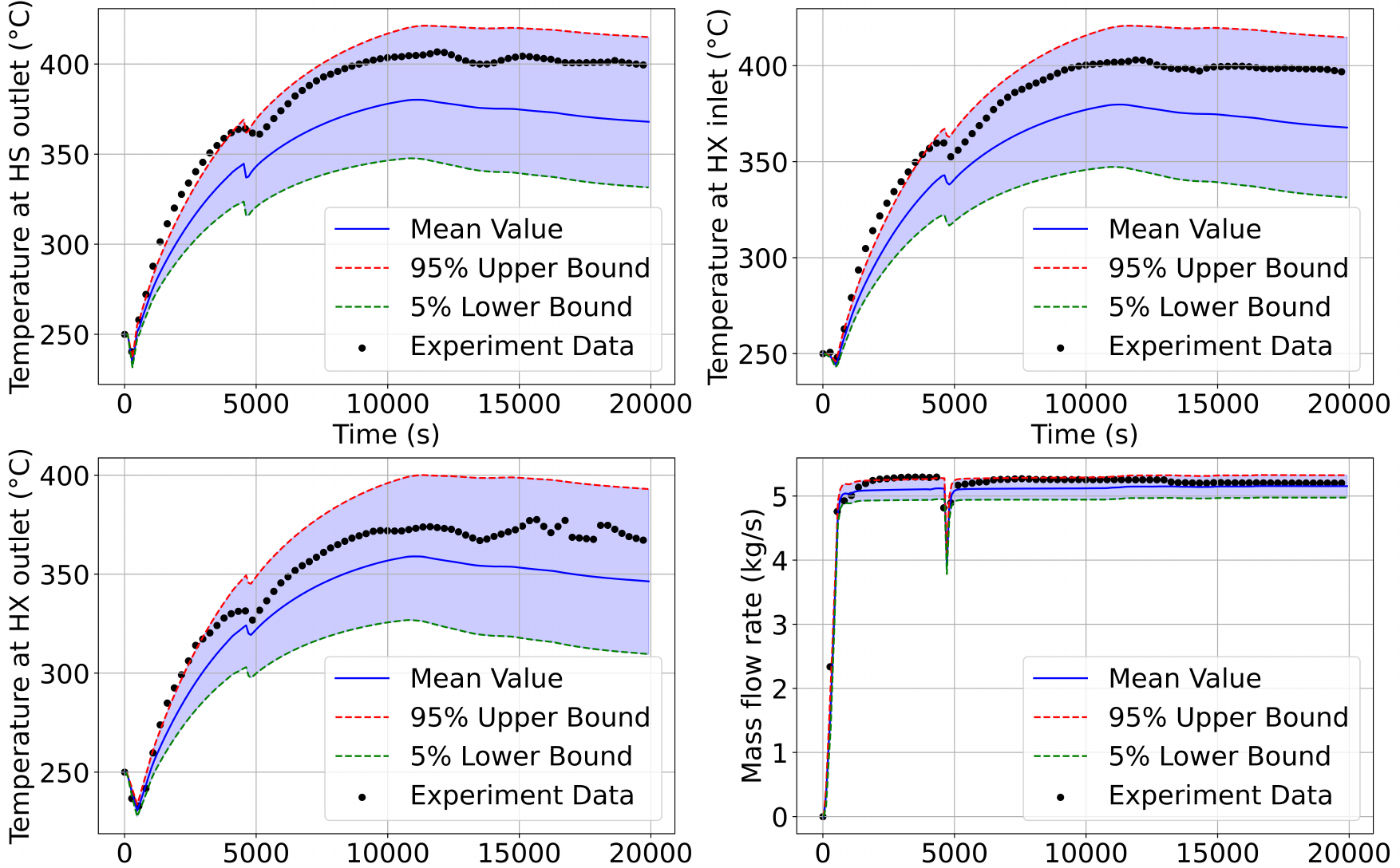}
    \caption{Prior predictive responses for the NACIE transient obtained by propagating samples from the prior parameter distribution \(p(\bm\theta\mid I)\) through the Flownex SE model. Experimental observations (black points) are compared with the prior predictive mean and 5th--95th percentile envelope for HS outlet temperature (top left), HX inlet temperature (top right), HX outlet temperature (bottom left), and LBE mass flow rate (bottom right). The broad intervals and visible mean deviations provide the baseline forward-UQ assessment before Bayesian calibration.}
    \label{fig:placeholderbase}
\end{figure}

Figure~\ref{fig:placeholderbase} shows that the prior predictive intervals are broad enough to cover many experimental observations, but the prior mean responses exhibit systematic bias. In particular, the temperature responses at the HS and HX outlets are underpredicted over substantial portions of the transient, and the predictive envelopes remain wide even where the experimental trend is well defined. This behavior is typical of FUQ based only on engineering priors: enlarging prior uncertainty can increase coverage, but it does not identify which parameter combinations are physically consistent with the observed transient. The prior ensemble therefore provides uncertainty coverage at the expense of predictive sharpness and does not fully correct model bias.

\begin{figure}[t!]
    \centering
    \includegraphics[width=1\linewidth]{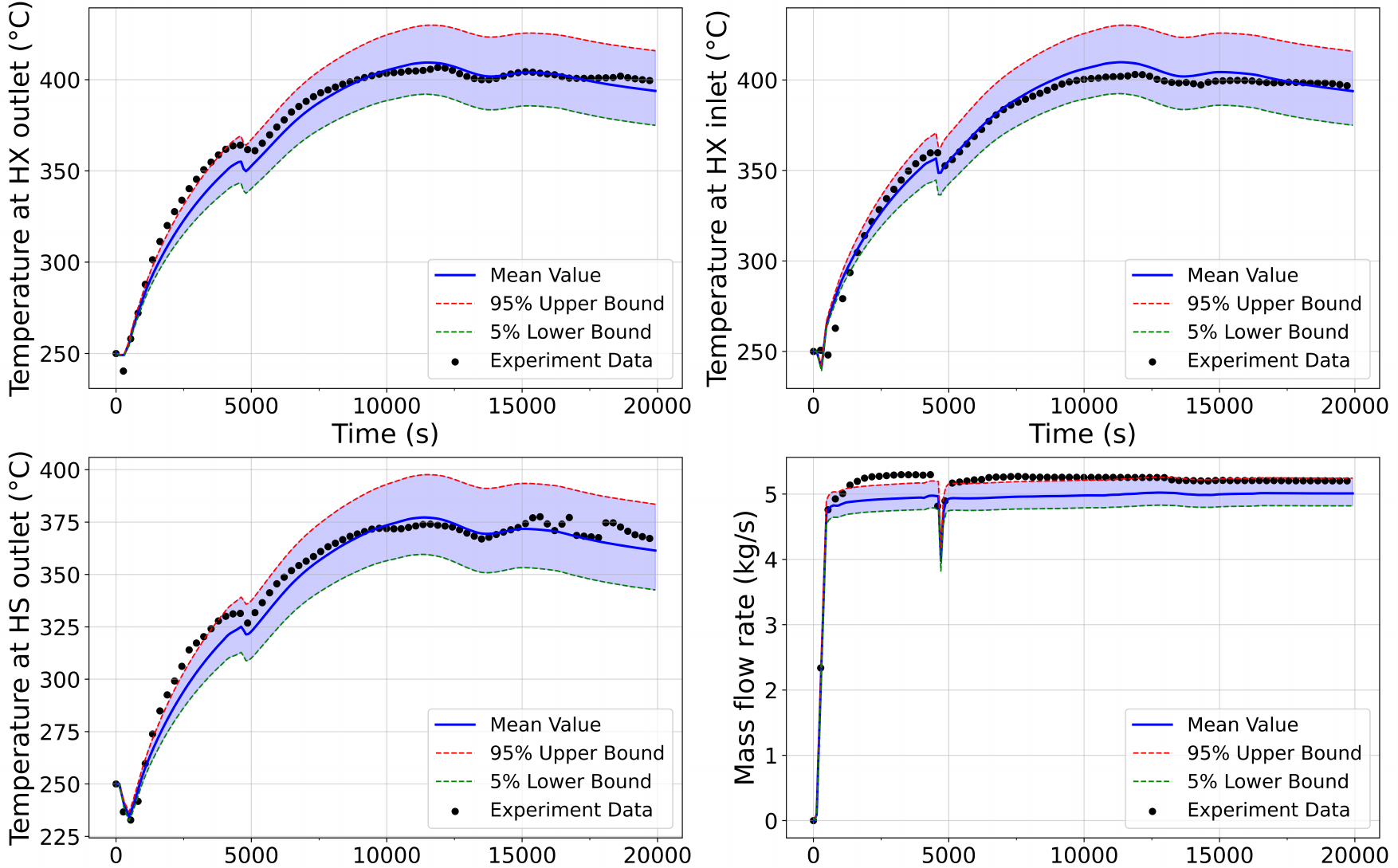}
    \caption{Posterior predictive responses for the NACIE transient after Bayesian inverse uncertainty quantification. Experimental observations (black points) are compared with the posterior predictive mean and 5th--95th percentile envelope obtained by propagating MCLMC posterior samples of \(\bm\theta\) through the predictive model. Panels show HS outlet temperature (top left), HX inlet temperature (top right), HX outlet temperature (bottom left), and LBE mass flow rate (bottom right), illustrating the bias reduction and uncertainty contraction achieved after calibration.}
    \label{fig:iuq_update}
\end{figure}

Bayesian inverse uncertainty quantification was then used to condition the uncertain parameters on the experimental transient data. Posterior samples were propagated through the predictive model to obtain posterior predictive distributions, shown in Fig.~\ref{fig:iuq_update}. The solid curves represent posterior predictive means, and the shaded or percentile bands represent the corresponding 5th--95th percentile intervals. The predictions in Fig.~\ref{fig:iuq_update} were evaluated at time locations not used directly in the likelihood construction: calibration data were assimilated at 66~s intervals, whereas the plotted posterior predictions were evaluated at intermediate 33~s locations excluding the calibration points. This separation provides a more stringent assessment of interpolation capability within the transient response, because the plotted comparisons are not merely reproductions of the data points used to construct the likelihood.

Comparison of Figs.~\ref{fig:placeholderbase} and~\ref{fig:iuq_update} demonstrates the effect of data assimilation. The posterior predictive means track the experimental trajectories more closely than the prior means for all four calibration QoIs, while the credible intervals contract substantially. This simultaneous reduction in bias and uncertainty indicates that the experimental data constrain physically relevant combinations of heat-transfer, hydraulic-resistance, and power-bias parameters. Importantly, the posterior intervals remain finite rather than collapsing to the observations, reflecting the continued presence of measurement uncertainty, residual model discrepancy, and parameter non-identifiability. The posterior predictive distribution therefore provides a sharper but still probabilistic representation of the transient system response.

\begin{figure}[t!]
    \centering
    \begin{subfigure}[b]{0.48\linewidth}
        \centering
        \includegraphics[width=\linewidth]{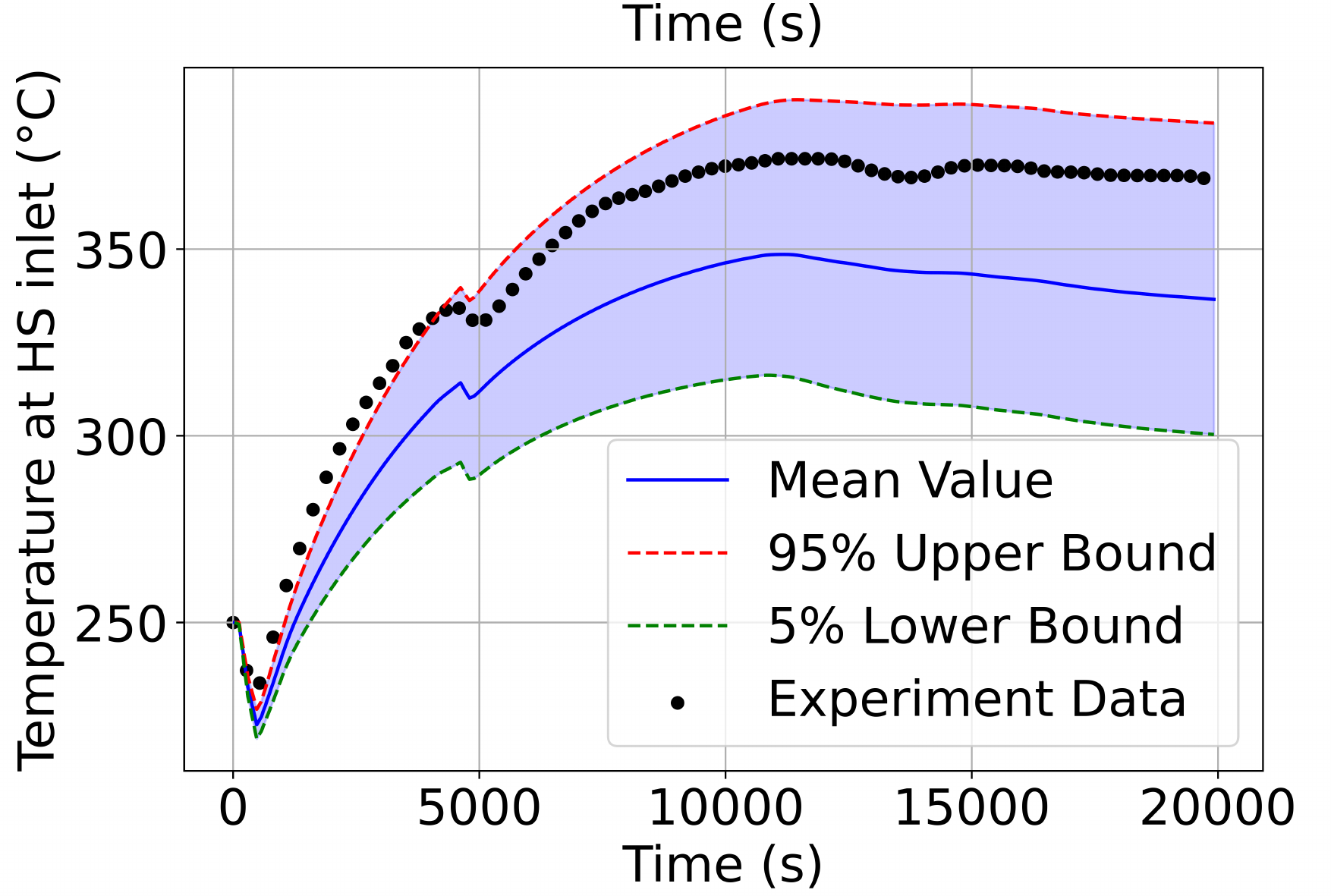}
        \caption{Without IUQ correction}
        \label{fig:nouq-v}
    \end{subfigure}
    \hfill
    \begin{subfigure}[b]{0.48\linewidth}
        \centering
        \includegraphics[width=\linewidth]{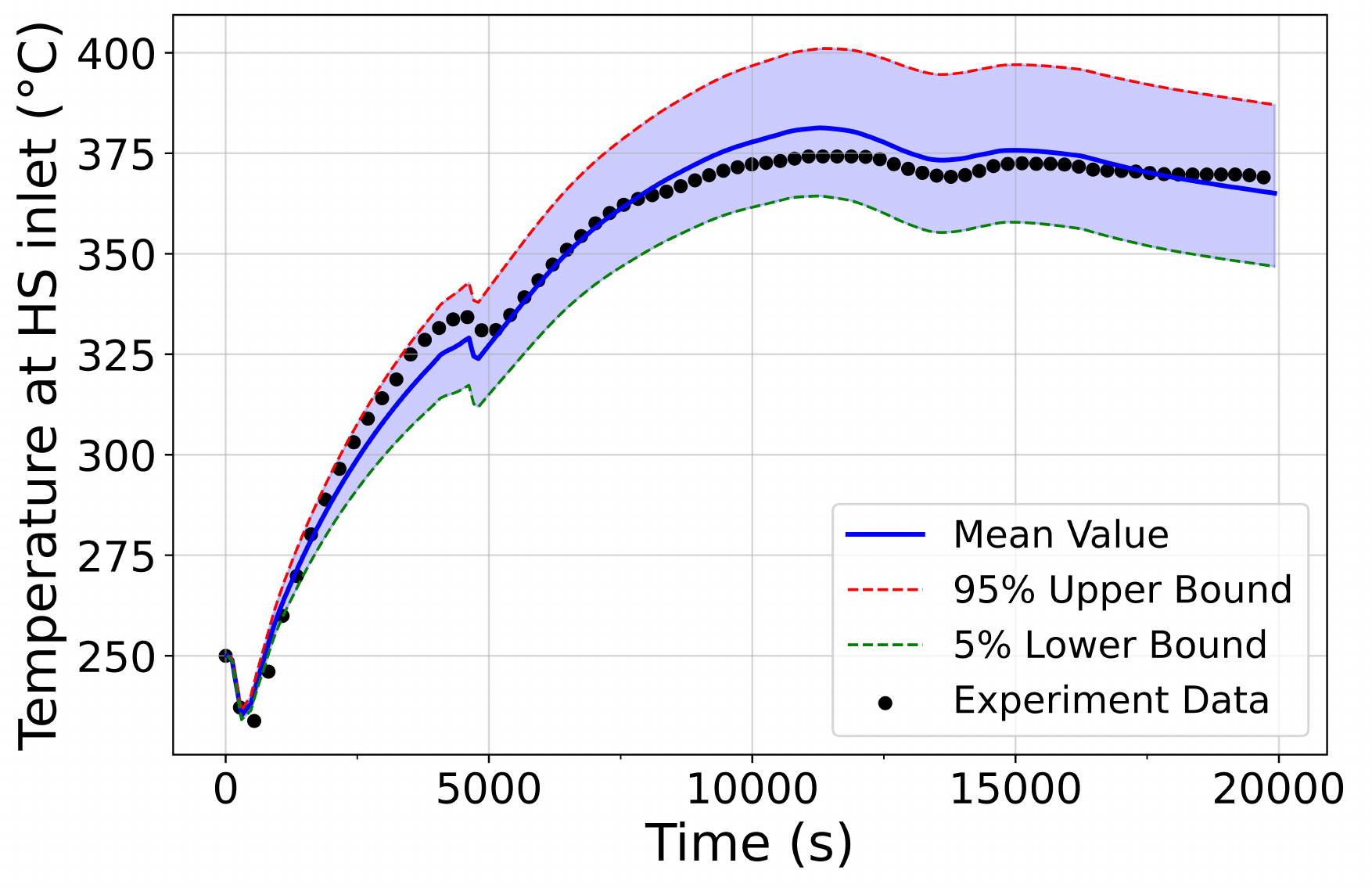}
        \caption{With IUQ correction}
        \label{fig:iuq-v}
    \end{subfigure}
    
    \caption{Prior (left) and posterior (right) predictive distributions for the HS inlet temperature, an output not used directly as a calibration target. Observed data (black dots) are compared with the predictive mean (solid blue line) and the 5th--95th percentile interval for the prior and posterior predictive distributions. Posterior samples were obtained using the MCLMC scheme.}
    \label{fig:iuq-validation}
\end{figure}

Figure~\ref{fig:iuq-validation} evaluates whether the calibrated parameters improve prediction of an output response that was not directly assimilated in the likelihood. The calibration used HS outlet temperature, HX inlet temperature, HX outlet temperature, and LBE mass flow rate, whereas the validation response in Fig.~\ref{fig:iuq-validation} is the HS inlet temperature. The prior prediction captures the broad scale of the transient but exhibits wider uncertainty and visible deviations from the measured trajectory. After IUQ, the posterior predictive mean aligns more closely with the HS inlet temperature measurements, and the 90\% credible interval narrows by up to approximately 60\%. This result indicates that the posterior parameters are not merely overfitting the four calibration QoIs; rather, they improve the internal consistency of the thermal--hydraulic model and transfer predictive information to an uncalibrated but dynamically coupled response.

\begin{figure}[t!]
    \centering
    \begin{subfigure}[b]{0.48\linewidth}
        \centering
        \includegraphics[width=\linewidth]{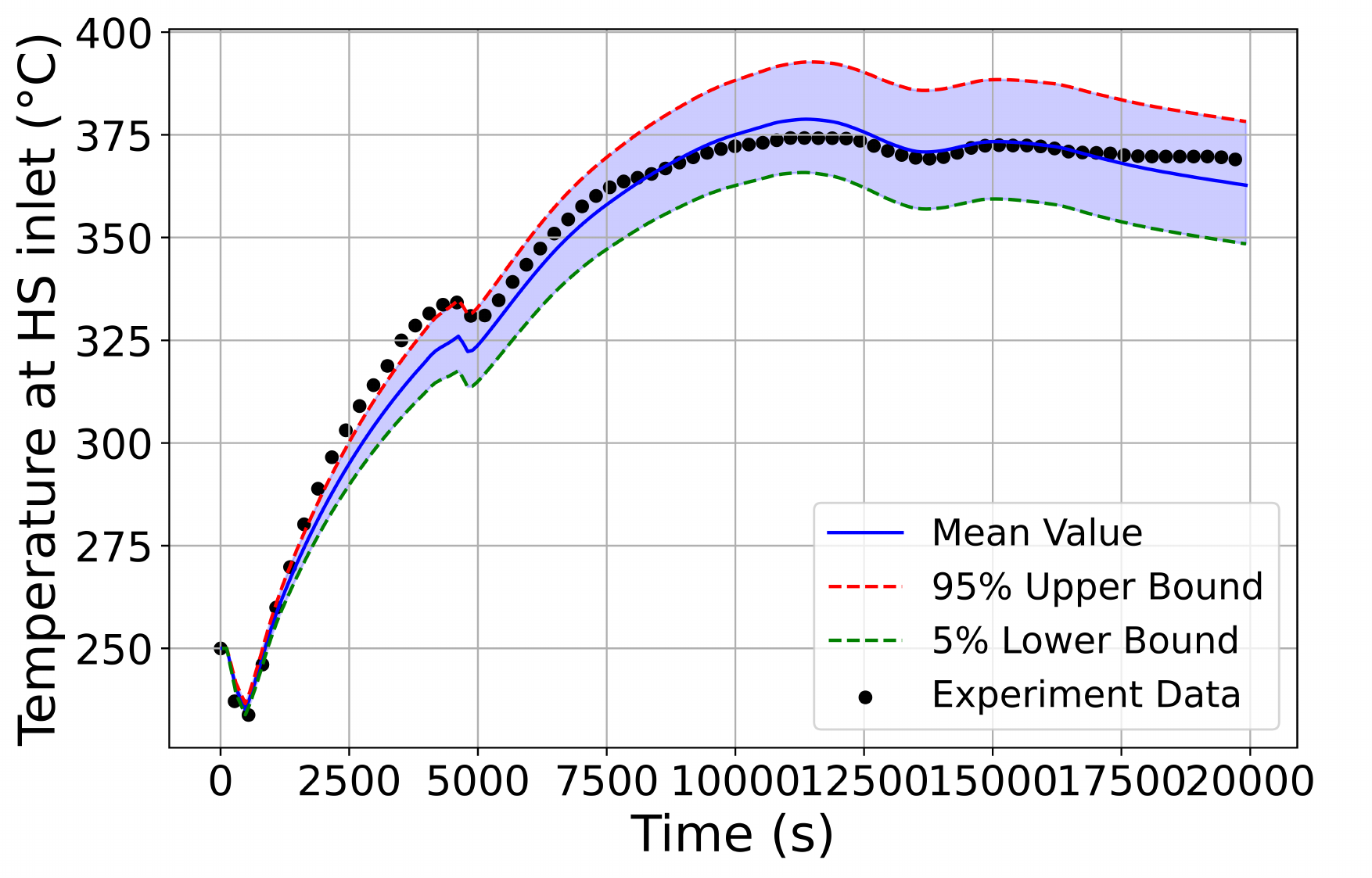}
        \caption{Flownex SE model}
        \label{fig:flownex_hs_in}
    \end{subfigure}
    \hfill
    \begin{subfigure}[b]{0.48\linewidth}
        \centering
        \includegraphics[width=\linewidth]{pic/IUQ-PCA-V.pdf}
        \caption{PCA--GPR model}
        \label{fig:PCA}
    \end{subfigure}
    
    \caption{Posterior predictive comparison for the HS inlet temperature using the high-fidelity Flownex SE/ENSS (left) and the PCA--GPR/MCLMC (right) combinations. Experimental observations (black dots) are compared with the posterior predictive mean (solid blue line) and 5th--95th percentile envelope in each panel. The close agreement between the two posterior predictive distributions indicates that the surrogate preserves the calibrated transient response and uncertainty structure of the reference simulator for this validation QoI.}
    \label{fig:model_comparison}
\end{figure}

Figure~\ref{fig:model_comparison} compares the posterior predictive HS inlet temperature obtained from the high-fidelity Flownex SE/ENSS and PCA--GPR/MCLMC combinations. The two model/sampler schemes produce closely aligned mean responses and uncertainty intervals, confirming that the surrogate preserves the calibrated predictive behavior of the reference simulator for this validation QoI. During the early portion of the transient, approximately 0--5000~s, the surrogate-based interval encloses the observations slightly more consistently. This difference should be interpreted cautiously: it does not imply that the surrogate is more physically accurate than the high-fidelity model, but rather that surrogate approximation, posterior uncertainty, and smoothing in the reduced PCA representation can produce a slightly broader or more regular posterior predictive envelope. The key result is that the surrogate maintains agreement with the high-fidelity posterior while achieving the computational acceleration needed for repeated Bayesian prediction.

To further assess generalization beyond the calibrated transient, an additional validation case was considered under a steady-state operating condition not included in the calibration dataset. The boundary conditions differed from those used in the transient calibration: the heating power was increased to 25~kW, the secondary-side HX cooling-water flow rate was set to 0.55~kg~s\(^{-1}\), and the inlet temperature was fixed at 25~\(^{\circ}\)C. Posterior parameter samples from the calibration stage (obtained from ENSS when using Flownex SE and from MCLMC when using PCA--GPR) were propagated without additional tuning to predict the steady-state LBE mass flow rate, and the resulting predictive distribution was compared with the corresponding experimental measurement.

\begin{figure}[t!]
    \centering
    \includegraphics[width=1\linewidth]{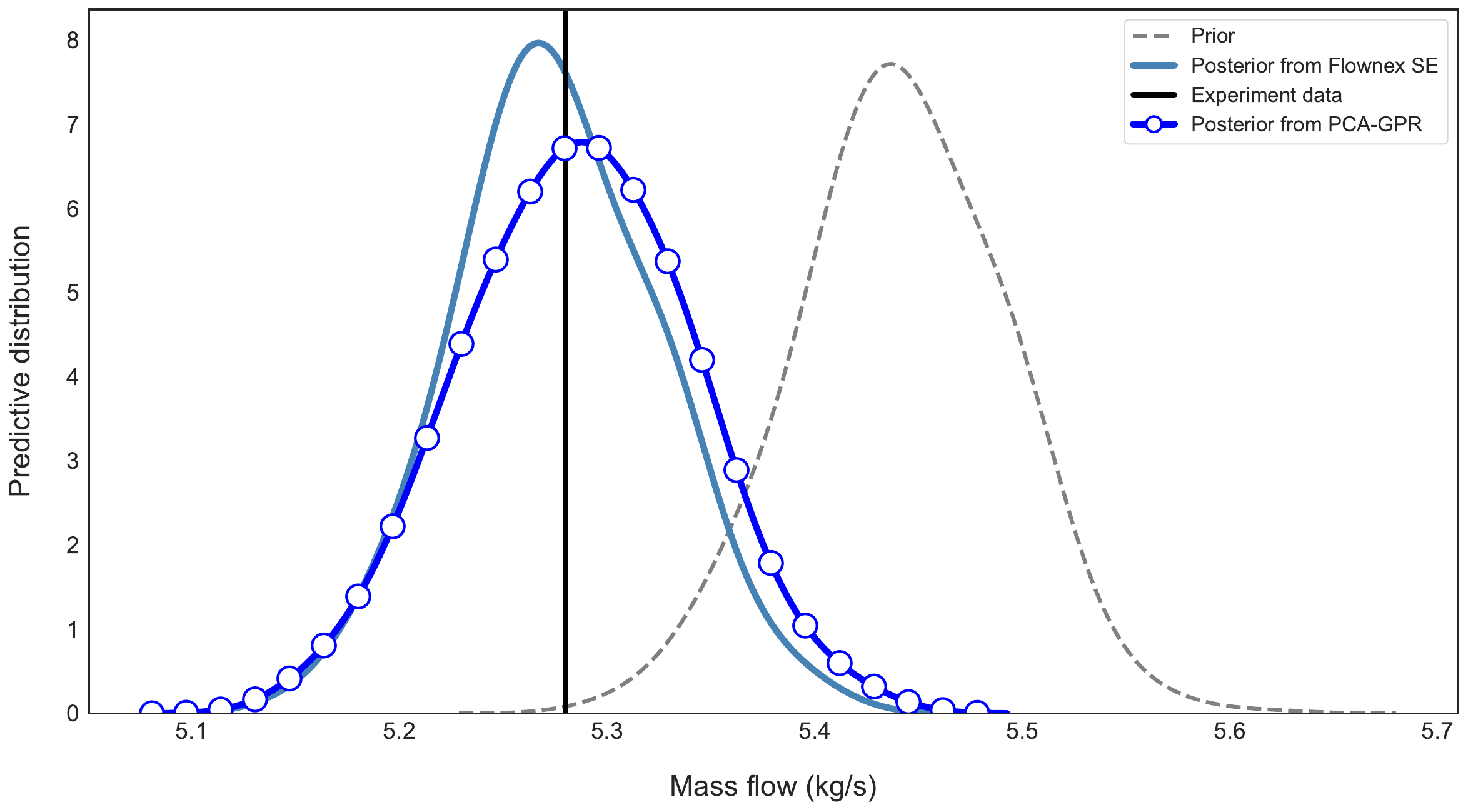}
    \caption{Prior and posterior predictive distributions for the steady-state LBE mass flow rate under an independent operating condition not used for calibration. The comparison shows how Bayesian updating shifts the predictive distribution from the prior forward-UQ baseline toward the measured mass flow rate, providing an out-of-sample validation of the calibrated parameter distribution and PCA--GPR-assisted IUQ workflow.}
    \label{FIG:steady_val}
\end{figure}

Figure~\ref{FIG:steady_val} shows that the prior predictive distribution substantially overestimates the measured steady-state mass flow rate; the experimental value lies near the lower tail of the prior distribution and outside its 90\% credible interval. After Bayesian calibration, the posterior predictive distribution obtained with Flownex SE shifts toward the experimental measurement, with the observation located near the central region of the posterior distribution. The PCA--GPR posterior predictive distribution follows the same shift and remains centered close to the experimental value, although it is slightly broader than the Flownex SE posterior. This modest broadening is acceptable and even conservative for safety-oriented UQ because it reflects residual surrogate and calibration uncertainty while preserving the calibrated mean response.

These predictive comparisons highlight the limitation of forward UQ when prior information is not sufficiently informative. A broad prior may increase the probability of covering observations, but it cannot by itself identify the physically plausible parameter combinations responsible for the measured response. In contrast, IUQ uses the likelihood to reweight parameter values according to their consistency with experimental data. The resulting posterior predictive distribution is therefore both better centered and more sharply constrained, while retaining uncertainty associated with measurement noise, model discrepancy, and imperfect parameter identifiability.

In summary, the posterior predictive analyses demonstrate that Bayesian calibration improves both calibrated and uncalibrated QoIs. The IUQ-cal\-i\-brated parameters reduce systematic bias, narrow predictive intervals, and improve agreement with experimental measurements for transient temperatures, transient mass flow rate, and an independent steady-state mass-flow validation case. The PCA--GPR surrogate reproduces the high-fidelity posterior predictive behavior with a slightly conservative uncertainty envelope, supporting its use as an efficient replacement for repeated Flownex SE evaluations in full-transient IUQ. Overall, these results confirm that the proposed surrogate-assisted Bayesian framework provides not only computational acceleration, but also improved predictive reliability for system-level nuclear thermal--hydraulic analysis.

\FloatBarrier   

\section{Conclusions}
\label{sec:conclusion}

This study developed and assessed a surrogate-assisted Bayesian inverse uncertainty quantification (IUQ) framework for full-transient calibration of the NACIE natural-circulation loop. The central objective was to make system-level Bayesian calibration computationally tractable without discarding the time-dependent information contained in the measured thermal--hydraulic responses. To this end, a PCA--GPR surrogate was constructed to emulate the high-fidelity Flownex SE simulator, and the surrogate was embedded in a Bayesian workflow combining forward UQ, IUQ, Bayesian-optimization-based initialization, and comparative MCMC sampling. The resulting framework provides a data-informed route for calibrating uncertain physical parameters in integral thermal--hydraulic transients while retaining full posterior uncertainty information.

The forward UQ analysis established the limitations of relying exclusively on prescribed prior parameter distributions. Although the prior predictive ensembles captured the broad qualitative trends of the NACIE transient, they produced wide uncertainty bands and systematic deviations in the mean responses. This behaviour shows that prior uncertainty propagation alone can provide coverage without necessarily providing predictive sharpness or bias correction. By contrast, Bayesian IUQ updated the prior distribution to the posterior distribution \(p(\bm{\theta}\mid D,I)\), thereby incorporating experimental evidence directly through the likelihood. The posterior distributions were more concentrated for identifiable parameters and revealed which quantities remained weakly informed by the available data, most notably the power-bias parameter \(P_b\). This distinction is important for reactor-safety applications because it separates uncertainty that can be reduced through calibration from uncertainty that requires more informative experiments or additional observables.

The PCA--GPR surrogate reproduced the dominant high-fidelity posterior structure while reducing the computational burden by approximately \(3000\)-fold relative to direct Flownex SE evaluations. Posterior summaries, marginal distributions, and bivariate correlation structures obtained with the surrogate remained consistent with those obtained using the Flownex SE/ENSS reference. This agreement demonstrates that the surrogate preserves not only the mean simulator response, but also the likelihood geometry required for Bayesian inference. The surrogate therefore enables repeated posterior evaluations, posterior predictive propagation, and sampler comparisons that would otherwise be prohibitively expensive for a full-transient system model.

The sampler comparison showed that computational acceleration and statistical efficiency are distinct but complementary requirements. Replacing Flownex SE with the PCA--GPR surrogate primarily reduced the cost of each likelihood evaluation, whereas gradient-based samplers reduced the number of correlated samples required to characterize the posterior. Among the evaluated algorithms, NUTS and MCLMC achieved substantially faster autocorrelation decay and larger effective sample sizes than the gradient-free ENSS sampler. MCLMC provided the most favorable overall performance, combining rapid decorrelation with low time per effective sample. At the same time, ENSS remained a robust gradient-free option and provided a useful reference for black-box settings where gradients are unavailable. These results indicate that differentiable surrogate models are especially valuable because they make gradient-informed Bayesian calibration practical for complex nuclear thermal--hydraulic systems.

The posterior predictive analyses confirmed that Bayesian calibration improved both calibrated and uncalibrated QoIs. Relative to prior predictions, the posterior predictive distributions showed reduced systematic bias, narrower credible intervals, and closer agreement with experimental measurements for transient temperatures and LBE mass flow rate. The calibrated parameters also improved prediction of the HS inlet temperature, which was not directly included as a calibration target, indicating that the posterior update improved the internal consistency of the coupled thermal--hydraulic model rather than merely fitting the assimilated outputs. In addition, validation under an independent steady-state operating condition showed that the posterior predictive mass-flow distribution shifted toward the experimental measurement, whereas the prior distribution substantially overestimated the observed value. These results support the generalizability of the calibrated parameter distributions across related operating conditions.

The data-size analysis further demonstrated that posterior contraction depends not only on the number of observations, but also on their sensitivity to each parameter. Increasing the number of assimilated transient data points reduced posterior uncertainty for \(h_a\), \(\lambda_p\), \(C_k\), and \(\log f\), indicating improved identifiability of heat-transfer, hydraulic-resistance, and residual-discrepancy effects. In contrast, \(P_b\) remained weakly constrained because the selected QoIs did not strongly excite its influence over the considered transient window. This finding has direct implications for experimental design: additional data are most valuable when they reduce compensatory parameter correlations and provide sensitivity to weakly identifiable mechanisms, rather than when they simply add redundant time points to an already informative response.

Overall, the proposed PCA--GPR-assisted Bayesian IUQ framework provides a practical pathway for data-informed calibration of system-level nuclear thermal--hydraulic models. The framework combines the physical interpretability of a high-fidelity simulator, the computational efficiency of a reduced-order probabilistic surrogate, and the rigor of posterior uncertainty quantification. For best-estimate-plus-uncertainty and risk-informed safety analysis, this capability is significant because it enables transient model calibration, uncertainty reduction, and posterior predictive assessment within a computational budget compatible with engineering analysis.

Several avenues for future work follow naturally from the present study. First, the framework should be extended to multiple transients and broader operating envelopes so that posterior parameter distributions can be tested for transferability across different power levels, cooling conditions, and nat\-u\-ral-circulation regimes. Second, future calibration should incorporate additional measurements that are more directly sensitive to weakly identifiable parameters such as \(P_b\), thereby reducing compensatory correlations and improving parameter interpretability. Third, surrogate uncertainty could be incorporated explicitly into the likelihood to distinguish emulator error from experimental noise and physical model discrepancy. Fourth, hierarchical or multi-fidelity Bayesian formulations could combine separate-effect, component-scale, and integral-effect data within a single calibration structure. Finally, the gradient-based sampling strategies evaluated here should be explored for larger parameter spaces and more complex reactor transients, where posterior geometry may be more nonlinear, multimodal, or strongly correlated. Addressing these issues would further strengthen the role of surrogate-assisted IUQ as a quantitative foundation for predictive reactor safety assessment.

\section*{CRediT authorship contribution statement}
Zikang Yao: Writing - original draft, Formal analysis, Methodology, Investigation, Visualization, Data curation.  
Eugene Yee: Supervision, Conceptualization, Visualization, Methodology, Writing - review, revision \& editing, Resources, Validation.  
Fue-Sang Lien: Funding acquisition, Project administration, Supervision, Writing - review \& editing.

\section*{Declaration of competing interest}
The authors declare that they have no known competing financial interests or personal relationships that could have appeared to influence the work reported in this paper.

\section*{Acknowledgments}
We acknowledge the support of the Natural Sciences and Engineering Research Council of Canada (NSERC) and Canadian Nuclear Safety Commission (CNSC), [funding reference number ALLRP 580447-2022].

\section*{Data availability}
The authors do not have permission to share data.

\bibliographystyle{elsarticle-num}   

\bibliography{references}

\end{document}